\documentclass[twocolumn]{aastex631}

\usepackage{diagbox}
\usepackage{graphicx}
\usepackage{natbib}
\usepackage{hyperref}

\usepackage{xspace}

\usepackage{txfonts}

\DeclareRobustCommand{\ION}[2]{%
\relax\ifmmode
\ifx\testbx\f@series
{\mathbf{#1\,\mathsc{#2}}}\else
{\mathrm{#1\,\mathsc{#2}}}\fi
\else\textup{#1\,{\mdseries\textsc{#2}}}%
\fi}

\newcommand{\hi}{\ION{H}{i}}
\newcommand{\hii}{\ION{H}{ii}}

\newcommand{\nii}{[\ION{N}{ii}]}
\newcommand{\oi}{[\ION{O}{i}]}
\newcommand{\oii}{[\ION{O}{ii}]}
\newcommand{\oiii}{[\ION{O}{iii}]}
\newcommand{\sii}{[\ION{S}{ii}]}
\newcommand{\siii}{[\ION{S}{iii}]}
\newcommand{\caii}{[\ION{Ca}{ii}]}
\newcommand{\feii}{[\ION{Fe}{ii}]}
\newcommand{\ariii}{[\ION{Ar}{iii}]}
\newcommand{\neiii}{[\ION{Ne}{iii}]}

\newcommand{\ha}{$\rm{H}\alpha$}
\newcommand{\hb}{$\rm{H}\beta$}
\newcommand{\hd}{$\rm{H}\delta$}
\newcommand{\hg}{$\rm{H}\gamma$}

\newcommand{\Ha}{$\rm{H}\alpha$}
\newcommand{\Hb}{$\rm{H}\beta$}

\newcommand{\kms}{km\,s$^{-1}$}

\def\pyf{\texttt{pyFIT3D}\xspace}
\def\pyp{\texttt{pyPipe3D}\xspace}

\begin{document}

\title{The SDSS-V Local Volume Mapper (LVM): Data Analysis Pipeline}

\newcommand{\OSU}{\label{OSU} Department of Astronomy, The Ohio State University, 140 West 18th Avenue, Columbus, Ohio 43210, USA}

\newcommand{\Alberta}{\label{Alberta} Department of Physics, University of Alberta, Edmonton, AB T6G 2E1, Canada}

\newcommand{\ANU}{\label{ANU} Research School of Astronomy and Astrophysics, Australian National University, Canberra, ACT 2611, Australia}

\newcommand{\IPAC}{\label{IPAC} Caltech-IPAC, 1200 E. California Blvd. Pasadena, CA 91125, USA}

\newcommand{\Carnegie}{\label{Carnegie} Observatories of the Carnegie Institution for Science, 813 Santa Barbara Street, Pasadena, CA 91101, USA}

\newcommand{\CCAPP}{\label{CCAPP} Center for Cosmology and Astroparticle Physics, 191 West Woodruff Avenue, Columbus, OH 43210, USA}

\newcommand{\CfA}{\label{CfA}Harvard-Smithsonian Center for Astrophysics, 60 Garden Street, Cambridge, MA 02138, USA}

\newcommand{\CITEVA}{\label{CITEVA} Centro de Astronomía (CITEVA), Universidad de Antofagasta, Avenida Angamos 601, Antofagasta, Chile}

\newcommand{\CNRS}{\label{CNRS} CNRS, IRAP, 9 Av. du Colonel Roche, BP 44346, F-31028 Toulouse cedex 4, France}

\newcommand{\ESO}{\label{ESO} European Southern Observatory, Karl-Schwarzschild Stra{\ss}e 2, D-85748 Garching bei M\"{u}nchen, Germany}

\newcommand{\ESOChile}{\label{ESOChile} European Southern Observatory, Avenida Alonso de Cordoba 3107, Casilla 19, Santiago 19001, Chile}

\newcommand{\HD}{\label{HD} Astronomisches Rechen-Institut, Zentrum f\"{u}r Astronomie der Universit\"{a}t Heidelberg, M\"{o}nchhofstra\ss e 12-14, D-69120 Heidelberg, Germany}

\newcommand{\ICRAR}{\label{ICRAR} International Centre for Radio Astronomy Research, University of Western Australia, 35 Stirling Highway, Crawley, WA 6009, Australia}

\newcommand{\IRAM}{\label{IRAM} Institut de Radioastronomie Millim\'{e}trique (IRAM), 300 Rue de la Piscine, F-38406 Saint Martin d'H\`{e}res, France}

\newcommand{\ITA}{\label{ITA} Universit\"{a}t Heidelberg, Zentrum f\"{u}r Astronomie, Institut f\"{u}r Theoretische Astrophysik, Albert-Ueberle-Str 2, D-69120 Heidelberg, Germany}

\newcommand{\IWR}{\label{IWR} Universit\"{a}t Heidelberg, Interdisziplin\"{a}res Zentrum f\"{u}r Wissenschaftliches Rechnen, Im Neuenheimer Feld 205, D-69120 Heidelberg, Germany}

\newcommand{\JHU}{\label{JHU} Department of Physics and Astronomy, The Johns Hopkins University, Baltimore, MD 21218, USA}

\newcommand{\Leiden}{\label{Leiden} Leiden Observatory, Leiden University, P.O. Box 9513, 2300 RA Leiden, The Netherlands}

\newcommand{\Maryland}{\label{Maryland} Department of Astronomy, University of Maryland, College Park, MD 20742, USA}

\newcommand{\MPE}{\label{MPE} Max-Planck-Institut f\"{u}r extraterrestrische Physik, Giessenbachstra{\ss}e 1, D-85748 Garching, Germany}

\newcommand{\MPIA}{\label{MPIA} Max-Planck-Institut f\"{u}r Astronomie, K\"{o}nigstuhl 17, D-69117, Heidelberg, Germany}

\newcommand{\Nagoya}{\label{Nagoya} Department of Physics, Nagoya University, Furo-cho, Chikusa-ku, Nagoya, Aichi 464-8602, Japan}

\newcommand{\NRAO}{\label{NRAO} National Radio Astronomy Observatory, 520 Edgemont Road, Charlottesville, VA 22903-2475, USA}

\newcommand{\OAN}{\label{OAN} Observatorio Astron\'{o}mico Nacional (IGN), C/Alfonso XII, 3, E-28014 Madrid, Spain}

\newcommand{\ObsParis}{\label{ObsParis} Sorbonne Universit\'{e}, Observatoire de Paris, Universit\'{e} PSL, CNRS, LERMA, F-75014, Paris, France}

\newcommand{\Princeton}{\label{Princeton} Department of Astrophysical Sciences, Princeton University, Princeton, NJ 08544 USA}

\newcommand{\UToledo}{\label{UToledo} University of Toledo, 2801 W. Bancroft St., Mail Stop 111, Toledo, OH, 43606}

\newcommand{\Toulouse}{\label{Toulouse} Universit\'{e} de Toulouse, UPS-OMP, IRAP, F-31028 Toulouse cedex 4, France}

\newcommand{\UBonn}{\label{UBonn} Argelander-Institut f\"ur Astronomie, Universit\"at Bonn, Auf dem H\"ugel 71, 53121 Bonn, Germany}

\newcommand{\UChile}{\label{UChile} Departamento de Astronom\'{i}a, Universidad de Chile, Camino del Observatorio 1515, Las Condes, Santiago, Chile}

\newcommand{\UConn}{\label{UConn} Department of Physics, University of Connecticut, Storrs, CT, 06269, USA}

\newcommand{\UCSD}{\label{UCSD}Center for Astrophysics and Space Sciences, Department of Physics,  University of California, San Diego, 9500 Gilman Drive, La Jolla, CA 92093, USA}

\newcommand{\UGent}{\label{UGent} Sterrenkundig Observatorium, Universiteit Gent, Krijgslaan 281 S9, B-9000 Gent, Belgium}

\newcommand{\ULyon}{\label{ULyon} Univ Lyon, Univ Lyon 1, ENS de Lyon, CNRS, Centre de Recherche Astrophysique de Lyon UMR5574,\\ F-69230 Saint-Genis-Laval, France}

\newcommand{\UMass}{\label{UMass} University of Massachusetts—Amherst, 710 N. Pleasant Street, Amherst, MA 01003, USA}

\newcommand{\UWyoming}{\label{UWyoming} Department of Physics and Astronomy, University of Wyoming, Laramie, WY 82071, USA}

\newcommand{\LAM}{\label{LAM} Aix Marseille Univ, CNRS, CNES, LAM (Laboratoire d’Astrophysique de Marseille), Marseille, France}

\newcommand{\UHawaii}{\label{UHawaii} Institute for Astronomy, University of Hawaii, 2680 Woodlawn Drive, Honolulu, HI 96822, USA}

\newcommand{\UCM}{\label{UCM} Departamento de F\'{\i}sica de la Tierra y Astrof\'{\i}sica, Universidad Complutense de Madrid, E-28040, Spain}

\newcommand{\IPARC}{\label{IPARC} Instituto de F\'{\i}sica de Part\'{\i}culas y del Cosmos IPARCOS, Facultad de Ciencias F\'{\i}sicas, Universidad Complutense de Madrid, E-28040, Spain}

\newcommand{\STScI}{\label{STScI} Space Telescope Science Institute, 3700 San Martin Drive, Baltimore, MD 21218, USA}

\newcommand{\McMaster}{\label{McMaster} Department of Physics and Astronomy, McMaster University, 1280 Main Street West, Hamilton, ON L8S 4M1, Canada}

\newcommand{\INAF}{\label{INAF} INAF -- Osservatorio Astrofisico di Arcetri, Largo E. Fermi 5, I-50157, Firenze, Italy}

\newcommand{\Sydney}{\label{Sydney} Sydney Institute for Astronomy, School of Physics A28, The University of Sydney, NSW 2006, Australia}

\newcommand{\CITA}{\label{CITA} Canadian Institute for Theoretical Astrophysics (CITA), University of Toronto, 60 St George St, Toronto, ON M5S 3H8, Canada}

\newcommand{\ASIAA}{\label{ASIAA} Institute of Astronomy and Astrophysics, Academia Sinica, No. 1, Sec. 4, Roosevelt Road, Taipei 10617, Taiwan}

\newcommand{\TKU}{\label{TKU} Department of Physics, Tamkang University, No.151, Yingzhuan Rd., Tamsui Dist., New Taipei City 251301, Taiwan}

\newcommand{\PSMA}{\label{PSMA} Penn State Mont Alto, 1 Campus Drive, Mont Alto, PA  17237, USA}

\newcommand{\ILL}{\label{ILL} ILL}

\newcommand{\stromlo}{\label{stromlo} Research School of Astronomy and Astrophysics, Australian National University, Mt Stromlo Observatory, Weston Creek, ACT 2611, Australia}

\newcommand{\UCatolica}{\label{UCatolica} Instituto de Astronom\'ia, Universidad Cat\'olica del Norte, Av. Angamos 0610, Antofagasta, Chile}

\newcommand{\UT}{\label{UT} McDonald Observatory, The University of Texas at Austin, 1 University Station, Austin, TX 78712-0259, USA}

\newcommand{\Vanderbilt}{\label{Vanderbilt} Department of Physics and Astronomy, Vanderbilt University, VU Station 1807, Nashville, TN 37235, USA}

\newcommand{\UNF}{\label{UNF} Department of Physics, University of North Florida, 1 UNF Dr. Jacksonville FL 32224}

\newcommand{\NAOC}{\label{NAOC} Chinese Academy of Sciences South America Center for Astronomy, National Astronomical Observatories, CAS, Beijing 100101, China}

\newcommand{\CASA}{\label{CASA} Center for Astrophysics and Space Astronomy, University of Colorado, 389 UCB, Boulder, CO 80309-0389, USA}

\newcommand{\UNAM}{\label{UNAM} Universidad Nacional Aut\'onoma de M\'exico, Instituto de Astronom\'ia, AP 106, Ensenada 22800, BC, M\'exico}

\newcommand{\UDP}{\label{UDP} Instituto de Estudios Astrof\'isicos, Facultad de Ingenier\'ia y Ciencias, Universidad Diego Portales, Av. Ej\'ercito Libertador 441, Santiago, Chile}

\newcommand{\Steward}{\label{Steward} Steward Observatory, University of Arizona, 933 N. Cherry Ave., Tucson, AZ 85721-0065, USA}  

\newcommand{\APO}{\label{APO} Apache Point Observatory and New Mexico State University, P.O.\ Box 59,
Sunspot, NM 88349-0059, USA}

\newcommand{\UNAMCU}{\label{UNAMCU} Universidad Nacional Aut\'onoma de M\'exico, Instituto de Astronom\'ia, AP 70-264, CDMX 04510, M\'exico}

\newcommand{\UWash}{\label{UWash}Department of Astronomy, University of Washington, Seattle, WA, 98195}

\newcommand{\CC}{\label{CC}Department of Physics, Colorado College, Colorado Springs, CO 80903}

\newcommand{\Utah}{\label{Utah}Department of Physics and Astronomy, University of Utah, 115 S. 1400 E., Salt Lake City, UT 84112, USA}

\newcommand{\UConcepcion}{\label{UConcepcion}Departamento de Astronom\'ia, Universidad de Concepci\'on, Casilla 160-C, Concepci\'on, Chile}

\newcommand{\FCLA}{\label{FCLA}Franco-Chilean Laboratory for Astronomy, IRL 3386, CNRS and Universidad de Chile, Santiago, Chile}

\newcommand{\Oklahoma}{\label{Oklahoma}Homer L. Dodge Department of Physics and Astronomy, University of Oklahoma, Norman, OK 73019, USA}

\newcommand{\UIUC}{\label{UIUC}Department of Astronomy, University of Illinois, Urbana, IL 61801, USA}

\newcommand{\Harvard}{\label{Harvard}Harvard-Smithsonian Center for Astrophysics, Cambridge, MA 02138, USA}

\newcommand{\caltech}{\label{caltech}Department of Astronomy, California Institute of Technology, Pasadena, CA 91125, USA}

\author[0000-0001-6444-9307]{Sebasti\'an F.\ S\'anchez}
\affiliation{Instituto de Astronom\'{\i}a, Universidad Nacional Aut\'onoma de M\'exico, A.P.\ 106, Ensenada 22800, BC, M\'exico}
\affiliation{Instituto de Astrof\'\i sica de Canarias, V\'\i a L\'actea s/n, 38205, La Laguna, Tenerife, Spain}
\email{sfsanchez@astro.unam.mx}
\correspondingauthor{Sebastian F. Sanchez}

\author[0000-0001-6444-9307]{Alfredo Mej\'{\i}a-Narv\'aez}
\affiliation{Departamento de Astronom\'{i}a, Universidad de Chile, Camino del Observatorio 1515, Las Condes, Santiago, Chile}

\author[0000-0002-4755-118X]{Oleg V. Egorov}
\affiliation{Astronomisches Rechen-Institut, Zentrum f\"ur Astronomie der Universit\"at Heidelberg, M\"onchhofstr.\ 12-14, D-69120 Heidelberg, Germany}

\author[0000-0001-6551-3091]{Kathryn Kreckel}
\affiliation{Astronomisches Rechen-Institut, Zentrum f\"ur Astronomie der Universit\"at Heidelberg, M\"onchhofstr.\ 12-14, D-69120 Heidelberg, Germany}

\author[0000-0002-7339-3170]{Niv Drory}
\affiliation{McDonald Observatory, The University of Texas at Austin, 1 University Station, Austin, TX 78712-0259, USA}

\author[0000-0003-4218-3944]{Guillermo A.\ Blanc}
\affiliation{Observatories of the Carnegie Institution for Science, 813 Santa Barbara Street, Pasadena, CA 91101, USA}
\affiliation{Departamento de Astronom\'{i}a, Universidad de Chile, Camino del Observatorio 1515, Las Condes, Santiago, Chile}








%
%

\author[0000-0002-6972-6411]{J. Eduardo M\'endez-Delgado}
\affiliation{Astronomisches Rechen-Institut, Zentrum f\"ur Astronomie der Universit\"at Heidelberg, M\"onchhofstr.\ 12-14, D-69120 Heidelberg, Germany}
\affiliation{Instituto de Astronom\'{\i}a, Universidad Nacional Aut\'onoma de M\'exico, A.P.\ 70-264, 04510 M\'exico D.\ F., M\'exico}

\author{Jorge K.\ Barrera-Ballesteros}
\affiliation{Instituto de Astronom\'{\i}a, Universidad Nacional Aut\'onoma de M\'exico, A.P.\ 70-264, 04510 M\'exico D.\ F., M\'exico}

\author[0000-0002-9790-6313]{Hector Ibarra}
\affiliation{Instituto de Astronom\'{\i}a, Universidad Nacional Aut\'onoma de M\'exico, A.P.\ 70-264, 04510 M\'exico D.\ F., M\'exico}

\author[0000-0002-3601-133X]{Dmitry Bizyaev}
\affiliation{Apache Point Observatory and New Mexico State University, Sunspot, NM, 88349, USA}

\author[0000-0002-8586-6721]{Pablo Garc\'ia}
\affiliation{Chinese Academy of Sciences South America Center for Astronomy, National Astronomical Observatories, CAS, Beijing 100101, China}
\affiliation{Instituto de Astronom\'ia, Universidad Cat\'olica del Norte, Av. Angamos 0610, Antofagasta, Chile}

\author{Aida Wofford}
\affiliation{Instituto de Astronom\'{\i}a, Universidad Nacional Aut\'onoma de M\'exico, A.P.\ 106, Ensenada 22800, BC, M\'exico}

\author{Alejandra Z. Lugo-Aranda}
\affiliation{Instituto de Astronom\'{\i}a, Universidad Nacional Aut\'onoma de M\'exico, A.P.\ 70-264, 04510 M\'exico D.\ F., M\'exico}

\begin{abstract}

We introduce the Data Analysis Pipeline (DAP) for the Sloan Digital Sky Survey V (SDSS-V) Local Volume Mapper (LVM) project, referred to as the LVM-DAP. { We outline our methods for recovering both stellar and emission line components from the optical integral field spectroscopy, highlighting the developments and changes implemented to address specific challenges of the data set.} The observations from the LVM project are unique because they cover a wide range of physical resolutions, from approximately 0.05 pc to 100 pc, depending on the distance to the targets. This, along with the varying number of stars sampled in each aperture (ranging from { zero, just one of a few, }to thousands), presents challenges in using previous spectral synthesis methods and interpreting the spectral fits. We provide a detailed explanation of how we model the stellar content and separate it from the ionized gas emission lines.  To assess the accuracy of our results, we compare them with both idealized and more realistic simulations, highlighting the limitations of our methods. We find that { the DAP} robustly correct for stellar continuum features and recover emission line parameters (e.g. flux, equivalent width, systemtic velocity and velocity dispersion) { with a precision and accuracy that fulfill the requirements of the primary goal of the analysis. In addition,} the recovered stellar parameters are reliable for single stars, the recovery of integrated populations is less precise. We conclude with a description of the data products we provide, instructions for downloading and using our software, and a showcase illustrating the quality of the data and the analysis on a deep exposure taken on the Huygens region at the center of the Orion Nebula.

\end{abstract}


\keywords{Sky surveys (1464), Milky Way Galaxy (1054), Large Magellanic Cloud (903), Small Magellanic Cloud (1468), Interstellar medium (847), Stellar feedback (1602)}


\section{Introduction} \label{sec:intro}

The Local Volume Mapper \citep[LVM,][]{drory24} is one of the mappers that comprises the Sloan Digital Sky Survey V \citep[SDSS-V,][Kollmeier in prep.]{koll17}. The core science goal for the LVM is to quantify stellar feedback physics at the energy injection scale ($<$10~pc), connecting ionizing stars with ionized gas diagnostics of the physical conditions in the interstellar medium (ISM).  To achieve this goal, it is currently
conducting a unique Integral Field Spectroscopy (IFS) survey of the Milky Way, the Large and Small Magellan Clouds, and nearby galaxies in the Local Volume using a new facility comprising 4 alt-alt mounted celiostats, 16 cm refractive telescopes, a lenslet-coupled fiber-optic system, and an array of spectrographs covering 3600-9800\,\AA\ at $R\sim4000$. The ultra-wide field science Integral Field Unit (IFU) has a diameter of 0.5 degrees with 1801 35.3\arcsec\ clear apertures in a hexagonal arrangement of 25 rings. The celiostat design ensures stationary optics and a stable fiber system, mitigating issues in traditional fiber systems \citep[e.g.][]{law21}. This innovative instrument will provide over $55\times10^6$ spectra, covering the Milky Way at spatial resolutions of 0.05 to 1 pc, the Magellanic Clouds at 10 pc resolution, and very nearby galaxies of large apparent diameter at 100 pc.

Survey operations began in November 2023 and will continue through May 2027. Following the precedent of previous SDSS surveys, all raw and reduced data, and the data products from dedicated analyses, as well as the codes, will be made publicly available, starting with SDSS DR20 in late 2025. Early science data have demonstrated the LVM's capability to deliver high-quality spectral mapping of Galactic nebulae \citep{kreckel24}, crucial for addressing its core science goals and providing new insights into the energy injection scales within Local Group galaxies.

To achieve these goals and digest the amount of data produced in these unique observations, it is required to create dedicated pipelines to reduce and analyze the data. In this article we present the first version of the LVM data analysis pipeline (DAP, { version 1.0.0}). The main goal of this tool is to automatically analyze the reduced, extracted and calibrated spectra provided by the Data Reduction Pipeline (Mej\'\i a-Narv\'aez et al., in prep), decoupling the stellar and ionized gas emission line components, and providing observational and physical parameters of both components. There are several tools and pipelines already available in the literature that perform similar analyses \citep[e.g., PPxF, MaNGA-DAP, PyCASSO, Fit3D, pyPipe3D,][for citing just a few]{ppxf, amorim17, dap, pypipe3d}. However, none of them has been developed to explore spectra that sample targets at the small physical scales of the LVM. At those scales it cannot be guaranteed that the number of stars within a single aperture is large enough to apply stellar population spectral synthesis, the basic physical reasoning behind all these tools \citep[][]{walcher:2011, conroy:2013}. We have implemented in the DAP a new approach that we consider valid when the requirement to sample a fully populated IMF is not fulfilled. In this approach, instead of modelling the stellar component using single stellar populations (SSP), we adopt combinations of representative stellar spectra valid for these resolved stellar populations (RSP). 

{ Our primary goal with the DAP is to robustly recover emission line properties (flux, equivalent width, systematic velocity, velocity dispersion), accurately correcting for underlying stellar continuum features. Our secondary goal is to recover stellar parameters associated with the stellar continuum (T$_{eff}$, log($g$), [Fe/H], [$\alpha$/Fe]), flexibly allowing for either small or large combinations of individual stars. 
In this article we aim to: (i) present the new methodology required to model the stellar spectrum when the number of stars in the considered aperture is not sufficiently large to apply the stellar synthesis method; (ii) present the procedure adopted to create the RSP template library propagating the probability distribution functions (PDFs) of the physical parameters that are represented by each template; (iii) describe and distribute the first operational version of the DAP, that is able to recover the ionized emission line properties (its primarily goal) and to obtain a preliminary model for the stellar component; (iv) present the proposed format with which we will distribute the dataproducts, distributing some example codes of how to use them; (v) characterize the quality of the recovered parameters based on  simulated and real data, highlighting the current limitations and suggesting possible future improvements; and (vi) show the implementation on real LVM data. The code we present here is working end-to-end, however we acknowledge that it will require additional updates throughout the course of the survey, as it is usually the case in this kind of project. For example, a refinement in the input parameters (e.g., RSP templates) can flexibly be implemented in future applications of the DAP. } 

The structure of this article is as follows: a summary of the data to be analyzed is presented in Sec. \ref{sec:data}, with the implemented methodology being described in Sec. \ref{sec:method}, including  an overview of the analysis sequence in Section \ref{sec:proc}, details on the adopted stellar component analysis and how the stellar templates are created in Sec. \ref{sec:rsp}, and a description of the delivered data products in Sec. \ref{sec:DP};  the reliability of the methodology is tested with extensive simulations described in Sec. \ref{sec:code}, including both physically motivated (Sec. \ref{sec:sim_idea}) and purely empirical (Sec. \ref{sec:sim_real}) simulations; the accuracy and prevision in the recovery of the emission lines and the stellar population properties are described in Sec.  \ref{sec:sim_el}, \ref{sec:sim_MaStar}, and \ref{sec:sim_el}; an example of the use of the DAP is presented in Sec. \ref{sec:show}, in which we analyze a deep exposure centred in the Huygens region of the Orion nebula, including a description of this particular dataset in Sec. \ref{sec:O_dataset}, with a summary of the performed analysis results in Sec. \ref{sec:O_results}, a detailed exploration of the generated deep integrated spectrum in Sec. \ref{sec:O_int}, the list of detected emission lines in Sec. \ref{sec:O_el}, the physical properties derived from those emission lines in Sec. \ref{sec:ori_fprop}, and the spatial distribution of the emission line fluxes presented in Sec. \ref{sec:ori_spa}; how to download the code is presented in Sec. \ref{sec:dist}, and finally a summary of the main results of this study is presented in Sec. \ref{sec:conc}.

\section{DATA} \label{sec:data}

Observational data for the Local Volume Mapper (LVM) project are obtained using a newly constructed facility at Las Campanas Observatory. As indicated before, the instrumental setup comprises four telescopes: one devoted to the acquisition of the science exposures (using the ultra-wide IFU comprising 1801 fibers described before), two monitoring the night-sky spectrum towards the east and the west (using 59 and 60 fibers, respectively), and one dedicated to the acquisition of spectrophotometric calibration stars (using 24 fibers). This setup feeds into three DESI-like spectrographs covering the wavelength range from 3600 to 9800 Angstroms with a resolution of R $\sim$ 4000 at H$\alpha$. Each spectrograph features a dichroic system, which divides the light received from each fiber into three wavelength ranges: (b) blue, from 3600 to 5800 \AA, (r) red, from 5750 to 7570 \AA, and (z) infrared, from 7520 to 9800 \AA. The size of the LVM's ultra-wide field IFU projected in the sky ensures detailed spectral and spatial coverage across the survey's targets, including the Milky Way, Magellanic Clouds, and a selection of local volume galaxies.

\begin{figure*}
 \minipage{0.99\textwidth}
 \includegraphics[width=17cm,clip,trim=0 230 5 0]{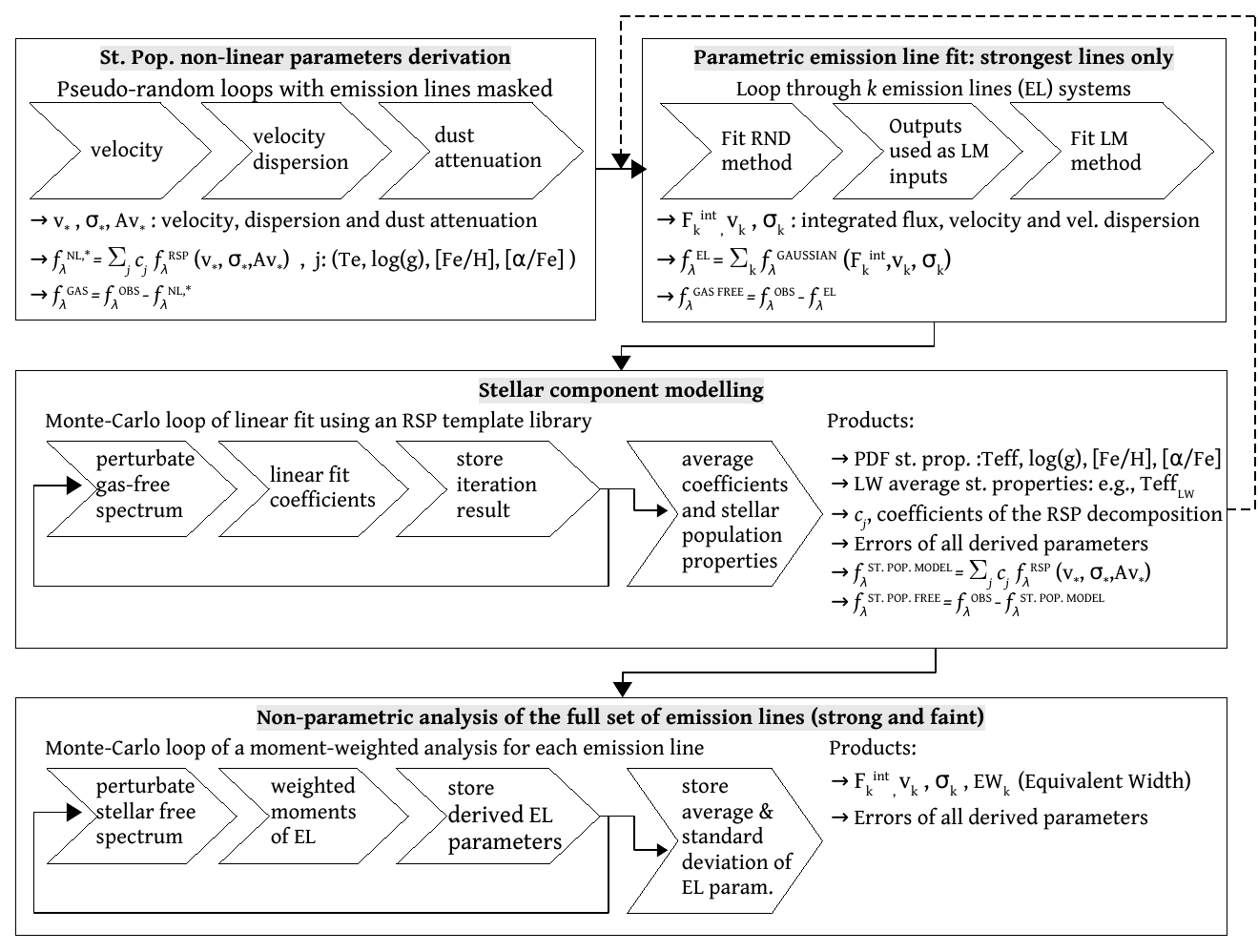}
 \endminipage
 \caption{Scheme of the LVM-dap analysis flow for a single fiber spectrum, including the main procedures: (i) derivation of the non-linear parameters of the stellar spectrum (v$_\star$, $\sigma_\star$ and A$_{\rm V,\star}$), (ii) parametric derivation of the properties of the ionized gas emission lines { (EL), including the flux intensity (f$_{\rm EL}$), velocity (v$_{\rm EL}$) and velocity dispersion ($\sigma_{\rm EL}$)}, (iii) stellar component synthesis, i.e., decomposition into a set of RSP templates and finally (iv) non-parametric derivation of the properties of the emission lines, including the equivalent width for each emission line (EW$_{\rm EL}$). { RND and LM stands for the two methods included in \pyf\ to fit the parametric models to the EL, as explained in the text.}}
 \label{fig:scheme}
\end{figure*}

The LVM survey's data reduction pipeline (DRP, Mejía-Narváez in prep.) follows the procedures described by \citet{sanchez06a}. The actual code is based on the {\sc py3D} reduction package, initially developed for the CALIFA survey \citep[][]{dr1}. Originally written in Python 2.7, it has been updated to Python 3.11, incorporating specialized procedures to cater to the unique features and requirements of the LVM dataset. The key steps performed by the LVM-DRP include: (i) initial raw data preprocessing to merge readings from different amplifiers into a single frame for each channel (b, r, and z) of each spectrograph (producing 9 different frames), followed by bias removal, gain correction, and cosmic ray identification and masking; (ii) identification and tracing of fiber spectra in each CCD of each spectrograph, including assessment of the FWHM along both dispersion and cross-dispersion axes; (iii) extraction of spectra using the established trace and width parameters, assuming a Gaussian shape for each fiber's spectrum projected along the cross-dispersion axis, and performing a concurrent stray-light correction; (iv) wavelength calibration and linear resampling of the extracted spectra; (v) differential correction for fiber-to-fiber transmission discrepancies; (vi) flux calibration based on the stars acquired by the spectrophotometric telescope simultaneously with the science observations; (vii) combination into a single spectrum of the spectra acquired in each channel (b, r, and z) for each science fiber, weighting by the inverse of the errors in the overlapping regions between arms; (viii) estimation and subtraction of the sky spectrum using the spectra obtained by the west and east telescopes devoted to observing the sky; (ix) implementation of an astrometric solution for each observation, using the information provided by the guiding cameras. Errors are propagated during each step of the data reduction.


The final product of this reduction is a FITS file that comprises a set of row-stacked spectra for each LVM pointing, in which each row corresponds to a science fiber. Different extensions contain the flux intensity, the estimated error, and additional information (such as a mask of bad pixels, broken/low-quality fibers, estimated sky, etc.). An additional extension includes the mapping of each science fiber in the sky (position table), based on the astrometric solution, and additional information regarding each fiber (e.g., additional quality masks). This FITS file is the input for the data analysis pipeline described here.

\section{Analysis} \label{sec:method}

The LVM-DAP uses the fitting algorithms included in \pyf \citep{pypipe3d}\footnote{\url{http://iyfs.astroscu.unam.mx/pyPipe3D/pyFIT3D.html}}. This package is fully coded in Python 3, comprising classes, methods, and functions that allow one to decouple and extract physical information about the stellar and ionized gas components in the observed spectra. It models the stellar component by adopting a linear combination of spectra that comprises a template library, convolved by a line-of-sight velocity distribution and attenuated by a certain dust extinction. The properties of the ionized gas emission lines are derived using both a parametric fit (assuming a Gaussian shape for each emission line) and a non-parametric procedure (performing a weighted moment analysis). \pyf is the basis of the \pyp pipeline, a reliable tool that has been tested and vetted through analyses involving mock datasets and specialized simulations \citep{guidi18,ibarra19,sarm23}. \pyp has been widely used for the massive analysis of the observations of IFS galaxy surveys, such as those provided by the CALIFA survey \citep[e.g.,][]{mariana16,espi20}, the MaNGA project \citep[e.g.,][]{ibarra16,jkbb18,sanchez18b,bluck19,laura19}, the SAMI Galaxy Survey \citep[e.g.,][]{sanchez19}, and the AMUSING++ compilation \citep{laura18,carlos20}. We adopt \pyf due to its modularity, which allows the construction of independent analysis tools that fit the needs and requirements of a particular dataset, like the one provided by the LVM survey.

\subsection{Outline of the analysis} 
\label{sec:proc}


Figure \ref{fig:scheme} shows the analysis flow of the DAP for each individual spectrum within each LVM exposure { (i.e., no spatial binning is performed in general within the DAP)}. It mimics the one implemented in \pyp, as described in \citet{pypipe3d}, Fig.~1, slightly modified to adapt it to the peculiarities of the LVM data regarding the modeling of the stellar component (discussed in Sec \ref{sec:rsp}). We include here a brief summary of the main steps that comprise the analysis to avoid unnecessary repetition, as they have been extensively described in previous publications \citep[e.g.,][]{pipe3d,pypipe3d}:

\noindent { Non-linear fit:} First, we determine the systemic velocity ($v_\star$), velocity dispersion ($\sigma_\star$), and dust attenuation (A$_{\mathrm{V,\star}}$) of the stars (i.e., the non-linear parameters) by fitting the continuum with a linear combination of a few stellar template spectra. We refer to this template as the simplified RSP template library. { Following \pyf\ the RSP templates must be adjusted to the instrumental resolution and/or line spread function (LSF) of the observed dataset for a correct derivation of the kinematic parameters.} The actual number of adopted spectra in the template library is tested with simulations (Sec. \ref{sec:code}), with a range of four to twelve. { This rather small number of templates is similar to what is adopted by \pyp\ for the analysis of unresolved stellar populations, and has proven to be sufficient for a accurate estimation of the non-linear parameters discussed here \citep{pypipe3d,sanchez22}}. These template spectra are adjusted in velocity, broadened for dispersion, and attenuated by dust extinction, while we mask the expected locations of the brightest emission lines (defined by the user). A brute-force exploration of a pre-selected range of parameters is adopted to find the best combination of non-linear parameters that reproduce the observed spectrum. { }

{ The correct derivation of any physical parameter associated with the stellar component, and in particular the non-linear parameters, relies on the ability to extract that information from a limited S/N spectrum while accounting for the intrinsic properties of the spectrum itself (resolution, wavelength range). As already shown in previous studies \citep[e.g.][]{cid-fernandes:2014aa, pipe3d,pypipe3d}, if the S/N is below a certain threshold (that roughly corresponds to the number of templates in the library) any decomposition and any derived parameter is very unreliable. Based on these previous results, the DAP only performs the non-linear fit described beofre if the S/N level (at the normalization wavelength) is above a threshold pre-defined by the user. In this implementation of the code this S/N cut is set to 20. Below this limit the code assumes the default values for the non-linear parameters. }

Two main modifications have been introduced in the DAP compared to \pyp\ for this analysis. First, we introduce an optional binning of the observed data in the derivation of the kinematics parameters ($v_\star$ and $\sigma_\star$) to account for the possibility that the adopted stellar template may have a lower spectral resolution than the actual data as is currently the case (see Sec. \ref{sec:rsp}). In the implementation described here, three spectral pixels are binned (just for this analysis), matching the spectral sampling to that of the adopted template ($\sim$1.5\AA). Second, we include an optional broad binning and smoothing in both the observed data and the stellar templates in the analysis of the dust attenuation (A$_{\mathrm{V,\star}}$). It is well known that this non-parametric attenuation is traced by the overall shape of the observed spectra, without any significant information remaining in the detailed spectral features (absorptions) in the spectra. Thus, the detailed information provided by the high spectral resolution dataset is not crucial for deriving this parameter. Following \citet[e.g.][]{wilki17}, we generate a coarse version of the spectra by averaging the values in bins of 75\AA\ width for this particular section of the analysis. The virtue of both binning procedures is that they speed up this computationally demanding part of the analysis without affecting the final results.

\noindent { Parametric emission-line fit:} After deriving
$v_\star$, $\sigma_\star$ and A$_{\mathrm{V,\star}}$, we remove a preliminary model of the stellar component and fit a series of pre-defined strong emission lines with Gaussian functions. 
{ This modeling is agnostic of instrumental resolution and / or LSF. Thus, the widths of the Gaussian functions are provided in wavelength units (\AA, in our case), without correcting for the instrumental effects (which should be done a posteriori).} { As explained in \citet{pipe3d}, a subset of emission lines that are located at relatively nearby wavelengths can be fit together configuring a
{\sc system}, for which modeling is minimized together. In some cases, for instance nearby doublets, the kinematic parameters could be tied between different emission lines, limiting the number of free parameters and improving the fit quality \citep{pipe3d,pypipe3d}}. { The fitting is performed following two steps. First, it finds an approximate location of the global minimum of the figure of merit ($\chi^2$/$\nu$) using a pseudo-random exploration of the range of parameters (RND method), and the result of this exploration is used as guess for a Levenberg–Marquardt minimization (LM method), that provides a more precise estimation of the fitted parameters.}

The actual list { of analyzed emission lines} is included in Appendix \ref{app:elines}, { and may be modified throughout the duration of the survey, with a subset of the full list included in } Table \ref{tab:fe_long_list} { corresponding to the strongest emission lines in the considered wavelength range}. As a result of this analysis, an emission-line model spectrum including the best-fitted models for each considered emission line is created. 

{ As indicated before, the non-linear fit and the stellar component model are reliable only above a certain threshold. Thus, below this S/N limit no stellar decomposition is performed based on a multi-RSP fitting. Instead, if the S/N is above one, it selects the single template within the library that matches best the continuum spectrum, based on a simple flux normalization and $\chi^2$/$\nu$ minimization. Finally, if the S/N is below one then we assume there is no stellar continuum, and no stellar model is created or subtracted. This later case would either correspond to the lack of any star within the aperture or the presence of stars that are too faint to provide any significant contribution.

\pyf\ allows us to include a high order polynomial function as part of the emission line modelling to deal with a possible non-zero background. However, this procedure is time consuming and in some cases it does not produce reliable results. Instead of adopting this approach, the parametric fitting assumes a zero-background residual from the subtraction of the stellar model (or the background, in case no stellar model is created). { The almost zero background is provided by a procedure introduced in \pyf\ that will be described below.}



}

\noindent { Stellar component synthesis:} We subtract the best model of these emission lines from the original spectrum and fit the remaining continuum again, this time using a broader library of stellar templates that capture a wide variety of stellar features. The spectra in this library are adjusted in velocity, broadened for dispersion, and attenuated for dust extinction, using the non-linear parameters derived in the first step of this sequence. This analysis provides two main products: (i) the coefficients of the linear combination of stellar spectra that best reproduce the observed spectrum, and (ii) the luminosity-weighted (LW) average values of the physical parameters that correspond to this model spectrum. { Like in the case of the non-linear fit, this analysis is performed if the S/N at the normalization wavelength is above a certain pre-defined threshold. If the S/N is below this limit but above a S/N of one, the algorithm select a single template within the library that matches the continuum spectrum following the procedures described before. In this case all the coefficients are set to zero but the one corresponding to this particular template (set to one), and the LW average values of the physical parameters are set, by construction, to the values of this particular template. Finally, if the S/N is below one, meaning there is no star within the aperture or the stars are too faint to produce any significant contribution to the observed spectrum, all coefficients are set to zero and no LW parameters are derived.}

This process provides us with a stellar spectrum model (synthesis), which we then subtract from the original spectrum. This stellar-subtracted spectrum retains the residuals/inaccuracies of the modelling of the stellar component, and any possible difference between the spectrophotometric calibration between the data and the adopted stellar template. To minimize the impact of these differences on further analyses (for example, the analysis of emission lines), \pyf introduced a smoothing and subtraction of the low order component of this residual \citep{pipe3d,pypipe3d}.
The result of this analysis is a new spectrum (gas-only spectrum) that is used to refine the parametric analysis of the subset of emission lines, following an iteration (which usually converges after two or three loops), or it can be used in the next step of the analysis (see Fig. \ref{fig:scheme}).

{ The treatment of the residual of the subtraction of the stellar model guarantees an almost zero background residual, allowing the code to fit the emission lines without the potential problems that arise in the presence of an underlying structure (as discussed above). However, it does not improve the accuracy and precision in any way in the derivation of the properties of the stellar component. Indeed, a non-zero background residual due to the lack of the proper stellar ingredients, the presence of a non-stellar continuum, or inaccuracies in the spectrophotometric calibration would be most probably associated with an erroneous estimation of the physical parameters of the stellar component \citep[as already noticed in similar situations by previous studies, e.g., ][]{gomes17}. However, we must recall that the primary goal of the current version of the DAP is to recover in the best possible way the properties of the emission lines. We will discuss later the potential relevance of this issue in the analysis of the current and foreseen LVM dataset when we apply the DAP to real data.}

\noindent { Non-parametric analysis of the full set of emission lines:} We analyze the gas-only spectrum derived in the previous step to estimate the integrated flux, velocity, velocity dispersion, and equivalent width (EW) for a predefined set of emission lines using a method based on weighted moments. { Like in the case of the parametric analysis, no correction of the instrumental resolution or LSF is applied and the velocity dispersion is provided in wavelength units (i.e., Angstroms).} The current list of emission lines analyzed in this step, described in Appendix \ref{app:elines} (Table \ref{tab:fe_long_list}), contains 192 lines.

Errors are derived for all the estimated parameters by adopting a Monte Carlo iteration that uses the errors estimated by the DRP, as described in Sec. \ref{sec:data}. { Following \pyf,  the known location of defects in the spectra, either produced by the presence of cosmic rays, imperfect fiber transmission corrections, inaccurate sky subtraction and/or mismatches between the overlapping regions of the different spectral arms of the spectrographs have been masked using the information provided by the DRP.}

{ 
\subsection{Analysis of the integrated spectrum}

The previous analysis is performed for each spectrum within each LVM exposure, using as input parameters those defined by the user. Among those input parameters, it is required to define the range of non-linear parameters to be explored, the S/N limit adopted to perform the stellar decomposition, the adopted RSP templates, and the set of emission lines analyzed using the parametric and non-parametric procedures. Based on our previous experience analyzing large samples of IFS data \citep[e.g.][]{sanchez22,eDR2}, it is recommended to tune some of these parameters within each exposure/pointing using the values derived from the analysis of the integrated spectrum. This spectrum, being the combination of $\sim$1700 individual spectra, has a considerably higher S/N than each individual one. Thus, before running the full analysis on each individual spectrum we run it over an integrated spectrum created by coadding the individual ones, weighted by their errors. Then, the code redefines the input guess and explored range of the kinematic parameters for both the stellar component and ionized gas emission lines in the analysis of the individual spectra to reflect the values recovered for the integrated spectrum.
}

%
%


\begin{figure*}
 \minipage{0.99\textwidth}
 \includegraphics[width=8.5cm,clip,trim=0 0 0 0]{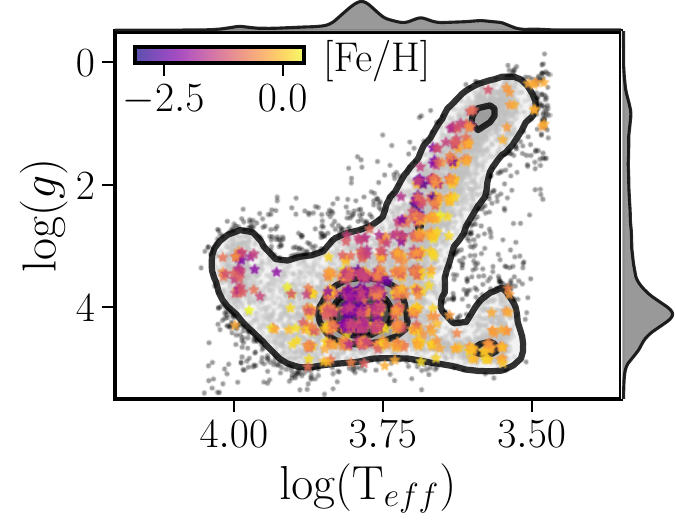}\includegraphics[width=8.5cm,clip,trim=0 0 0 0]{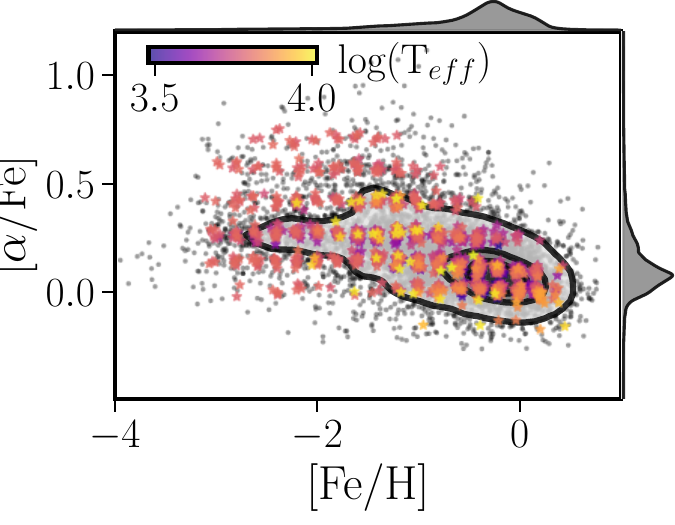}
 \endminipage
 \caption{Distribution of the physical properties derived using CoSha for 22773 stars extracted from the MaStar stellar library (grey-solid line and density contours), together with the average properties for the 1235 RSPs created by clustering those MaStar stars in the space of physical properties using a regular grid (colored stars). Left (right) panel shows the distribution in the log($g$)-T$_{eff}$ ([$\alpha$/Fe]-[Fe/H]) plane, with colors tracing [Fe/H] (T$_{eff}$). In each panel successive contours encircle 95\%, 65\%, and 25\% of the MaStar stars, respectively.}
 \label{fig:MaStar_cl}
\end{figure*}

\subsection{Stellar Population Analysis} \label{sec:rsp}

For the typical LVM observations, sampling the Milky Way (MW) and galaxies within the Local Group, individual apertures (fibers) often capture just a single star, a few stars, or a fraction of a stellar cluster. Only for the more distant galaxies observed within the survey is the physical aperture large enough to adopt a modeling of the stellar component based on stellar synthesis \citep{conroy:2013}, as usually adopted in IFS and/or single aperture observations of nearby galaxies \citep[e.g.,][]{starlight,cid-fernandes13}. When individual apertures encompass a number of stars large enough to sample the complete mass function (MF), and therefore, the initial mass function (IMF) at the moment of birth of a certain population, it is possible to model the stellar spectra using single-stellar populations \citep[SSP][]{1972a&a....20..383t,bruzual83,bruzualcharlot03}. It is usually accepted that this number is on the order of 10$^5$ -- 10$^6$ stars \citep[e.g.][]{or-du22}.
When the number of stars within the aperture is below this value, but it is of the order of some hundreds to a few thousands, it is possible to generate pseudo-SSPs assuming a stochastic (but incomplete) sampling of the IMF \citep[e.g.,][]{fuma11,cerv13a,cerv13b}. Using this methodology, instead of assigning a set of physical parameters to each (pseudo)-SSP and finally to the fitted spectra, a probability distribution function (PDF) of physical parameters (and IMFs), is derived.


\subsubsection{Resolved Stellar Population Templates}
\label{sec:rsp_der}

As stated before, in the case of the LVM data, the typical number of stars sampled within each aperture is far below the limits of a fully populated SSP. We cannot guarantee that we have just one single star within the aperture, either. If that were the case, we could have performed a tagging of the observed star, assigning a certain set of physical properties (T$_{\mathrm{eff}}$, log($g$), [Fe/H], and [$\alpha$/Fe]) by comparing its spectrum with those in a stellar library of known properties \citep[e.g.,][]{yan19,mejia21}. Therefore, we need to develop a new methodology valid for a wide dynamical range in the number of stars, ranging from one (or a few) to a few hundred, or even larger numbers in general. { By construction the methodology already deals with the case in which no stellar component is detected with sufficient S/N to extract any physical information and/or there is no star sampled within the considered aperture, as indicated in Sec. \ref{sec:proc}} .

As indicated in the introduction, we call this new method Resolved Stellar Population analysis. The central idea is to fit the spectra with a library of stellar spectra. This, per se, is not a new procedure. It has been adopted for decades. One of the most recent examples of this approach is the MaNGA DAP \citep{dap}, in which a template of 42 stellar spectra was created by performing a clustering procedure on the MILES stellar library \citep{miles,falc11}, increasing the S/N and reducing the size of the template library. This library was then used to fit and subtract the underlying stellar population to later analyze the emission lines in the IFS spectra.

In essence, we have followed a similar approach. However, our stellar library retains information on the physical properties of the stars adopted to construct the templates. In this way, it can be used to characterize the properties of the stellar population in the fitted spectra. However, contrary to previous methods, instead of assigning a particular set of physical parameters to a template, we associate a PDF of physical parameters. { The foreseen format for the RSP template library includes a set of projections of these PDFs for each template, as will be described later on.}

The methodology to create the RSP template can be applied to any existing stellar library. However, for the particular implementation in the current version of the LVM DAP, we adopt the MaStar stellar library \citep{yan19}. This library comprises approximately 30,000 individual spectra of stars covering the wavelength range from 3622 to 10354 \AA\ with a spectral resolution of $R\sim$1800 at H$\alpha$. These spectra were collected using the same instrument as the Mapping Nearby Galaxies at Apache Point Observatory \citep[MaNGA][]{bundy15} project. MaStar stands out from previous empirical libraries by offering a larger selection of stars and covering a wider range of stellar parameters. This includes a better representation of cool dwarfs, low-metallicity stars, and stars with varying [$\alpha$/Fe] levels, thanks to a well-designed target selection strategy that leverages existing stellar-parameter catalogs. Recently, \citet{mejia21} presented a new Code for Stellar Properties Heuristic Assignment (CoSHa) and applied it to the MaStar library. Using a sophisticated Gradient Tree Boosting algorithm, they determined the key physical properties (T$_{eff}$, log($g$), [Fe/H], and [$\alpha$/Fe]) for approximately 22,000 stars\footnote{\url{http://ifs.astroscu.unam.mx/MaStar/}}. This final stellar library with accurately derived physical properties covers a wide range of values for the considered physical parameters: effective temperatures between 2900 K and 12,000 K, surface gravity from -0.5 to 5.6, metallicity between -3.74 and 0.81, and $\alpha$-element abundance from -0.22 to 1.17.

We adopt the CoSha-MaStar stellar library as a starting point to generate the RSP templates to analyze the LVM data, { and  note that these are input parameters that can be adapted to each science case}. We acknowledge some limitations in { this specific library}. For instance, the spectral resolution of the LVM data is higher than that of the library. Therefore, any derivation of the stellar velocity dispersion would be very unreliable, being only valid at a first order for the very high values. Furthermore, the range of physical parameters covered by the stellar library does not reach the hottest stars (in particular the OB stars) that would be required to fit the observed spectra, and the number of binary stars in this library may not be representative of real population \citep{sana12} . { This would limit its use to explore the stellar component in detail when there is a significant fraction of these hot stars, however as we will see later on, this does not  significantly affect the derivation of the properties of the ionized gas emission lines, which is the primary goal of the DAP. } We will address these issues in future versions of the LVM-DAP by extending the library using stellar atmosphere models and observed hot star spectra, { or using a completely different initial stellar library. As indicated before the stellar template is not hard-coded within the code, and the DAP is designed to flexibly work with different stellar libraries without requiring any modification.}

\subsubsection{Clustering in the physical parameters space}
\label{sec:rsp_phy}

One primary goal of a useful RSP library is to sample the space of physical parameters in the most homogeneous way, only limited by the boundaries imposed by nature. This is not the case for any empirical stellar library, which is biased towards the stars accessible in the solar neighborhood \citep[$\sim$1 kpc;][]{mejia22}. Furthermore, the sample selection and the observational constraints imposed further biases in the final distribution of stars, as demonstrated by \citet{mejia22}. Figure~\ref{fig:MaStar_cl} shows the distribution of the CoSha-MaStar library in the T$_{eff}$-log($g$) and [Fe/H]-[$\alpha$/Fe] planes. As expected, there is an over-representation of stars with T$_{eff}\sim$6000-7000 K, log($g$)$\sim$4, and solar metallicities and abundances (i.e., [Fe/H]$\sim$[$\alpha$/Fe]$\sim$0). To mitigate this bias, we group the stars following a regular grid in the space of physical parameters, adopting a bin size of $\Delta$log(T$_{eff}$) = 0.03 dex, $\Delta$log($g$) = 0.03 dex, $\Delta$[Fe/H] = 0.2 dex, and $\Delta$[$\alpha$/Fe] = 0.15 dex. This binning is applied to those stars within the MaStar library with a S/N$>$5 in the blue range of the spectral range (3980.0-4090.0 \AA), and with no evidence of emission lines (due to their intrinsic variability). This limits the stellar library to 19,303 spectra. Finally, we average the spectra of all stars within the same multi-parametric bin, retaining only those in which there are at least three stars per bin.

The final RSP template comprises 1235 spectra. To each of these RSPs, we assign a PDF in the space of physical parameters that is created by propagating the values assigned by CoSha to each MaStar spectrum within the considered bin. The result of this analysis is included in Fig. \ref{fig:MaStar_cl}, showing the distribution in the T$_{eff}$-log($g$) and [Fe/H]-[$\alpha$/Fe] planes of the central values assigned to each RSP. By construction, they follow a regular distribution. This procedure has the additional advantage of increasing the S/N of the stellar templates, as a result of combining several original MaStar spectra.

\begin{figure*}
 \minipage{0.99\textwidth}
 \includegraphics[width=18cm,clip,trim=0 5 0 10]{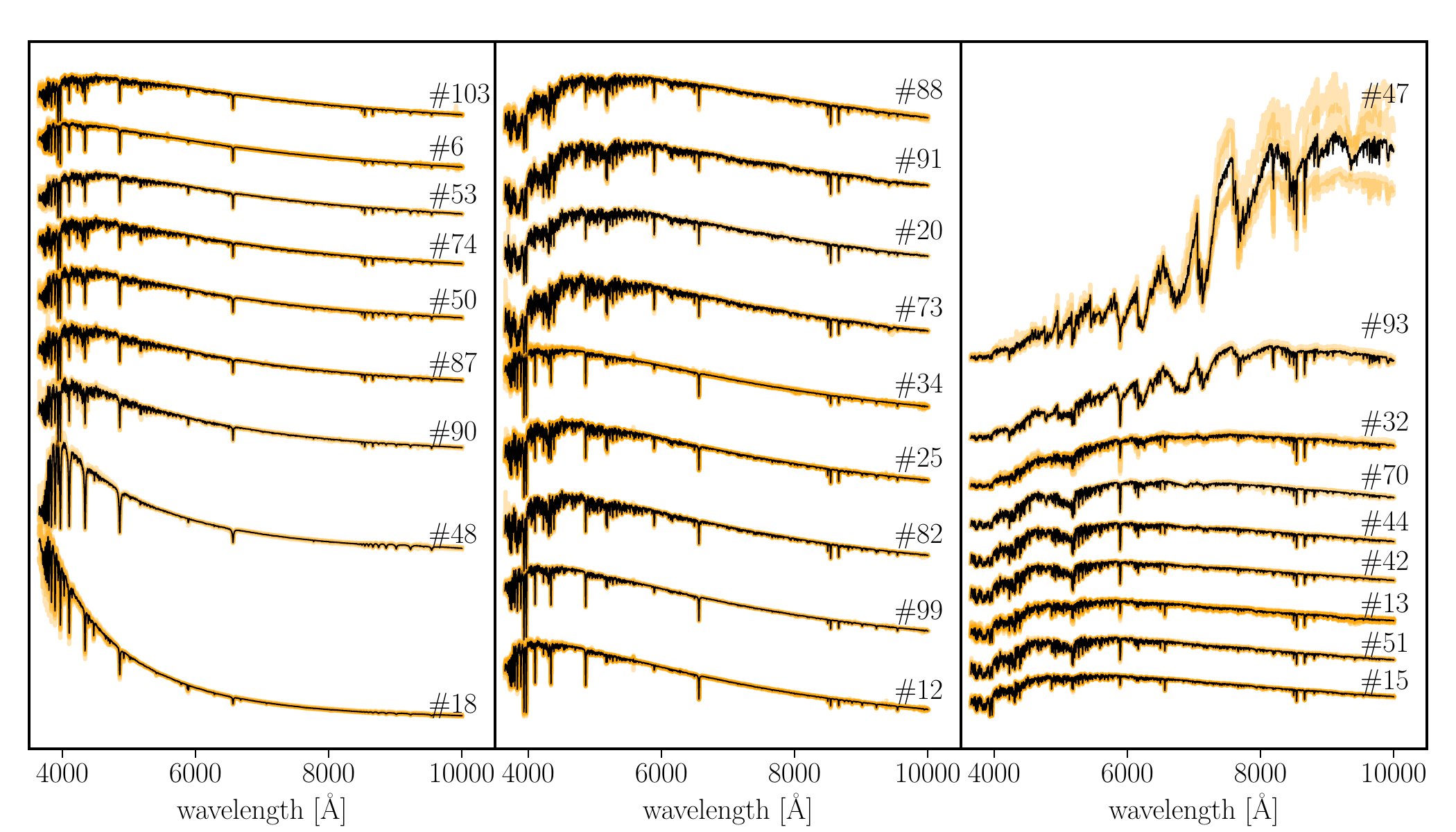}
 \endminipage
 \caption{Sub-set of the spectra included in the template comprising 108 RSPs created by clustering in the observational space the 1235 RSPs which properties are shown in Fig. \ref{fig:MaStar_cl}. Each black line shows the final RSP spectrum, created by averaging all the spectra belonging to the same cluster, represented by the orange lines. Spectra are ordered by blue-to-red color, from the bluest (bottom-left) to the reddest (top-right), arbitrarely scaled to avoid overlapping. The id number of each RSP within the considered template is included at the right-end of each spectrum. }
 \label{fig:MaStar_spec}
\end{figure*}

\subsubsection{Clustering in the observational parameters space}
\label{sec:rsp_obs}

The resulting RSP library is still not useful for the purpose of the analysis described above. First, the number of spectra is still too large to be used efficiently in the stellar decomposition described in Sec. \ref{sec:proc} \citep[e.g.,][]{cid-fernandes14,pipe3d}. Second, the library has strong degeneracies, i.e., RSP templates corresponding to very different sets of physical parameters may present very similar observational parameters (i.e., spectra). It may be the case that in other spectral ranges (e.g., ultraviolet or infrared) they are clearly distinguishable, however not at our level of signal-to-noise ratio (S/N) given the quality of the empirical library. To address both issues, we perform an additional clustering, this time in the observed spectra space. That is, we group those templates within the 1235 RSP library that are more similar and average them, irrespective of their physical properties. Following the same reasoning as before, we propagate the individual PDFs of the RSP templates within each cluster to generate the final PDFs.

The clustering is performed using the Spectral Clustering technique \citep[e.g.][]{ding10}, implemented in the Python library {\tt scikit-learn} under the module {\tt sklearn.cluster.SpectralClustering}. This function operates by constructing a similarity graph from the input data. Each node in this graph represents an observational dataset (i.e., an RSP template), with edges connecting nodes that are similar to each other. The edge weights can be defined using various metrics, such as the Gaussian kernel of the Euclidean distance between points. Subsequently, the algorithm computes the Laplacian of the similarity graph, and eigenvalues and eigenvectors of this Laplacian are used to project the data into a lower-dimensional space where traditional clustering algorithms, like $k$-means, can be effectively applied. This technique is particularly well suited for astronomical data in general, and spectroscopic data in particular, where the inherent clustering structure is complex and not well-defined by linear boundaries. The only two input parameters required to apply this procedure are the number of output clusters ($n_{cl}$) and the order of the $k$-neighbor to create the similarity matrix. In our analysis, we fixed the $k$ to three and selected different values for $n_{cl}$, including 4, 12, 36, 108, and 342. Prior to any clustering, the original RSPs are scaled, normalizing their fluxes to one in a spectral window of 50 \AA\ around 5000 \AA. This normalization is required for the clustering in the space of observed parameters.

\begin{figure}
 \minipage{0.99\textwidth}
 \includegraphics[width=8.5cm,clip,trim=15 0 20 0]{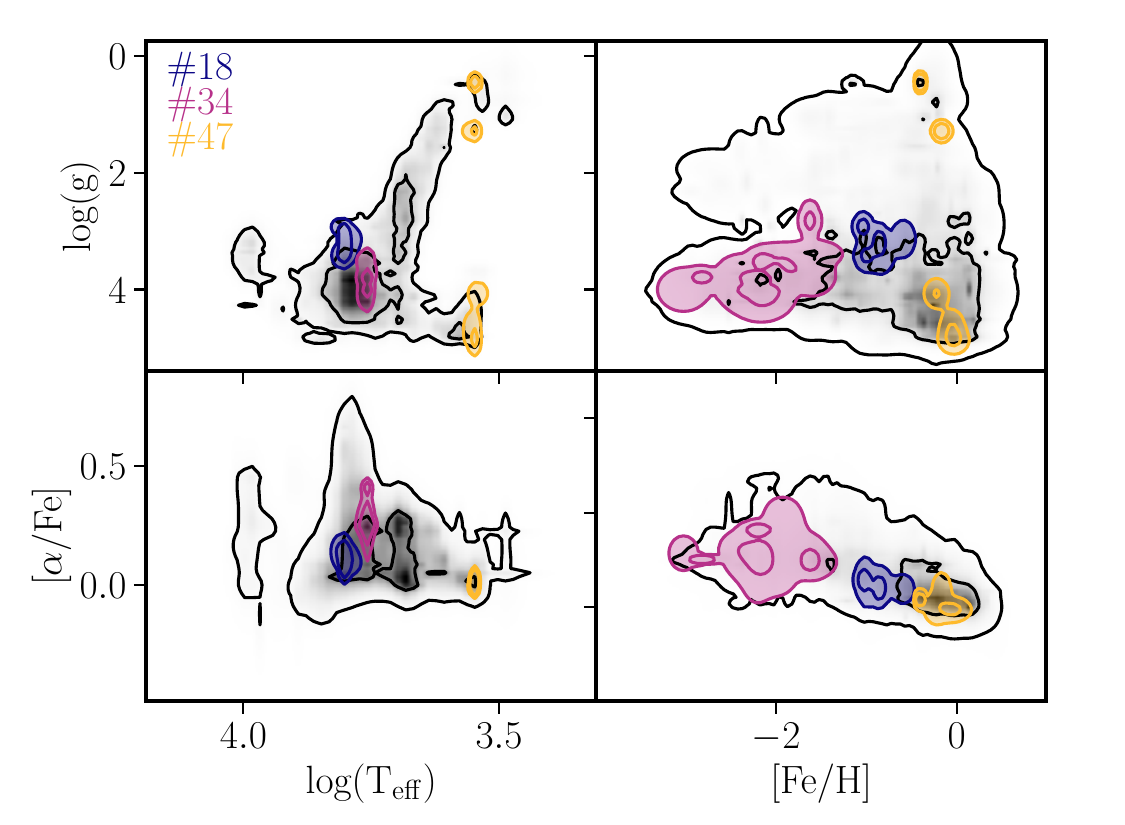}
 \endminipage
 \caption{Probability distribution function of the physical properties of stars (T$_{eff}$, log($g$), [Fe/H] and [$\alpha$/Fe]) for the full template comprising 108 RSPs (black contours), together with the same distribution for three selected RSPs within the template (colour contours) shown in Fig. \ref{fig:MaStar_spec}. Each panel shows the PDFs for a pair of physical properties: T$_{eff}$-log($g$) (top-left); [Fe/H]-log($g$) (top-right); T$_{eff}$-[$\alpha$/Fe] (bottom-left) and [$\alpha$/Fe]-[Fe/H] (bottom-right).  In each panel each successive contour corresponds approximately to 1, 2, and 3$\sigma$.}
 \label{fig:MaStar_PDF}
\end{figure}

Figure \ref{fig:MaStar_spec} illustrates the result of this analysis, showing the spectra of a subset of the RSP library generated by selecting 108 clusters, together with the spectra of the original RSPs included in each cluster. By construction, the shape of all spectra grouped in the same cluster is rather similar. However, as already indicated, this does not mean that each of those spectra corresponds to the same or similar physical parameters. This is evident when exploring the projected PDFs of each cluster in the space of observed parameters. Figure~\ref{fig:MaStar_PDF} shows three examples of PDFs corresponding to three different clusters whose spectra are shown in Fig.~\ref{fig:MaStar_spec} (\#18, \#34, and \#47), together with the overall distribution for the full RSP library comprising 108 spectra. A simple exploration of the distribution of each PDF in the space of physical parameters shows that for some parameters the RSP corresponds to almost a single value (e.g., $T_{\mathrm{eff}}$), while for other parameters they present a bi-valuated (or multi-valuated) distribution (e.g., log($g$)). Besides this multi-valuated distribution, the errors and degeneracies are considerably different for each parameter and pair of parameters, also varying from RSP to RSP. For instance, $T_{\mathrm{eff}}$ and log($g$) do not present any clear degeneracy, while [Fe/H] and [$\alpha$/Fe] present a degeneracy that is described by an anticorrelation of the PDFs of both parameters. We should stress that these multi-valuated distributions, degeneracies, and errors can be mitigated, for instance, by the selection of a larger number of clusters ($n_{cl}$). Obviously, the selection of a very low $n_{cl}$ number, e.g., 4-12, increases them. However, increasing this number beyond 108 does not produce any significant improvement in the results, as we will translate the degeneracy to the fitting procedure: the DAP algorithm would choose a different RSP among the degenerated ones in each MC iteration described in Sec. \ref{sec:proc}. This will only slow down the process without improving the quality of the modeling.  Using external information, like the distance provided by GAIA \citep{gaia2} and the absolute magnitude may break some of the degenerancies. . { We may explore how to implement them in future versions of the DAP, or as a part of the post-processing of the data required to interpret the results. Nevertheless this is out of the scope of the goals of the current manuscript, aimed to describe the overall procedure, without focusing on the physical interpretation of the results, which would require a much more detailed exploration.}

{ 

\begin{table}
\begin{center}
\caption{Description of the RSP template file.}
\begin{tabular}{clll}\hline\hline
HDU	&  EXTENSION & \# Format & \# Dimensions\\
  \hline
  0  & SPECTRA           &  RSS   & (\#WAVE, \#RSP) \\ 
  1  & PARAMETERS        &  Table & (\#PAR, \#RSP)\\   
  2  & ERROR             &  RSS   & (\#WAVE, \#RSP) \\    
  3  & ORG\_PARAM        &  Table & (\#PAR, \#RSP\_ORG)\\   
  4  & PDF\_TEFF\_LOGG   &  Cube & (NX, NY, \#RSP)\\ 
  5  & PDF\_TEFF\_MET    &  Cube & (NX, NY, \#RSP)\\        
  6  & PDF\_TEFF\_ALPHAM &  Cube & (NX, NY, \#RSP)\\        
  7  & PDF\_MET\_ALPHAM  &  Cube & (NX, NY, \#RSP)\\          
  8  & PDF\_MET\_LOGG    &  Cube & (NX, NY, \#RSP)\\       
  9  & PDF\_LOGG\_MET    &  Cube & (NX, NY, \#RSP)\\         
\hline
\end{tabular}\label{tab:hdu_rsp} 
\end{center}
{
Structure of extensions included in the each RSP file, where: (i) \#WAVE is the number of spectral pixels; (ii) \#RSP is the number of templates in the library; (iii) \#PAR is the number of parameters associated to each template, as described in the text; (iv) \#RSP\_ORG is the number of RSP templates or stellar spectra from the library adopted to build the described RSP template; and (v) (NX,NY) are the dimension of the projection of the PDF in the considered space of parameters.}
\end{table}

\subsubsection{RSP template library format}

We store the result of the clustering procedure described previously in a single FITS file with multiple extensions that comprises: (i) the RSP spectra themselves stored in a raw-stacked spectra (RSS) format (SPECTRA); (ii) the uncertainty spectra that retain the dependencies among the spectra coadded to create each RSP template, in the same format (ERROR); (iii) the LW average physical parameters that correspond to each RSP template (PARAMETERS), together with the same values for the original stellar library or RSP template from which the current template was built (ORG\_PARAM); and (iv) the PDFs projected in the space of physical parameters associated to each template, as shown in Fig. \ref{fig:MaStar_PDF}, required to create Fig. \ref{fig:MaStar_fit_PDF} and \ref{fig:dap_PDF} (PDF\_TEFF\_LOGG, PDF\_TEFF\_MET, PDF\_TEFF\_ALPHAM, PDF\_MET\_ALPHAM,, PDF\_MET\_LOGG, PDF\_LOGG\_MET). These PDFs are stored as cubes in which each slice along the z-axis corresponds to the individual PDF projection for the considered pair of parameters (e.g., PDF\_TEFF\_LOGG corresponds to T$_{eff}$ vs. log(g)) for each RSP template in the library. In this way,  the PDF corresponding to a particular stellar decomposition can be generated by multiplying each slice of the cube by the corresponding weight/coefficient of the decomposition and averaging all of them. A set of RSP-template files are distributed (\url{https://ifs.astroscu.unam.mx/sfsanchez/lvmdap/}) together with some python notebooks illustrating how to handle them (included in the DAP distribution).
}

\subsection{Data products} \label{sec:DP}

The full analysis outlined in the previous sections provides four different sets of parameters for each observed spectrum: (i) the average physical parameters that describe the physical properties of the stellar component, (ii) the coefficients or weights of the decomposition of the stellar population in the considered template library (COEFFS hereafter), (iii) the emission line properties derived by the parametric analysis (PM\_EL hereafter), once subtracted the stellar component model (see Fig. \ref{fig:scheme}), and (iv) the emission line properties derived by the non-parametric analysis (NP\_EL hereafter). All these parameters (and their corresponding errors) are stored in a single FITS file for each analyzed frame, which contains the result of the analysis of all the spectra observed in a single pointing. The file consists of a set of extensions with different FITS tables, one for each set of parameters provided by the analysis. Each row in each table corresponds to a science fiber in the original pointing/frame, and each column stores one of the derived parameters (or its corresponding error), plus a unique ID shared among all tables that allows us to identify, assign, and trace back the parameters to a certain fiber and a certain location on the sky. Two additional tables are included: one storing the position table, i.e., the mapping in the sky of each fiber of each pointing to a certain RA and DEC (PT hereafter), and another storing the set of parameters adopted for the current analysis (INFO hereafter). To facilitate traceability, the original header of the FITS file corresponding to the analyzed frame is stored in the file too (being the PRIMARY extension). { We adopted the following units for the different quantities stored in the different extensions of the DAP file: (i) all emission line flux intensities, either derived using the parametric or non-parametric analysis are given in 10$^{-16}$ erg s$^{-1}$ cm$^{-2}$ spaxel$^{-1}$; (ii) flux densities measured in the continuum are given in  10$^{-16}$ erg s$^{-1}$ cm$^{-2}$ spaxel$^{-1}$ $\AA^{-1}$; (iii) EWs are given in $\AA$; (iv) both stellar (v$_\star$) or emission line velocities (v$_{EL}$) are given in km s$^{-1}$; (v) stellar velocity dispersion ($\sigma_\star$) is given in  km s$^{-1}$, assuming that the resolution of the RSP templates have been matched to that of the data; and finally (vi) the velocity dispersion of the emission lines ($\sigma_{EL}$), is given in $\AA$, uncorrected for the instrumental resolution at the considered wavelength.}  The data format of this FITS file, which we name the DAP file, is summarized in Table \ref{tab:hdu} \footnote{ An example of a DAP file ({\sc  sim\_example.dap.fits}) is distributed in the following link: \url{https://ifs.astroscu.unam.mx/sfsanchez/lvmdap/}}.

\begin{table}
\begin{center}
\caption{Description of the DAP file.}
\begin{tabular}{clll}\hline\hline
HDU	&  EXTENSION & \# Rows & \# Columns\\
  \hline
  0  & PRIMARY     &  &  \\      
  1  & PT          &  \#spec &  6   \\
  2  & RSP         &  \#spec & 1 + \#RSP \\
  3  & COEFFS      &  \#spec $\times$ \#RSP & 13 \\
  4  & PM\_ELINES   &  \#spec $\times$ \#PM\_EL & 10 \\
  5  & NP\_ELINES\_B &  \#spec &  1+\#NP\_EL\_B $\times$8\\
  6  & NP\_ELINES\_R &  \#spec &  1+\#NP\_EL\_R $\times$8\\
  7  & NP\_ELINES\_I &  \#spec &  1+\#NP\_EL\_I $\times$8\\
  8  & INFO        &  \#param & 2 \\
\hline
\end{tabular}\label{tab:hdu} 
\end{center}
Structure of extensions included in the each DAP file, where: (i) \#spec is the number of science spectra (or fibers) included in the analyzed {\tt Tile} or RSS (row-stacked spectra) frame; (ii) \#RSP is the number of
templates/spectra included in the stellar library; (iii) \#PM\_EL is the number of individual
models (emission lines) included in the parametric analysis of the ionized gas emission lines, and
(iv) \#NP\_EL\_BAND is the number of emission lines included in the non-parametric analysis for each BAND (B, R and I) corresponding to each arm of the spectrograph. 
\end{table}

\section{Reliability of the method} \label{sec:code}

As indicated before, the core routines included in the DAP have been inherited from the \pyf\ algorithms. Thus, they have been extensively tested in previous studies \citep[e.g.,][]{pypipe3d}. However, we have introduced several modifications, particularly in the derivation of the non-parametric parameters of the stellar component (vel$_\star$, $\sigma_\star$, and A$_{V,\star}$), as described before (Sec. \ref{sec:proc}). In addition, the ability of the method to recover the properties of the stellar component using the newly introduced RSP methodology has never been tested. Finally, the algorithms were never used to analyze data of this high spectral resolution, neither for the stellar component nor for the emission lines. For all these reasons, we have applied the procedure to a set of simulations in order to determine the quality and accuracy of the recovered parameters.

\begin{figure*}
 \minipage{0.99\textwidth}
 \includegraphics[width=9cm,clip,trim=10 0 10 15]{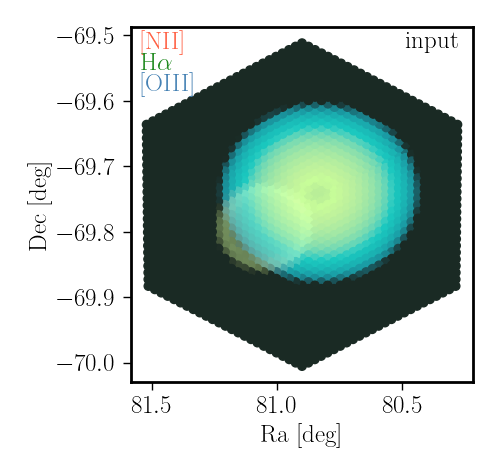}\includegraphics[width=9cm,clip,trim=10 0 10 15]{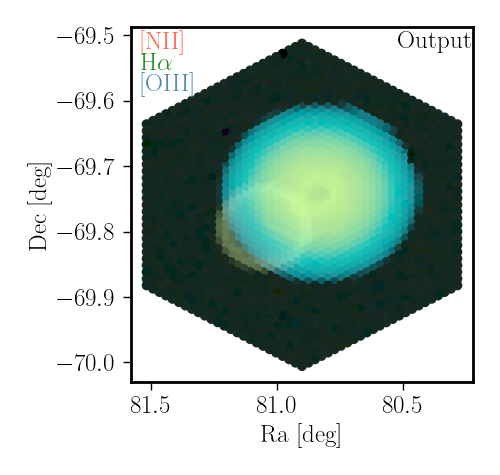}
 \endminipage
 
\minipage{0.99\textwidth}
 \includegraphics[width=18cm,clip,trim=30 0 50 10]{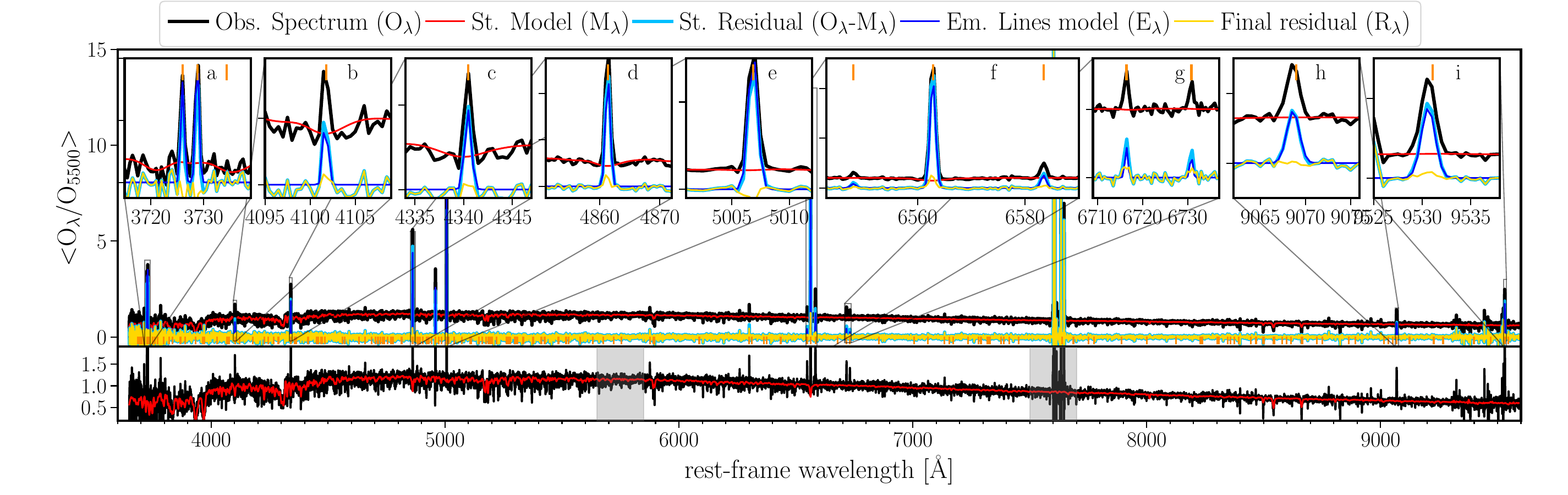}
 \endminipage
 \caption{Upper-panels: Spatial distribution of the input (left panel) and recovered (right panel) fluxes of the \oiii 5007 (blue), H$\alpha$ (green) and \nii 6583 (red) emission lines in an arbitrary scale. Lower-panel: Example of the spectral modelling, showing the spectrum integrated across the Field-of-View (FoV) of the simulated IFS dataset (black solid line), together the best model for the stellar component (red solid line) and the set of strong ionized emission lines (blue solid line). The residuals of the subtraction of the stellar model (cyan solid line), and both models (yellow solid line) are also included. All the spectra are presented in the original flux intensities (middle panel), and normalized to one at 5000 \AA~(bottom panel). The orange vertical marks indicate the wavelengths at which the algorithm would attempt to extract the properties of a predefined set of emission lines using a weighted moment analysis, as described in the text. Shaded regions correspond to the overlapping regimes between the three arms of the spectrographs. The panel insets show zoom-ins in the wavelength range covered by a set of strong emission lines, including the \oii\ doublet, H$\delta$, H$\gamma$, \Hb, \oiii, \Ha, \nii, \sii, and the \siii 9069, 9531 doublet. }
 \label{fig:lvmsim}
\end{figure*}

\begin{figure*}
 \minipage{0.99\textwidth}
 \includegraphics[width=9cm,clip,trim=20 10 50 20]{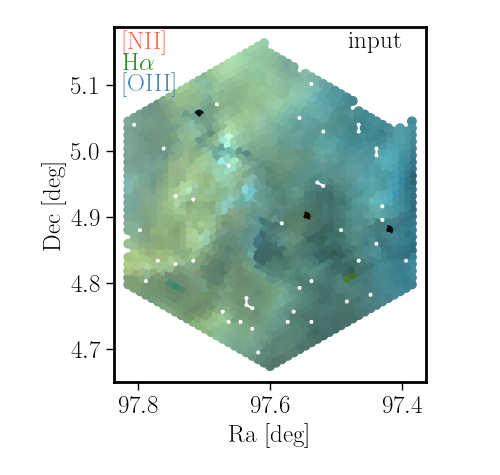}\includegraphics[width=9cm,clip,trim=20 10 50 20]{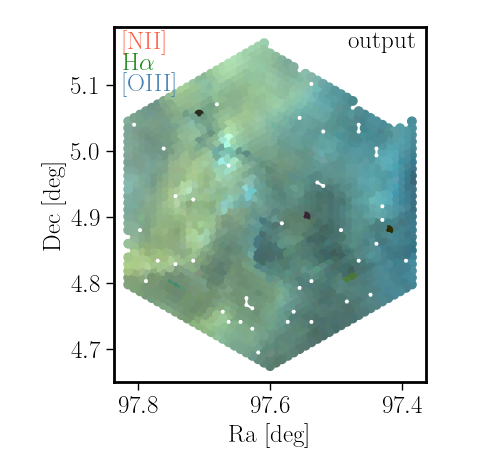}
 \endminipage
 
\minipage{0.99\textwidth}
 \includegraphics[width=18cm,clip,trim=30 0 20 10]{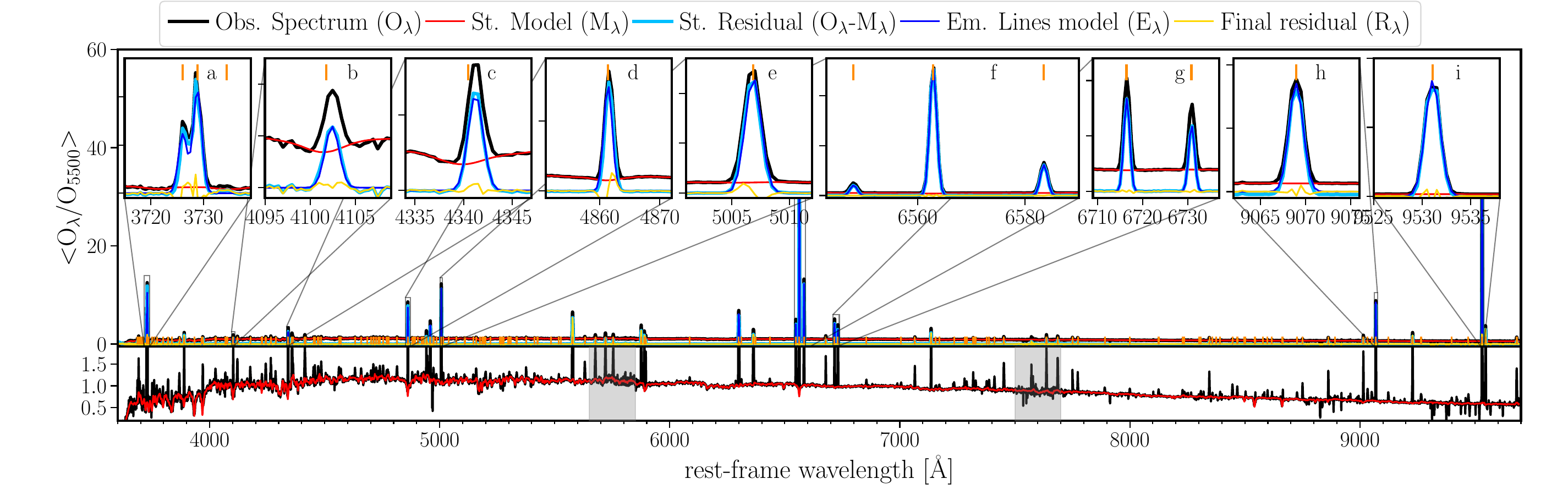}
 \endminipage
 \caption{Similar figure as the one shown in Fig. \ref{fig:lvmsim}, adopting the same nomenclature, symbols and color schemes, for the realistic simulation performed using one LVM pointing on the Rosetta nebula. White hexagons inside the IFU corresponds to broken and low-transmission fibers in the input data.}
 \label{fig:dapsim}
\end{figure*}

\subsection{Idealized simulations}
\label{sec:sim_idea}

We constructed two different sets of simulations to test the DAP. The LVM data simulator\footnote{\url{https://github.com/sdss/lvmdatasimulator}} creates a full simulation of an LVM observed frame. It simulates an idealized nebula with particular geometry, kinematics, physical conditions, and a known ionizing source (or set of sources). Based on the required parameters and pre-generated grids of photoionization models, it generates the corresponding emission lines that should be observed at different locations within the nebula in three-dimensional space. If needed, it can add not only the spectra of the ionizing source but also additional stellar spectra. Using this information, it reproduces the LVM observation by sampling the simulated target through a set of apertures that mimic the science fiber bundle, adapting the spectral resolution, resampling the spectra, and adding white noise to match the nominal values of the LVM observed spectra at the specified atmospheric conditions and sky brightness. Details on the code and the procedure to generate these mock observations will be presented elsewhere (Congiu, Egorov et al., in prep.).

Figure \ref{fig:lvmsim} shows an example of these idealized simulations, together with the results of its analysis by the DAP. In this particular case, the simulation encompasses a complex astrophysical scenario involving an expanding \hii\ region, a supernova remnant (SNR), diffuse ionized gas (DIG), and a set of foreground stars, observed at a distance of 50 kiloparsecs. This setup aims to provide insights into the interactions and emission characteristics of these various components under specific environmental conditions resembling the LMC.

The expanding \hii\ region in this simulation is characterized by a maximum H$\alpha$ brightness of $10^{-14} \text{ erg s}^{-1} \text{ cm}^{-2} \text{ arcsec}^{-2}$. To simulate this \hii\ region, we assume a gas metallicity of 0.6 $Z_\odot$, a hydrogen density of $100 \text{ cm}^{-3}$, an effective temperature for the ionizing source of 30,000 K, adopting a black-body spectrum, and a bolometric luminosity of $\log(L/L_\odot) = 6.5$. This source produces an ionizing photon flux of $\log Q = 50.0$. We consider that the nebula consists of a thick bubble expanding at 15 km s$^{-1}$, with a radius of 120 pc and an inner radius of 20 pc. This way, it fits within the Field-of-View (FoV) of the LVM IFU. A constant intrinsic velocity dispersion of 15 km s$^{-1}$ is assumed for all emission lines across the entire nebula.

Surrounding this region, the DIG exhibits a lower H$\alpha$ brightness of $5 \times 10^{-17} \text{ erg s}^{-1} \text{ cm}^{-2} \text{ arcsec}^{-2}$, reflective of less intense but widespread ionization typical in extended galactic structures. To simulate this component, we assume a similar metallicity, electron temperature, and density. However, a higher intrinsic velocity dispersion, 20 km s$^{-1}$, and a lower luminosity, $\log(L/L_\odot) = 5.5$, were considered. Thus, the resulting ionizing photon flux is slightly higher than $\log Q = 50$. The DIG component is assumed to be at the same distance as the nebula.

The emission lines produced by the \hii\ and the DIG components were estimated using the {\tt Cloudy} v17.02 photoionization code \citep{ferland98, ferland2017}, using the previously described parameters and considering a spherical distribution. The line ratios remain constant across the DIG component and correspond to the integrated 1D spectrum from the Cloudy model. For the \hii\ region component, line ratios in each volumetric element change with the radial distance according to the emissivities in the selected photoionization model. The 2D distribution in each line is obtained by integrating the 3D model along the line-of-sight.


The SNR component exhibits a maximum H$\alpha$ brightness of $2 \times 10^{-15} \text{ erg s}^{-1} \text{ cm}^{-2} \text{ arcsec}^{-2}$. The SNR spectrum is modeled using the \texttt{MAPPINGS} code \citep{dopita1996}, which includes shock ionization (not covered by \texttt{Cloudy}). It is assumed a pre-shock density of $1.0 \text{ cm}^{-3}$, a shock velocity of 500 km s$^{-1}$, a pre-shock temperature of 21,539 K, and a magnetic field strength of 1 Gauss. The SNR is located to the southeast of the center of the \hii\ region. It consists of a smaller (60 pc), thinner (0.2 pc) but faster expanding bubble (100 km s$^{-1}$). Like in the case of the \hii\ region, a constant (but slightly higher) velocity dispersion of 25 km s$^{-1}$ is adopted. Given that the radial distribution of the emisivities is not accessible for these MAPPINGS models, we assume a constant line ratio across the SNR component.

Finally, a set of foreground or background stars has been added to the simulated dataset. These stars were obtained from GAIA DR2 \citep{gaia2}. We selected all stars brighter than $g=14$ mag in the FoV, assuming that it is centered on R.A. $\sim 80.9^\circ$ and Dec. $\sim -69.75^\circ$. The spectra for these 151 stars were obtained from the Pollux stellar spectra database \citep{palacios2010} by selecting the model with the closest $T_{eff}$ to what is provided by GAIA.

The resulting simulated dataset is treated as indicated before to realize an observation through the LVM instrument. The resulting ``observed" spectra are analyzed using the LVM DAP. In Figure~\ref{fig:lvmsim}, the top panels show the spatial distribution of the input (left panel) and recovered (right panel) emission line fluxes, adopting a three-color scheme in which \oiii~corresponds to blue, H$\alpha$ to green, and \nii~to red. The two main components in the simulated dataset (i.e., the \hii\ region and the SNR) are clearly appreciated in both images. The similarities in the spatial distribution, flux intensities, and colors illustrate the quality of the fluxes recovered by the DAP. The bottom panel shows the integrated spectrum across the entire FoV of the simulation (black solid line), together with the best-fitted model, including the stellar component (red solid line) and the strong ionized gas emission lines (blue solid line) modeled as Gaussian distributions. The residuals from the subtraction of the stellar component (cyan solid line) and both components (yellow solid line) are also shown. A set of insets at particular wavelengths illustrates the quality of the modeling.

\subsection{Realistic simulations}
\label{sec:sim_real}

The second set of simulations, which we will call ``realistic", uses the outputs of the DAP analysis on real LVM observations. Essentially, it uses the properties of the emission lines extracted by the DAP based on the weighted-moment analysis on a real LVM frame (reference frame) to generate RSS spectra of the ionized gas emission. Each emission line in each spectrum is simulated using a Gaussian function with the integrated flux derived by the DAP for the reference frame. However, the velocity and velocity dispersion are selected randomly, following a normal distribution with a range of values between $\pm$100 km s$^{-1}$ centered at zero (for the velocity) and 40 km s$^{-1}$ (for the velocity dispersion). Thus, by construction, the spatial distribution of the flux intensities of the simulated emission lines is the same as the spatial distribution of those properties in the reference frame. However, the kinematic parameters cover a wider range, more similar to the expected values for the foreseen LVM dataset. In addition, for each fiber, a stellar spectrum is simulated by randomly combining spectra from a selected RSP-template, adding a certain dust attenuation (A$_{V,\star}$), and applying certain kinematics (vel$_\star$, $\sigma_\star$). In this step, the number of stars to include in the stellar component, the adopted RSP-template, and the range of values to pick a certain set of non-linear parameters are input parameters of the simulation.

\begin{figure*}
 \minipage{0.99\textwidth}
 \includegraphics[width=8.5cm,clip,trim=0 0 0 0]{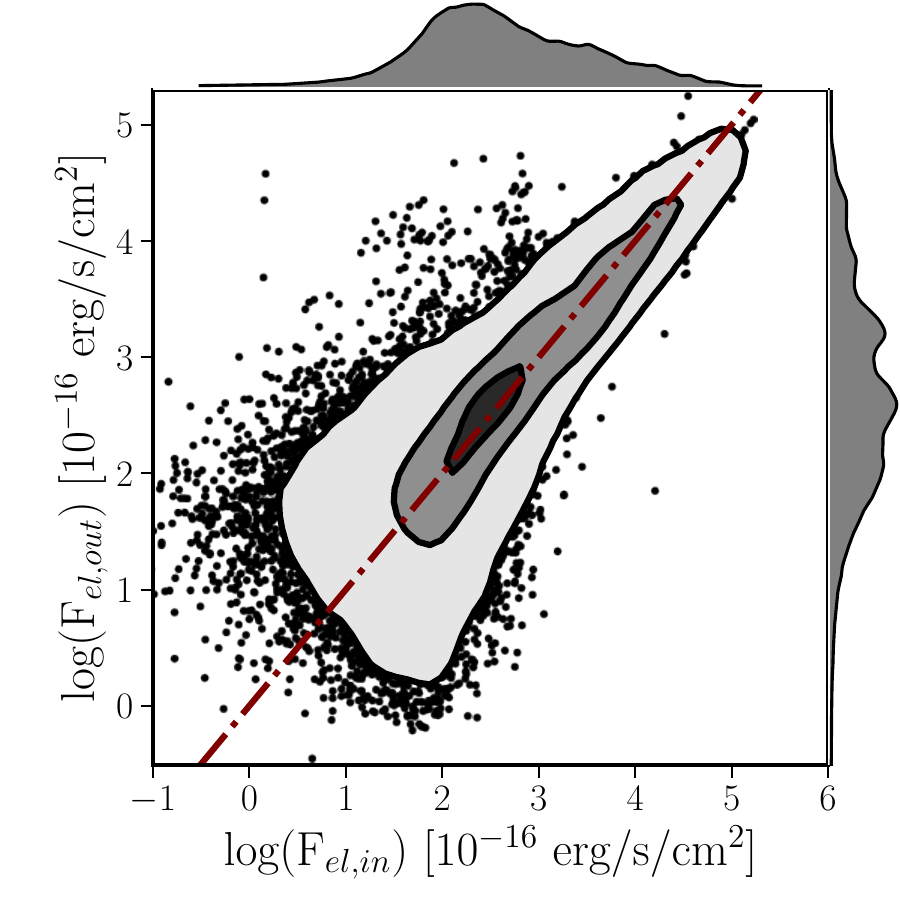}\includegraphics[width=8.5cm,clip,trim=0 0 0 0]{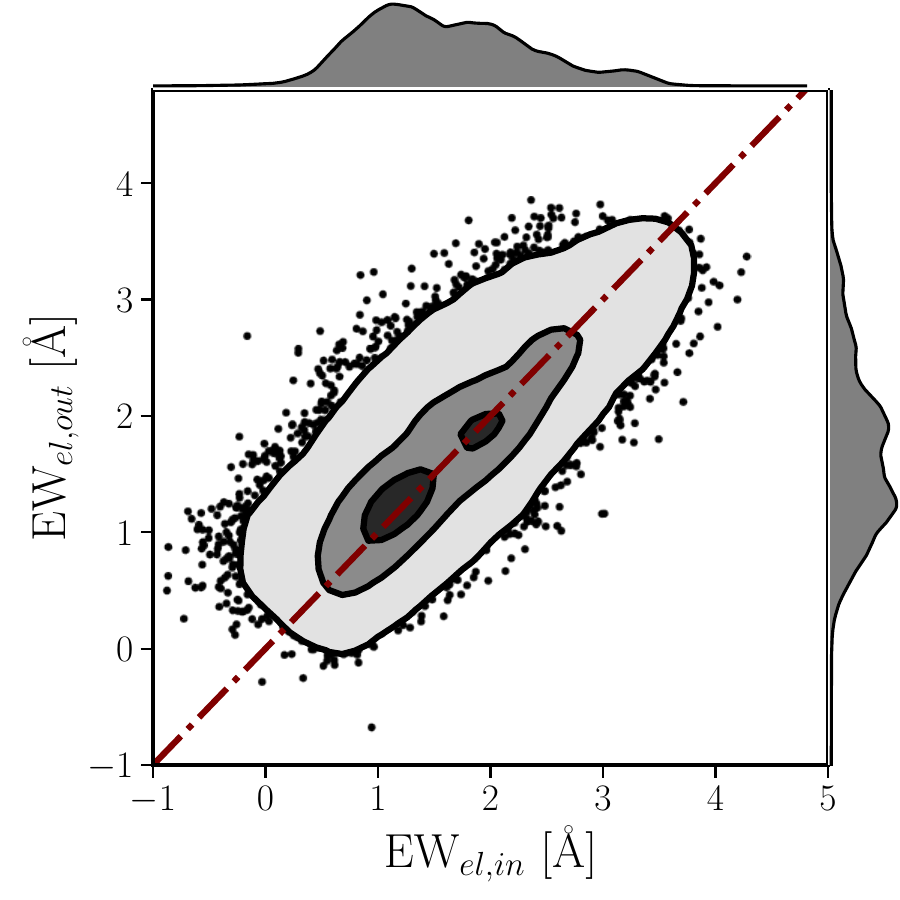}
 \includegraphics[width=8.5cm,clip,trim=0 0 0 0]{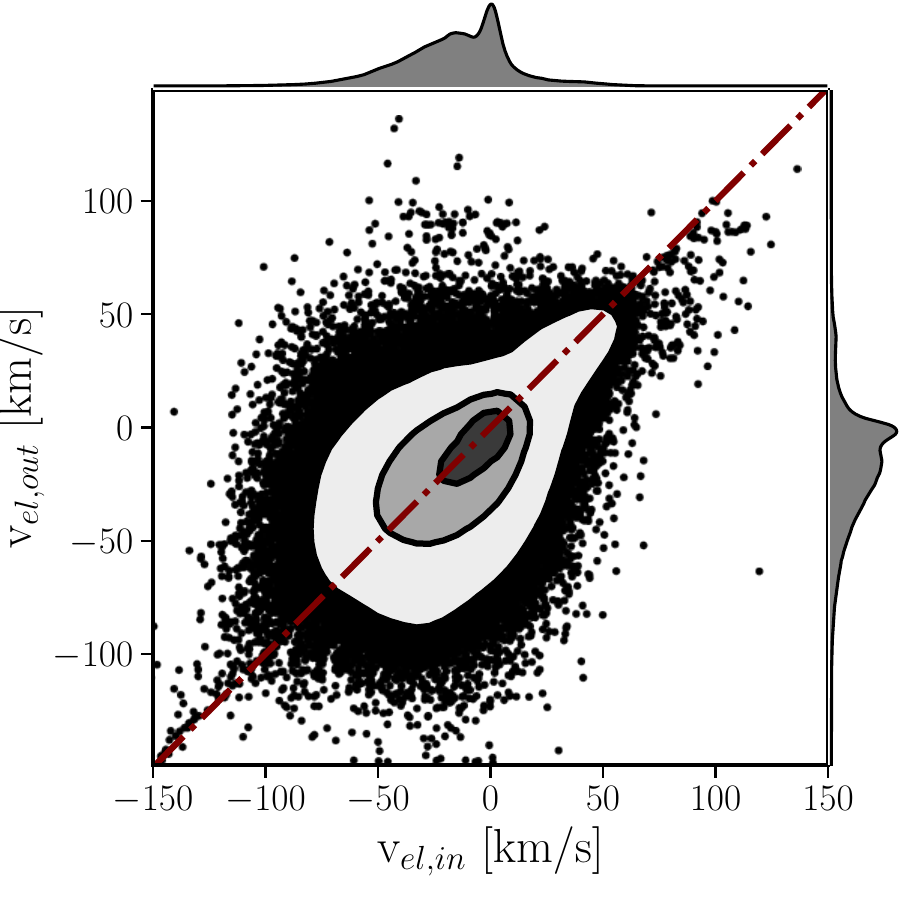}\includegraphics[width=8.5cm,clip,trim=0 0 0 0]{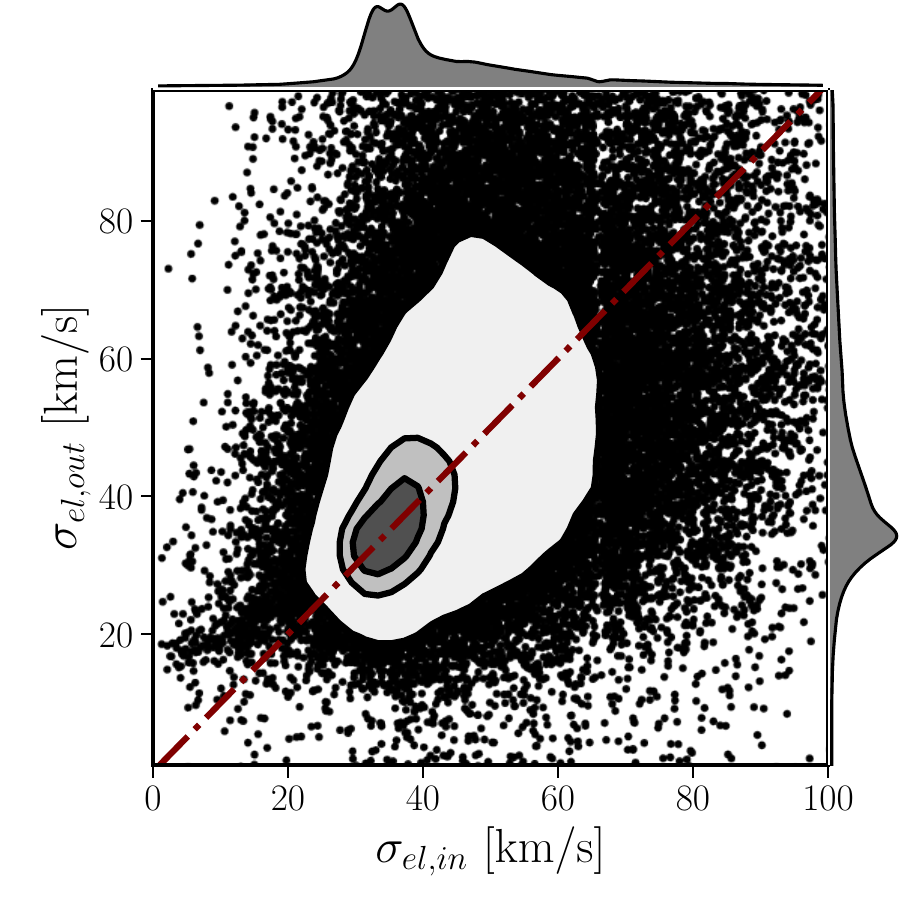}
 \endminipage
 \caption{Comparison between the simulated and recovered parameters for the emission lines based on the non-parametric analysis described in Sec. \ref{sec:proc}, for the ad-hoc realistic simulations described in Sec. \ref{sec:sim_real}. A total of 326400 emission lines were included in the simulation. We represent here the results for the 44979 ones with a S/N$>$3. Each panel shows the distribution the flux intensity (top-left), EW (top-right), velocity (bottom-left),  and velocity dispersion (bottom-right). { In each panel, black solid circles represent the individual values and contours represent the density, encircling 95\%, 65\% and 15\% of the points, respectively. Dark-red dashed-dotted lines represent the one-to-one relation.
 }} 
\label{fig:dap_el}
\end{figure*}

The spectral range, spectral resolution, and sampling are the same as those in the original dataset. To simulate a realistic noise distribution, including both white and systematic effects, we used the residuals of the DAP modeling of the reference frame to construct the noise spectra. For each aperture, a noise spectrum is built by picking, for each spectral pixel, the value at the same wavelength in one of the spectra within the residual RSS. Spectral pixel by spectral pixel and aperture by aperture, the spectrum from which the residual value is chosen randomly. In this way, the resulting noise spectrum for each aperture corresponds to a possible realization of the noise distribution described by the residual RSS. Finally, two scaling factors are introduced to control the relative strength of the emission line intensities and the stellar component with respect to the average noise level. In this way, we can control the S/N of both components for the simulated data. It is worth noting that in almost any simulation, due to the large dynamical range of flux intensities covered by all the pool of emission lines analyzed by the DAP (192 emission lines in the current implementation), there is a wide range of S/N covered for the bulk of emission lines. By construction, this procedure generates a noise distribution that is more similar to the real one than the pure white noise implemented in the idealized simulations discussed in the previous section. However, contrary to the previous procedure, it does not allow us to control the real (input) physical conditions of the explored data, being only valid for a particular observing setup and DRP version. Thus, it lacks the generality offered by the LVM simulator.

Figure \ref{fig:dapsim} is equivalent to Fig. \ref{fig:lvmsim} for a realistic simulation. In this particular case, we adopted as a reference frame one pointing near the center of the Rosette nebula. Like in the case of the idealized simulation, it is clear that the distribution of emission line intensities for the represented emission lines is very well recovered by the code. The accuracy of the model of the stellar and ionized emission line components is qualitatively illustrated by the spectra shown in the middle and lower panels.

\subsection{Recovery of emission line parameters} 
\label{sec:sim_el}

In the previous sections, we described the set of simulations adopted to assess the quality of the parameters recovered by the DAP, presenting a qualitative comparison between the input and output values. In this section, we present a more quantitative statement on the accuracy of the recovered parameters.

Figure \ref{fig:dap_el} shows a comparison between the simulated and recovered parameters for the simulation shown in Fig. \ref{fig:dapsim}: flux intensity (F$_{el}$), equivalent width (EW$_{el}$), systemic velocity (v$_{el}$), and velocity dispersion ($\sigma_{el}$). We cover a wide range of S/Ns of the emission lines naturally by including all the set of emission lines explored by the non-parametric procedure described in Sec. \ref{sec:proc}, comprising S/N values well below 1 and well above 100, depending on the particular emission line. We must remind the reader that this procedure was introduced in \citet{pipe3d_ii} to extract the flux intensities and EWs of weak emission lines tying (or even fixing, if desired) the kinematics parameters to those of the strongest emission lines. Simulations demonstrate that for low-S/N this non-parametric approach is able to recover those parameters with a better precision and accuracy \citep{pypipe3d}. The S/N of the stellar-component continuum ranges between 1 and 100 for the full set of simulations, comprising a mix of 5 to 125 individual stars. By construction, the contrast between the emission lines and the stellar continuum changes naturally with the scaling of the stellar component. The four explored parameters are distributed along the one-to-one relation, without an evident offset, bias, or deviation from that trend.


\begin{figure*}
 \minipage{0.99\textwidth}
 \includegraphics[width=8.5cm,clip,trim=0 0 0 0]{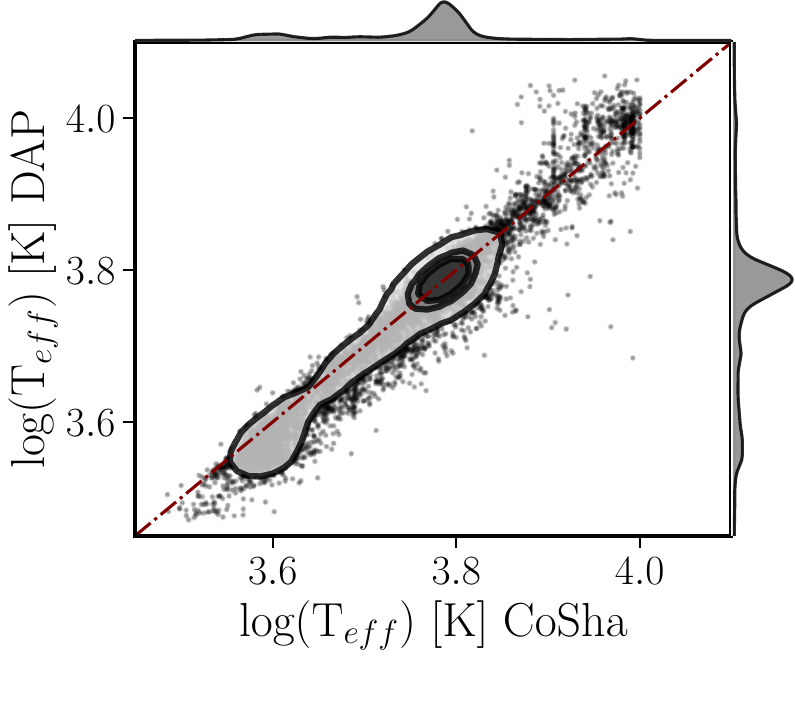}\includegraphics[width=8.5cm,clip,trim=0 0 0 0]{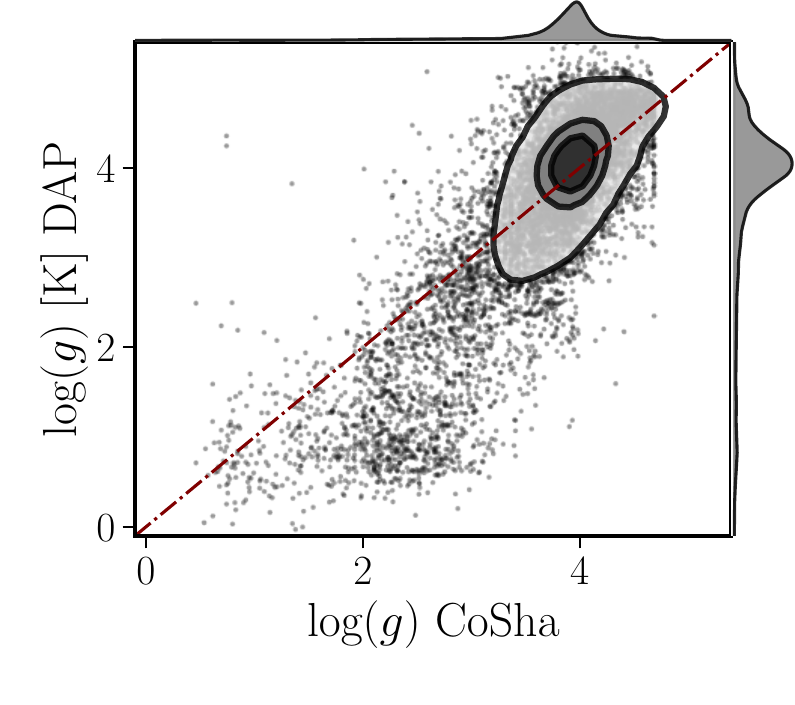}
  \includegraphics[width=8.5cm,clip,trim=0 0 0 0]{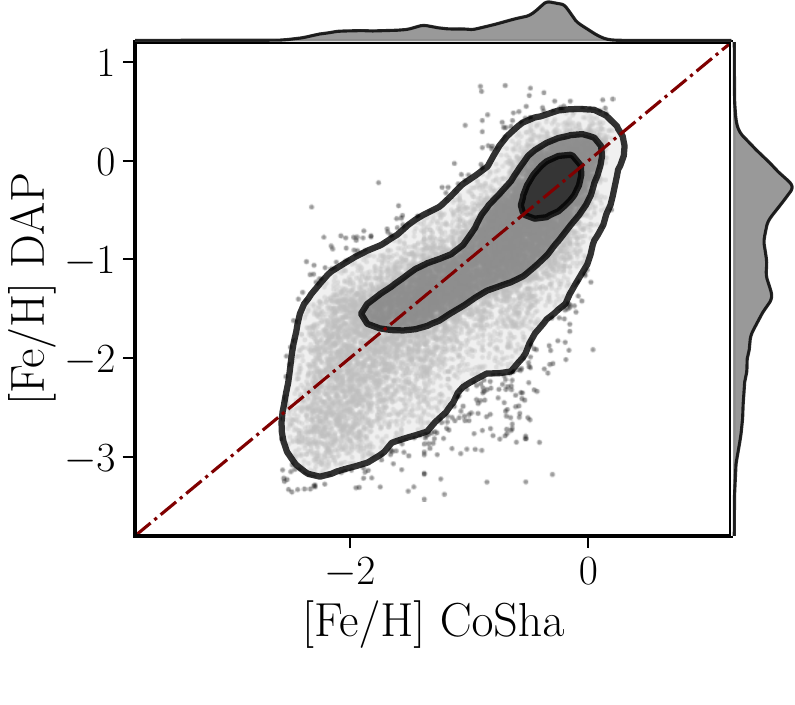}\includegraphics[width=8.5cm,clip,trim=0 0 0 0]{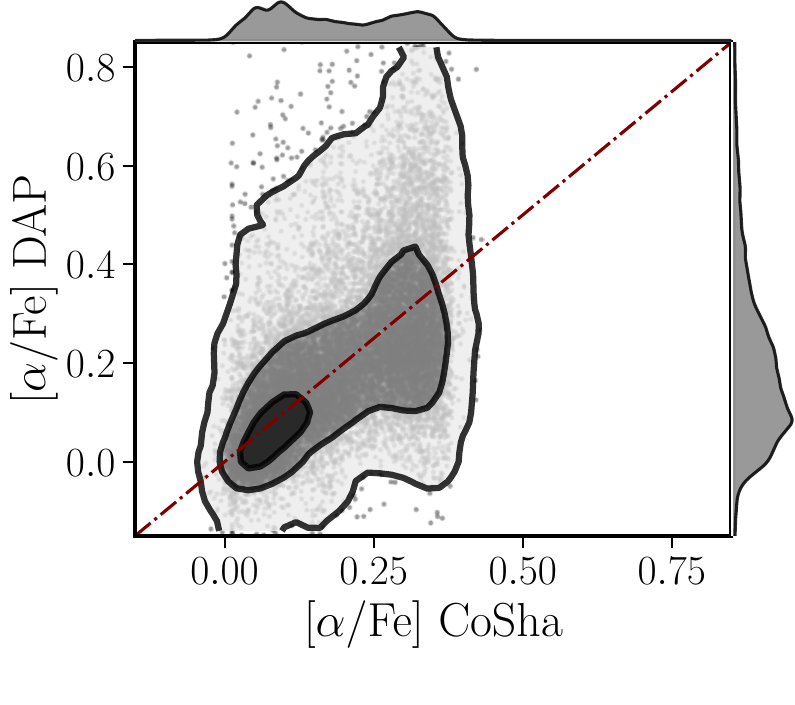}
 \endminipage
 \caption{Comparison between the stellar physical parameters (T$_{eff}$: top-left, log($g$): top-right ,[Fe/H]: bottom-left, and [$\alpha$/Fe]: bottom-right) estimated by the LVM-DAP using a RSP-library with just 108 templates, and the values reported by CoSha for the $\sim$19,000 individual MaStar stars with S/N$>$10. { Contours, symbols and lines in each panel have the same meaning as those in Fig. \ref{fig:dap_el}.}}
 \label{fig:cosha}
\end{figure*}

{ 
%
%
\begin{table}[t]
\begin{center}
\caption{Accuracy and precision in the recovery of the emission line parameters}
\begin{tabular}{lcccc}\hline\hline
S/N  & $\Delta$F$_{el}$/F$_{el}$ & $\Delta$EW$_{el}$/EW$_{el}$  & $\Delta$v$_{el}$ & $\Delta \sigma_{el}$\\ 
\hline
      & (\%)           & (\%)            & (km/s)            & (km/s)\\
$>$5  & 17.2$\pm$24.1  &  5.4$\pm$17.3   &  -7.3$\pm$23.1    & -0.4$\pm$13.3 \\
$>$10  &  0.0$\pm$12.9  &  1.6$\pm$10.6   &   0.3$\pm$3.0     & -0.3$\pm$3.1  \\
$>$15 & -0.3$\pm$6.7   &  1.3$\pm$9.5    &   0.2$\pm$2.0     &  0.1$\pm$0.9  \\
$>$20 & -0.5$\pm$8.0   &  1.8$\pm$9.4    &   0.1$\pm$1.2     &  0.0$\pm$0.8  \\
$>$30 & -0.1$\pm$3.0   &  3.3$\pm$12.1   &  -0.8$\pm$1.5     & -0.1$\pm$0.6  \\
\hline
\end{tabular}\label{tab:el_sim}\\
\end{center}
\end{table}


Each parameter is recovered with a different level of quality, that we characterize by the precision (average offset between output and input value for a particular parameter) and accuracy (standard deviation of offset between output and input value for a particular parameter): (i) $\Delta$ log(F$_{el}$) = 0.09$\pm$0.29 dex; (ii) $\Delta$ EW$_{el}$ = 0.10$\pm$0.37 $\AA$; (iii) $\Delta$ v$_{el}$ = -0.51 $\pm$ 20.78 km s$^{-1}$; and (iv) $\Delta$ $\sigma_{el}$ = 0.45$\pm$10.97 km s$^{-1}$. As expected, the quality of the recovery of the parameters depends the S/N, as illustrated by Table \ref{tab:el_sim}, with the largest offsets (worst precision) and deviations (inaccuracies) for the lowest S/N, and the lowest offsets (best precision) and deviations (accuracies) for the highest S/N. It is worth noting that there is a maximum accuracy even in the case of the highest S/N, that is usually associated with systematic effects (such as resolution), as already noted in previous studies \citep[e.g.][]{pipe3d,pypipe3d}.


}

{ We adopt as a figure of merit of the accuracy of the recovery of a certain parameter the fraction of simulations that presents deviations lower than 10\% (20\%) of the original value.
This deviation corresponds to the typical error for an emission line with S/N$\sim$10 ($\sim$5), fully dominated by its own photon noise. For the flux intensities this fraction corresponds to  82\% (93\%) of the simulations. This value is reduced } to 42\% (68\%) for the EW, 30\% (54\%) for the velocity, and 52\% (74\%) for the velocity dispersion. Note that the apparently large relative errors in the velocity are due to the fact that most of the input velocities are clustered around zero. Thus, the best recovered parameter is the flux intensity, and the worst ones being the kinematic parameters { for the full set of emission lines}. Similar results (for the emission lines) are found for the complete set of simulations { ,  consistent with the results from previous studies} \citep[e.g.,][]{pypipe3d}. { This result was indeed} expected as, in general, the zeroth moment of a distribution is better recovered than the subsequent ones as the order increases for the same signal-to-noise. The novelty of these results resides in the much better spectral resolution of the simulated data, which is a factor of 2-3 higher than that adopted in the previous simulations.

We repeated the analysis for the idealized simulations. Despite the differences between both simulation methods, we find similar results regarding the recovery of the properties of the emission lines. { Finally, it is worth mentioning that similar results are obtained when using the values derived using the parametric procedure, in particular for the flux intensities. In fact, the values derived using both methods are compatible in most cases (see Appendix \ref{sec:ec}).}



\subsection{Recovery of the parameters for a single star} \label{sec:sim_MaStar}

The RSP-based methodology adopted by the DAP to analyze the stellar component in the spectra was developed to deal with the wide dynamic range of stars captured by the LVM fibers in each different pointing. At one extreme, the stellar component could be fully dominated by a single star. Before analyzing how well this component is recovered when mixing different stars within the same aperture, the scenario intended to be replicated by the simulations described before, it is important to determine how well the procedure recovers the properties of individual stars. To do so, we ran the DAP for the 
$\sim$20,000 star spectra within the MaStar library adopted to generate the RSP templates as described in Sec. \ref{sec:rsp_der}, fitting each spectrum with the RSP library comprising 108 templates. We recall that for all those stars we have the input stellar parameters provided by the CoSha tagging procedure \citep{mejia21}, and therefore we can compare them with the parameters recovered by the DAP. In this particular implementation, no emission lines are considered in the fitting process. However, the kinematic parameters and the dust attenuation for the stellar component are allowed to vary within the following values: -150 $<$ v$_\star$ $<$ 150 km s$^{-1}$, 0 $<$ $\sigma_\star$ $<$ 30 km s$^{-1}$, and 0 $<$ A$_{\rm V,\star}$ $<$ 1.5 mag.

Figure \ref{fig:cosha} shows the results of this analysis, including, in each panel, the distribution of the stellar parameters recovered by the DAP (T$_{eff}$, log($g$), [Fe/H], and [$\alpha$/Fe]) as a function of the values assigned by CoSha. Although in general the values are distributed around the one-to-one relation for the vast majority of the analyzed stars, there are clear differences in the precision and accuracy of the recovery of each parameter. The effective temperature is by far the best-recovered parameter, without any clear deviation from the one-to-one relation in the full dynamical range covered by the dataset ($\Delta$T$_{eff}$ = -75 $\pm$ 564 K). 80\% (71\%) of the temperatures are recovered within 10\% (5\%) of the original value. The case of log($g$) is quite different. The values, more clearly concentrated around a certain value of $\sim$4 dex, do not present a clear bias only in this regime that dominates the global statistics ($\Delta$log($g$) = -0.02 $\pm$ 0.64 dex). However, for values below $<$ 3 dex, the DAP clearly underestimates this parameter by almost an order of magnitude, with the general trend following an S-shaped distribution. The low number statistics in the range of parameters of the MaStar library, biased towards the vicinity of the Sun \citep[e.g.,][]{mejia22}, could give the erroneous impression that this effect is not relevant. However, the fraction of stars for which the DAP recovers log($g$) within 10\% (5\%) of their real value is just 62\% (40\%). The two remaining parameters that trace the chemical composition of the stars are also distributed along the one-to-one relation without significant biases in a statistical sense: $\Delta$[Fe/H] = -0.03 $\pm$ 0.34 dex and $\Delta$[$\alpha$/Fe] = 0.004 $\pm$ 0.134 dex. However, none of them are as well recovered as the former ones, with just 26\% of [Fe/H] (54\% of [$\alpha$/Fe]) values recovered within 50\% of the input values.

\begin{figure}
 \minipage{0.99\textwidth}
 \includegraphics[width=8.5cm,clip,trim=30 10 80 80]{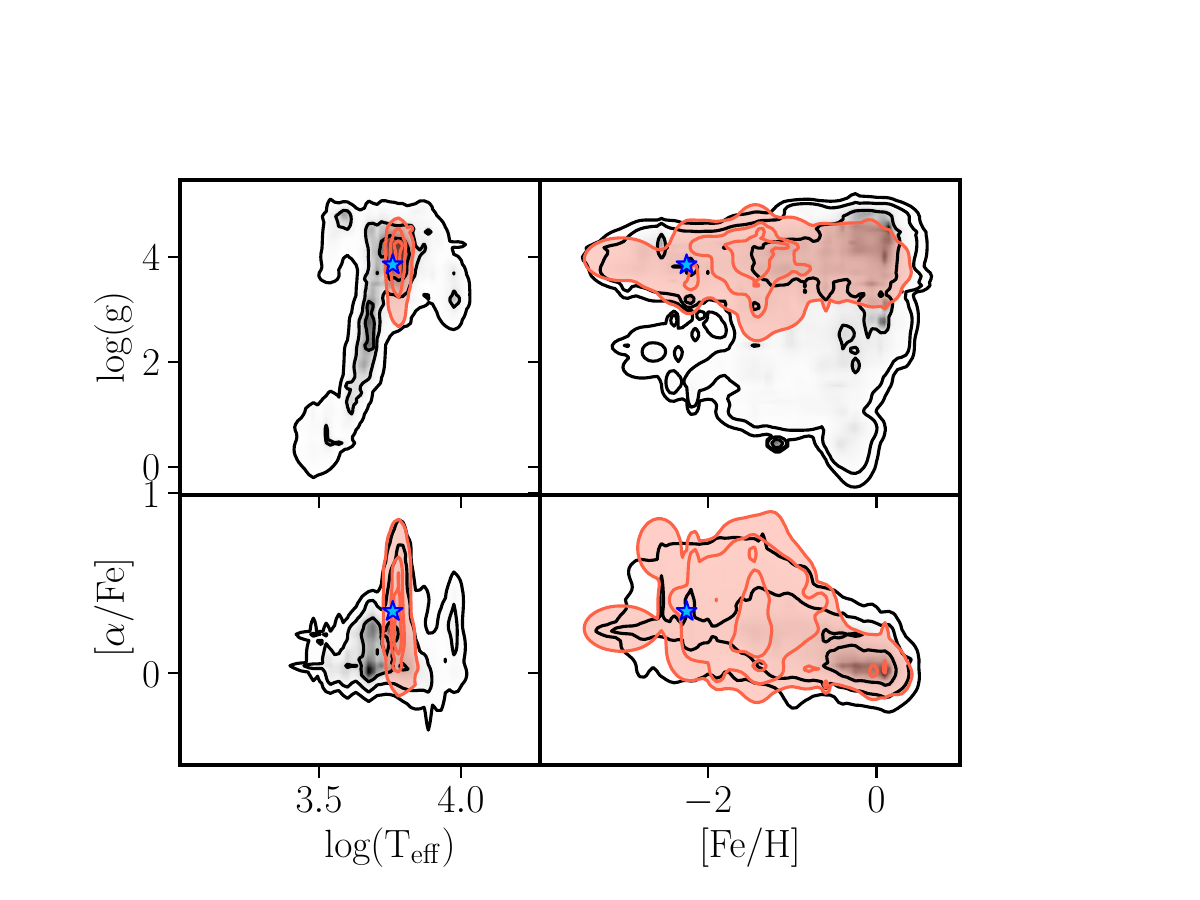}
 \endminipage
 \caption{Comparison between the value of the physical properties  (T$_{eff}$, log($g$), [Fe/H], and [$\alpha$/Fe]) of one arbitrarily selected star of within MaStar catalog as derived by CoSha (blue star) and the PDF recovered by the DAP (red contours). The PDFs for the full template comprising 108 RSPs, already shown in Fig. \ref{fig:MaStar_PDF} has been included for reference (black contours).
 Each panel shows the PDFs for a pair of physical properties: T$_{eff}$-log($g$) (top-left); [Fe/H]-log($g$) (top-right); T$_{eff}$-[$\alpha$/Fe] (bottom-left), and [$\alpha$/Fe]-[Fe/H] (bottom-right). In each panel each successive contour corresponds approximately to 1, 2, and 3$\sigma$.}
 \label{fig:MaStar_fit_PDF}
\end{figure}

It is important to note that the primary goal of the current version of the DAP was not to tag the physical properties of individual stars, but to provide a good model of the underlying stellar spectra to recover the main properties of the emission lines. Indeed, the method does not recover just the best physical parameters corresponding to the stellar component, but the probability distribution functions (PDFs) of those parameters that best represent them. An example of these PDFs for one arbitrary star within the MaStar catalog is included in Fig.~\ref{fig:MaStar_fit_PDF}. The original values for the set of physical parameters assigned to this star by CoSha have been included in the figure as a reference. It is apparent that the original values are located within the boundaries of the 85\% (65\%) percentage for all the projected PDFs (for the log(g)-T$_{eff}$ projected PDF). However, it is also evident that the PDFs recovered by the procedure do not correspond to a simple/symmetrical distribution around the original/input values. This illustrates the limitations of the method and/or the data themselves to distinguish between different physical properties of the stars using the current spectra.

\begin{figure*}
 \minipage{0.99\textwidth}
 \includegraphics[width=8.5cm,clip,trim=0 0 0 0]{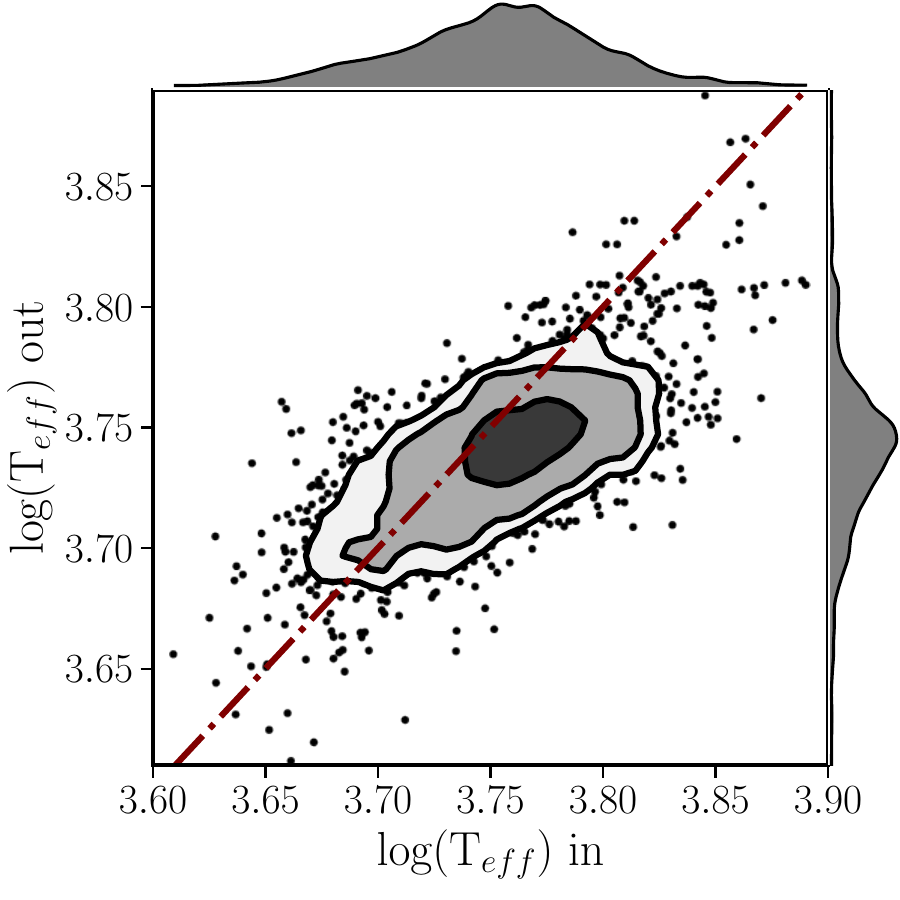}\includegraphics[width=8.5cm,clip,trim=0 0 0 0]{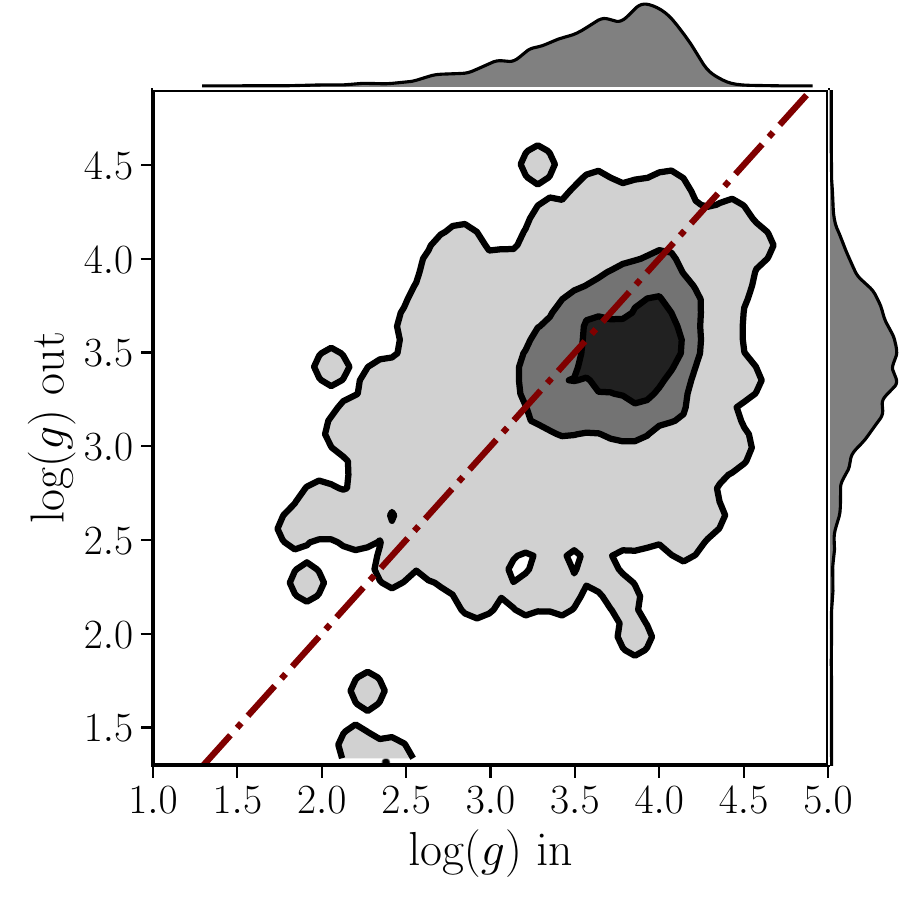}
  \includegraphics[width=8.5cm,clip,trim=0 0 0 0]{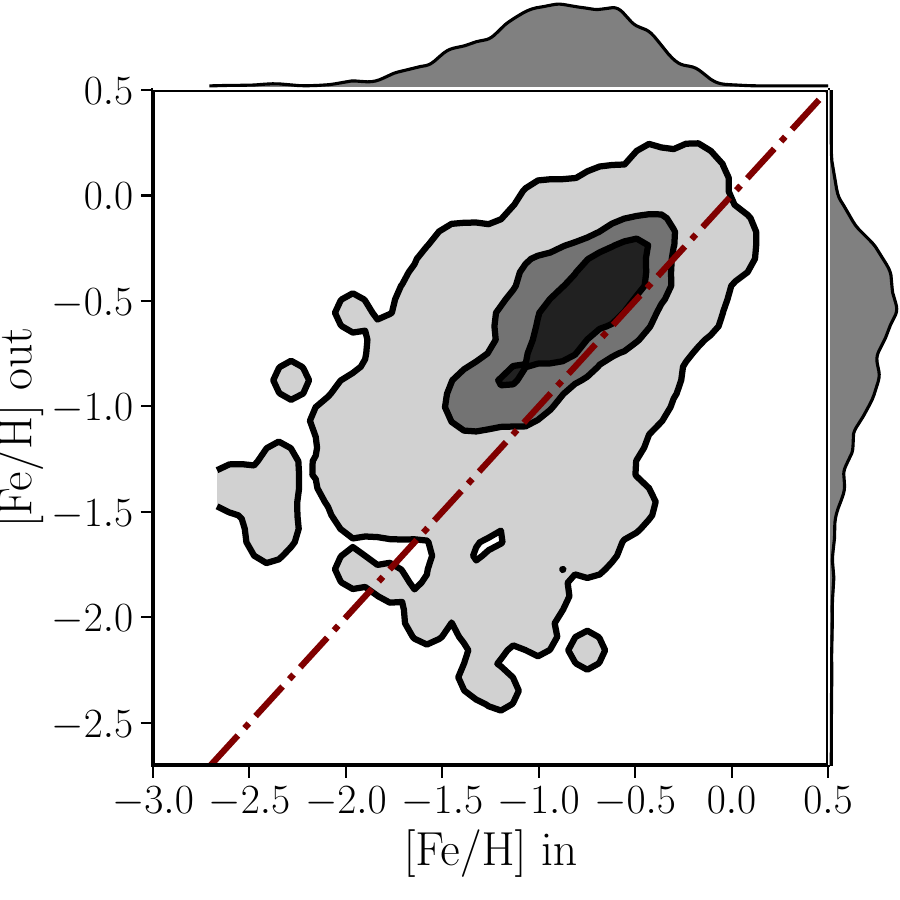}\includegraphics[width=8.5cm,clip,trim=0 0 0 0]{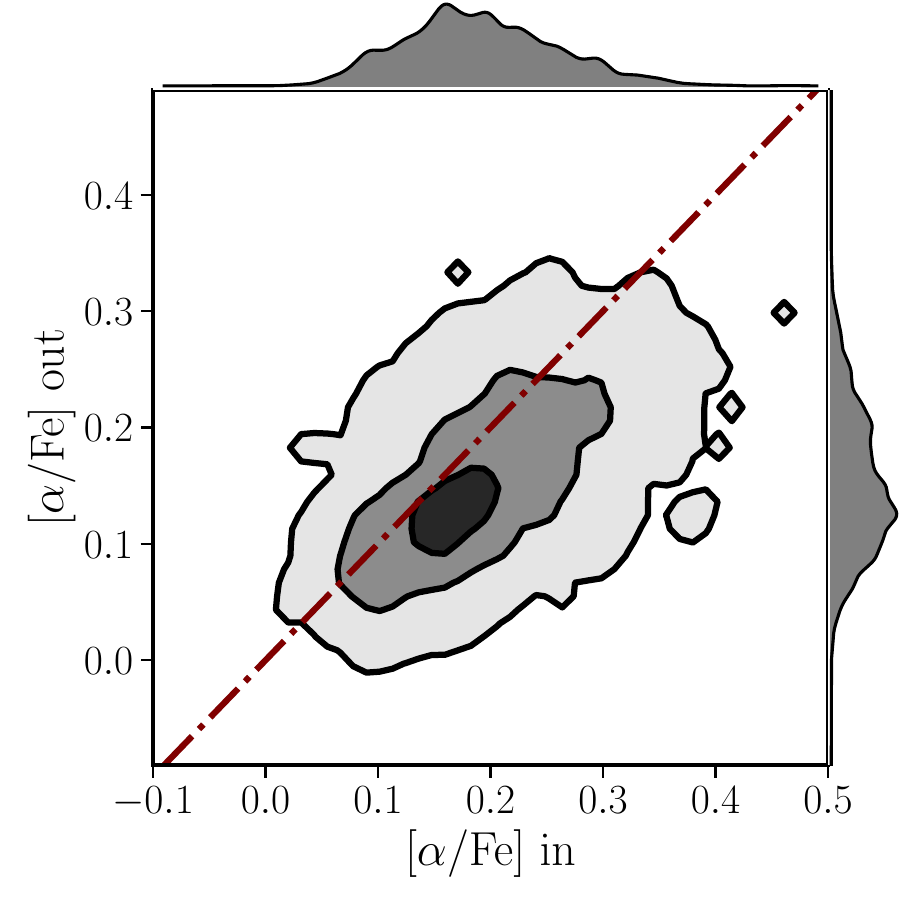}
 \endminipage
 \caption{Comparison between the recovered and simulated stellar physical parameters (T$_{eff}$: top-left, log($g$): top-right, [Fe/H]: bottom-left, and [$\alpha$/Fe]: bottom-right) using LVM-DAP for a set of 1700 realistic simulations in which it was a assumed a number of stars between 5 and 125. Contours, symbols, and lines in each panel have the same meaning as those in Fig. \ref{fig:dap_el}.}
 \label{fig:dap_cl}
\end{figure*}

Another important remark regarding the current exploration is that we have performed this test using the same stellar library and spectra adopted to generate the RSP templates used in the fitting process. Therefore, we are not fully testing the ability of the procedure to recover physical properties of individual stars in the most general sense, but rather exploring the ability to recover them within the boundaries of properties covered by the MaStar library. 
{ This is a limitation frequently found in this kind of explorations \citep[e.g.][]{cid-fernandes:2014aa,walcher15}.}
However, it is encouraging that the method recovers those properties well, within the limitations indicated before. If that were not the case, we would have to re-evaluate the full procedure. We will re-evaluate the capabilities of the adopted procedure in the future when other stellar libraries with known physical properties become available, for instance, as a result of the Milky-Way Mapper survey and/or incorporating libraries optimized to contain low-metallicity massive stars \citep[ULLYSES+XShootU][]{ro-du20,vink23}. 

In summary, we conclude that the currently adopted procedure is good enough to model the spectra and recover the physical parameters of the stellar component when it is dominated by a single star (with uncertainties ranging from $\sim$5\%\ for T$_{eff}$ to $\sim$30\%\ for [Fe/H]). We should remark that in this procedure we have not adopted any cut in the S/N of the star spectra provided by MaStar or in the quality of the values reported by CoSha.

\subsection{Recovery of the parameters of a stellar population} \label{sec:sim_cl}

Once the quality in the recovery of the stellar parameters when the spectra are dominated by a single star is determined, along with their corresponding limitations, we explore how well these parameters are recovered when the aperture encircles an arbitrary number of stars. To do so, we use the set of realistic simulations following the procedure described in Sec.~\ref{sec:sim_real}, in which we recall that we mixed the spectra of a number of stars. A total of 5100 simulations are created by mixing 5, 25, or 125 stars, with spectra extracted from the 1235 RSP library created by averaging spectra of stars with similar physical properties (maximizing both the range and coverage within the space of physical parameters). Then, for each number of mixed mixing stars, 15,300 simulated spectra are created, with an individual S/N for each mixed star of 5, 20, and 100.

\begin{figure}
 \minipage{0.99\textwidth}
 \includegraphics[width=8.5cm,clip,trim=15 0 20 0]{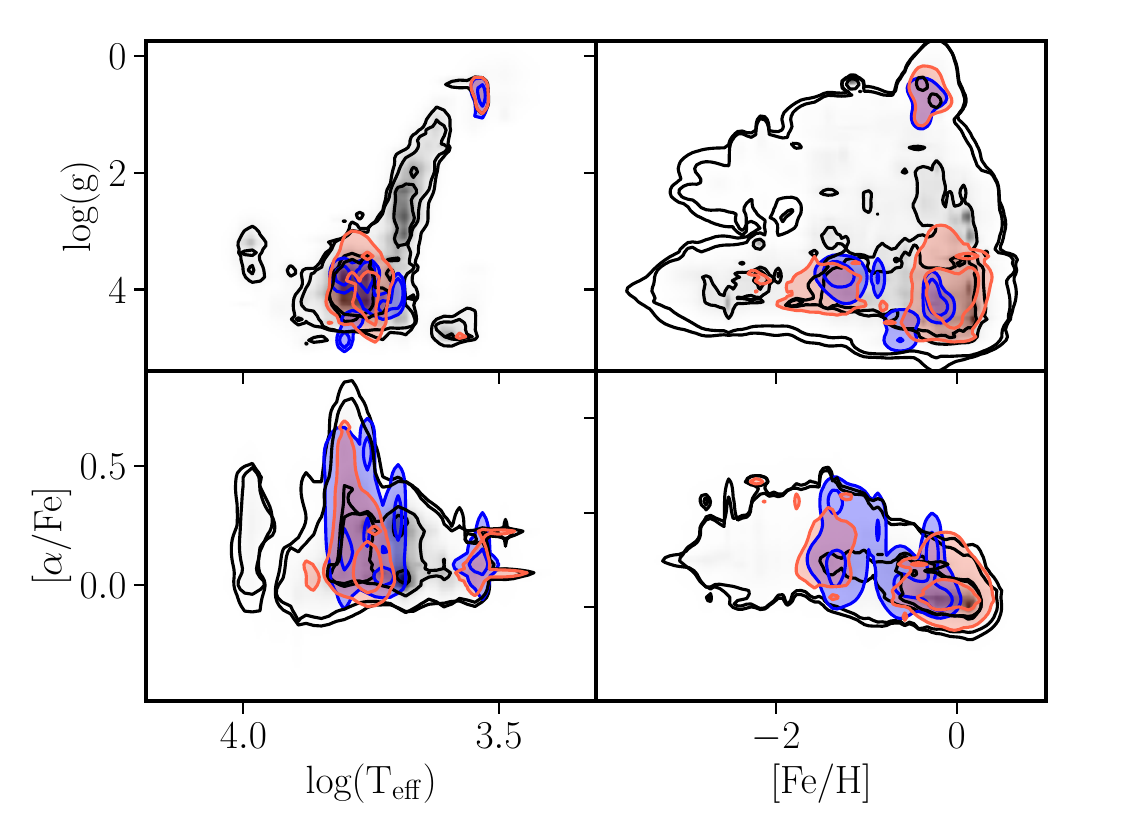}
 \endminipage
 \caption{Comparison between the simulated (blue contours) and recovered (red contours) PDF of the physical properties of stars (T$_{eff}$, log($g$), [Fe/H], and [$\alpha$/Fe]) derived for one of the 1700 simulations generated by mixing randomly the spectra of 5 stars within the RSP template comprising 1253 stars with an average S/N$\sim$20. The black contours in each panel, the meaning of each contour, and the distribution of panels is similar to the one shown in Fig. \ref{fig:MaStar_fit_PDF}}
 \label{fig:dap_PDF}
\end{figure}

Figure \ref{fig:dap_cl} shows the comparison between the input and recovered average values for the set of physical parameters of the stars (i.e., T$_{eff}$, log(g), [Fe/H], and [$\alpha$/Fe]), following a similar scheme as the one adopted for the comparison shown in Fig.~\ref{fig:cosha}. It is important to recall that in this case we are not comparing the properties of individual stars, but the average properties of a certain (arbitrary) distribution of randomly selected different stars. Thus, by construction, the range of values covered by the input parameters cannot cover the full range of values of the input library, being less spread and more concentrated around the mean values. Furthermore, the average value is weighted by the contribution of each star to the stellar luminosity in the normalization wavelength at 5000$\AA$. There is no physical reason that justifies the selection of this wavelength, and it is not clear how this may affect the results. This normalization is a feature of any stellar decomposition procedure in the literature that adopts a multi stellar-model template decomposition \citep[e.g.][]{starlight,ppxf,pipe3d}. It indeed deserves a detailed exploration that it is beyond the scope of this article.

With these caveats in mind, we explore the distributions shown in Fig~\ref{fig:dap_cl}. In contrast with the results reported when analyzing single stars, when we mix stars, we find considerable offsets and much larger deviations between the input and output values, following different patterns depending on the explored parameter: (i) in the case of T$_{eff}$, the procedure systematically recovers lower average values than the original ones, departing not only from the one-to-one relation but also from a simple constant offset ($\Delta$T$_{eff}$ = -174 $\pm$ 349 K). Thus, although the output values scale with respect to the input ones following a monotonic, essentially linear relation, the slope of the relation is lower than one. The bias ranges from almost zero at a temperature of $\sim$3000 K, to nearly 0.25 dex at $\sim$6000 K. Once corrected, the precision is improved ($\Delta$T$_{eff,corr}$ = 0 $\pm$ 255 K), with 96\% (70\%) of the output values within 10\% (5\%) of the input ones; (ii) for log(g), the input average parameters are too concentrated around a single value, and the differences between the output and input parameters are so large that it is not possible to describe a linear relation between them. However, the underestimation of the values is evident ($\Delta$ log(g) = -0.18 $\pm$ 0.37). Once corrected for this offset, 70\% (40\%) of the output values are within 10\% (5\%) of the input ones; (iii) the metallicity, [Fe/H], is overestimated with an almost constant offset ($\Delta$[Fe/H] = 0.23 $\pm$ 0.29 dex). After applying this correction, only 51\% (28\%) of the output values are within 10\% (5\%) of the input ones; finally (iv) the relative abundance of alpha elements, [$\alpha$/Fe], is underestimated following a constant offset as well ($\Delta$[$\alpha$/Fe] = -0.04 $\pm$ 0.05 dex). By taking this offset into account, 21\% (8\%) of the output values are within 10\% (5\%) of the input ones.

%
%

\begin{figure*}
 \minipage{0.99\textwidth}
 \includegraphics[width=6cm,clip,trim=0 0 0 0]{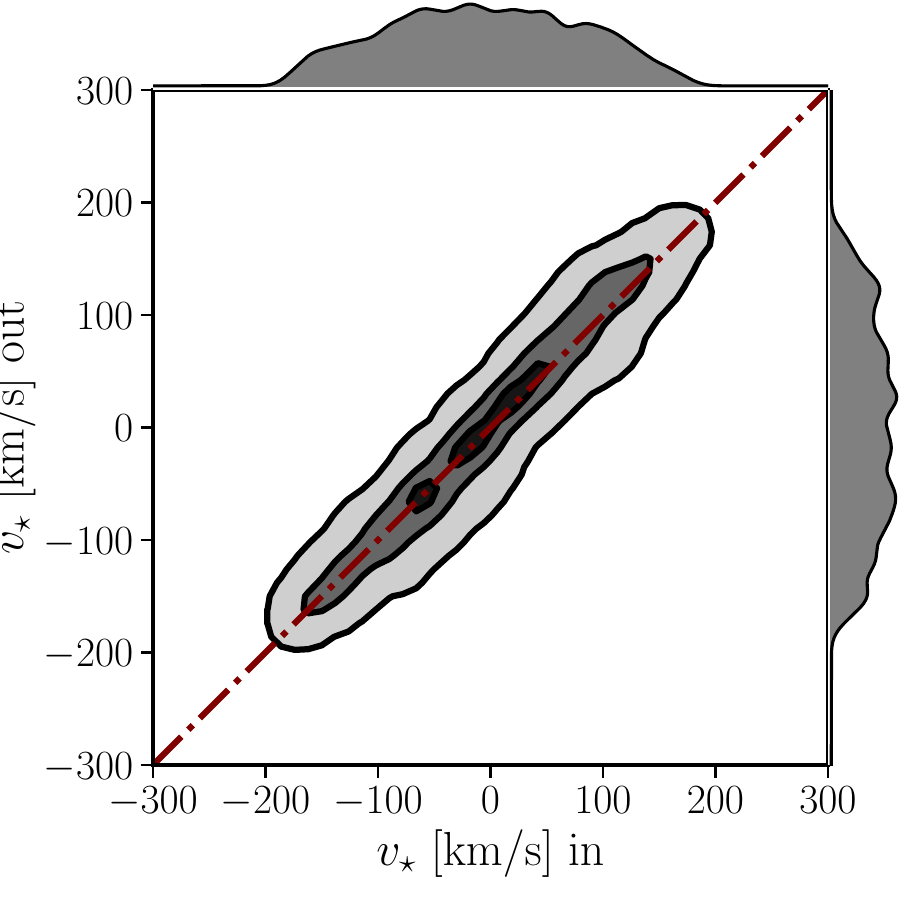}\includegraphics[width=6cm,clip,trim=0 0 0 0]{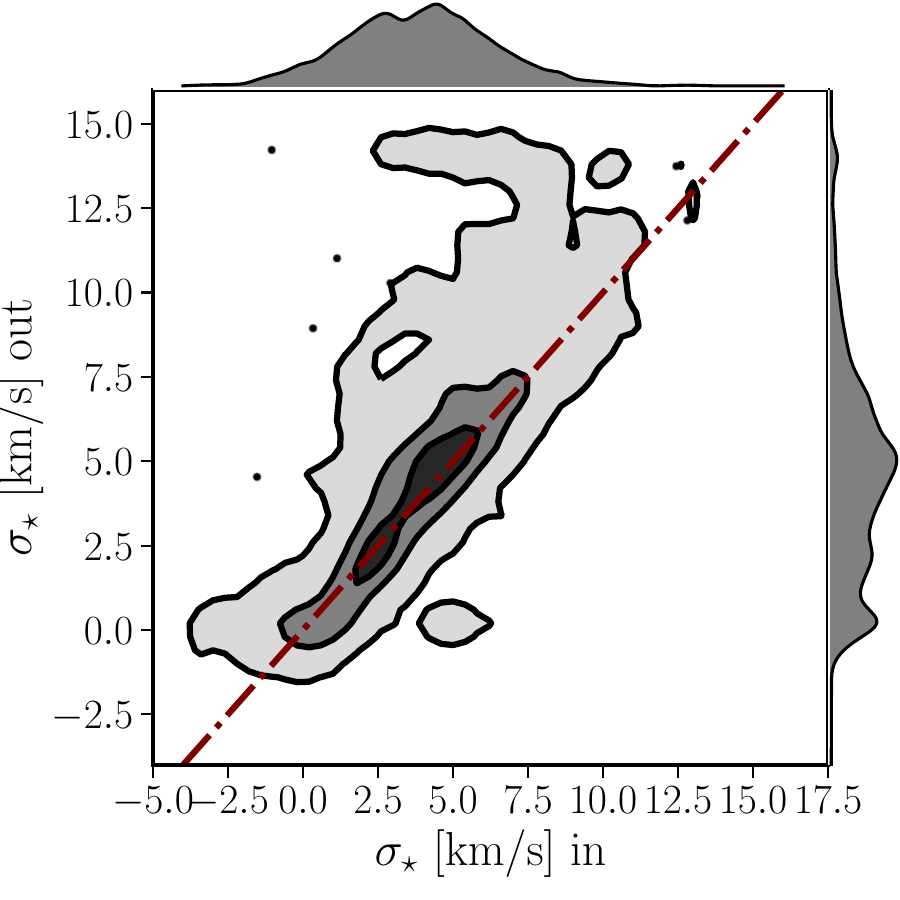}\includegraphics[width=6cm,clip,trim=0 0 0 0]{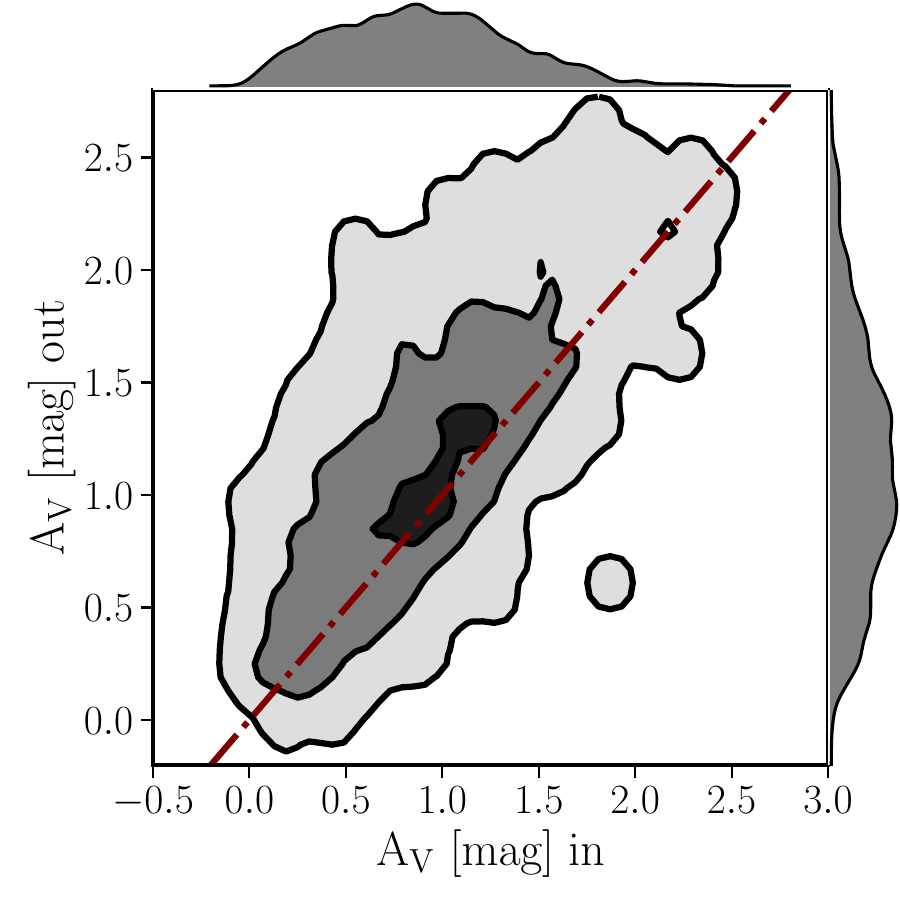} \endminipage
 \caption{ Comparison between the recovered and simulated non-linear parameters of the stellar component: v$_\star$ (left panel), $\sigma_\star$ (middel panel) and A$_{\rm V,\star}$ (right panel). Contours, symbols and lines in each panel have the same meaning as those in Fig. \ref{fig:dap_el}.}
 \label{fig:dap_nl}
\end{figure*}

As in the case of the experiments done above using single star spectra, it is important to remark that the procedure is not intended to recover just the average (luminosity-weighted) parameters that represent a combination of stars observed through a certain aperture, but the PDF that best represents the distribution of parameters covered by those stars. As described in Sec.~\ref{sec:rsp}, using our methodology, one star or an RSP does not have just a set of physical parameters but a PDF within the space of parameters that takes into account the errors and the inability to distinguish between physical parameters given the set of templates and observed spectra. Finally, the analysis provides a recovered PDF that ideally corresponds to the original one. Figure~\ref{fig:dap_PDF} shows an example of the comparison between the simulated and recovered PDFs for one particular simulation out of the 1700 explored in this subsection. As expected on the basis of the results above, there is no one-to-one correspondence between the simulated and recovered distributions. However, it is also clear that within the limitations of the method, there is a considerable degree of agreement between both distributions. { Statistically this matching should be enough for the main objectives of the LVM survey, i.e., to explore the feedback process at different spatial scales. However, for other more specific scientific cases that may use the LVM data in the future, we recommend the use of ad hoc simulations}, in which the mix of stars does not correspond to a simple randomly (and somehow arbitrary) selection of stellar templates within the RSP library, { to determine if this matching fulfill their requirements.}

{ Despite the limitations of the method highlighted above, it is important to stress that both the precision and accuracy of the DAP in the recovery of the physical parameters of the stellar component are of the order of what has been reported in the literature in the analysis of single stars \citep[e.g.][]{mejia21} and unresolved stellar populations \citep[e.g.][]{walcher15}. We lack examples to compare with of partially resolved stellar populations, as this is indeed one of the novelties of the described procedure. In the case of \citet{mejia21}, comparing with the results shown in their Figures 6 and 7, our method is worse at recovering some of the parameters (e.g., T$_{eff}$, with larger offsets and slightly larger dispersion), while others are recovered equally well (e.g., log(g) and [$\alpha$/Fe]). Finally, \citet{walcher15} reported similar systematic offsets and dispersions around them when comparing input and output metallicities ([Fe/H]) and abundances ([$\alpha$/Fe]) for a mock sample of $\sim$3000 simulated galaxy spectra}

{

\subsection{Recovery of the non-linear parameters}
\label{sec:sim_nl}

Along the previous sections we characterize the quality in the recovery of the properties of the emission lines (Sec. \ref{sec:sim_el}) and the stellar component (Sec. \ref{sec:sim_cl} and \ref{sec:sim_MaStar}). However, the analysis sequence described in Sec. \ref{sec:proc} and summarized in Fig. \ref{fig:scheme} requires that prior to the recovery of those properties it is needed to obtain the so-called non-linear parameters of the stellar population. These comprise the kinematic parameters (v$_\star$ and $\sigma_\star$) and the dust attenuation affecting the stellar component (A$_{\rm V,\star}$). The procedure for obtaining those parameters was directly inherited from \pyp, with little modifications (described in Sect. \ref{sec:proc}). Thus, we expect that
these parameters are recovered with the same quality as reported in \citet{pypipe3d} for the S/N cut adopted to derive them (i.e., with an accuracy of a few percent).

To explore the quality in the recovery of the non-linear parameters, we selected a simulation from the pool of realistic simulations described in Sec. \ref{sec:sim_real}, in which all the stellar components have S/N$>$20 for all the fibers within the FoV (i.e., the regime in which these parameters are robustly estimated by the DAP). The average S/N of this simulation was $\sim$50 at $\sim$5000\AA. Figure \ref{fig:dap_nl} shows the comparison between the recovered and input values for the non-linear parameters for this particular simulation. In general the recovered values for the three parameters follow an almost one-to-one relation with respect to the input values. The velocity (v$_\star$) is recovered with an insignificant systematic offset $\sim$2.3 \kms, and an accuracy of $\pm$16.7 \kms, which corresponds to  half of the expected velocity resolution at the wavelength considered. The velocity dispersion ($\sigma_\star$) is recovered with even better precision ($\sim$0.6 \kms) and accuracy ($\pm$1.6 \kms, $\sim$1/20th of the spectral resolution). Finally, the largest systematic offset ($\sim$0.23 mag) and scatter ($\sim$0.34 mag) between the recovered and input values are found for dust attenuation, A$_{\rm V,\star}$.

}

{

\begin{figure*}
 \minipage{0.99\textwidth}
 \includegraphics [width=8.75cm,clip,trim=10 20 25 10]{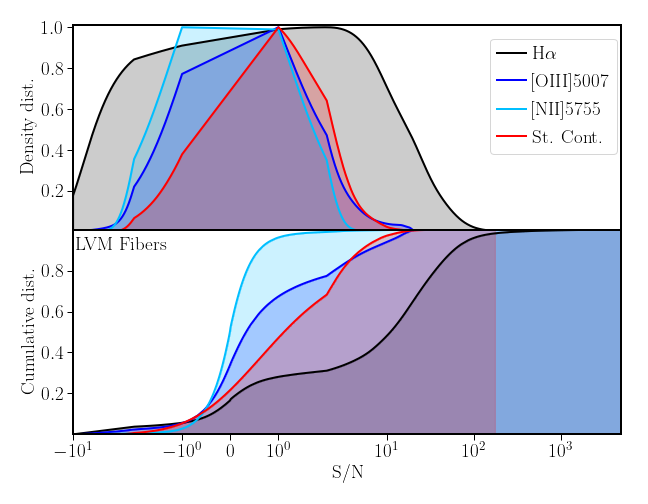}
 \includegraphics [width=8.75cm,clip,trim=10 20 25 10]{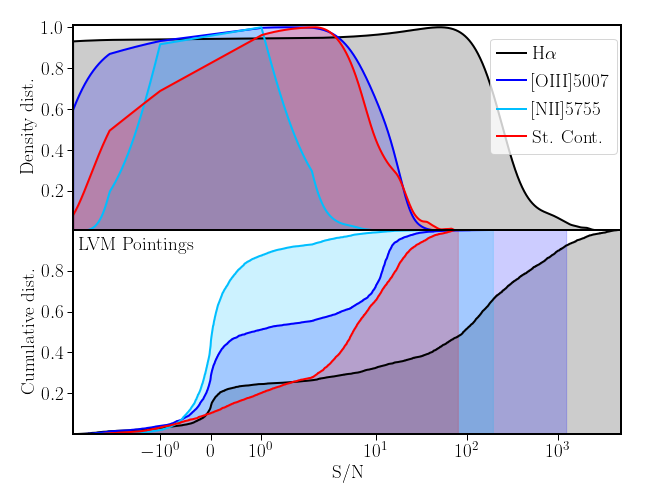}
 \endminipage
 \caption{ Normalized histograms of the distribution of S/N ratios (top panels) and cumulative distributions (bottom panels) for three representative emission lines (\Ha: black, \oiii 5007: blue, and \nii 5755: cyan) and the stellar continuum around $\sim$5000\AA\ (red) for $\sim$7 million individual LVM fiber spectra (left panels) and $\sim$4000 spectra integrating each LVM pointing (right panels). The X-axis is shown on a logarithmic scale for a better visualization of the distributions.}
 \label{fig:real}
\end{figure*}

\section{Application on real data}
\label{sec:real}

The LVM started its survey operations in summer 2023, and so far it has observed $\sim$1/5th of the entire foreseen pointings. The $\sim$4000 poitings observed before December 2023  have been used as benchmarks for the DRP (Mej\'\i a-Narv\'ez in prep.) and the DAP. Although both codes will be continually revised and upgraded throughout the course of the survey, and therefore the current data and dataproducts are far from being representative of the final expected data quality, they are useful to determine the expected content in the LVM data. Furthermore, as described in Sec \ref{sec:proc}, the DAP performs different analyzes based on the S/N of the continuum (multi-RSP decomposition or single RSP selection) and the expected S/N of the emission lines (parmetric and non-parametric procedures). To have a first overview of the future content of the LVM dataset and to determine how relevant those different procedures are, we have explored the distribution of S/N ratios for three representative emission lines (\Ha, \oiii 5007 and \nii 5755) and the continuum at the normalization wavelength of the RSP template (5000\AA). The three selected lines are representative of the strongest (\Ha), the average intensity (\oiii) and weakest emission lines (\nii). On the other hand, the continuum  traces how frequently it would be required to perform a detailed stellar synthesis and when that analysis would be irrelevant. Finally, as the current procedure performs the described analysis fiber by fiber (among those included in the science fiber bundle) and for the integrated spectrum pointing by pointing, we can explore the S/N distributions in both cases, illustrating the importance of possible spatial binnings in future explorations.

\begin{table}[t]
\begin{center}
\caption{Results of the analysis of the distribution of S/N ratios}
\begin{tabular}{lrrrr}\hline\hline
Feature & \multicolumn{2}{c}{Fibers} & \multicolumn{2}{c}{Pointings}\\
           & S/N$>$3  & S/N$>$20 & S/N$>$3 & S/N$>$20 \\
\hline
\Ha        &  66.5\% &  34.8\%  & 72.1\% & 64.2\% \\
\oiii 5007 &  17.4\% &  0.2\%   & 42.4\% & 4.1\%  \\
\nii 5755  &  0.7\%  &  0.1\%   &  5.0\% & 0.4\%  \\
St. Cont.  & 19.9\%  &  0.4\%   & 66.3\% & 14.1\% \\
\hline
\end{tabular}\label{tab:real}\\
\end{center}
\end{table}

Figure \ref{fig:real} shows the results of this analysis, including the normalized and cumulative distributions of the S/N for the three emission lines and the continuum, for both the individual fibers and the integrated pointings. By selection, there is a clear hierarchy in the distribution of S/N ratios, which is more evident in the case of the results based on the integrated spectra, with S/N$_{\rm H\alpha}$ $>$ S/N$_{\rm [OIII]}$ $>$ S/N$_\star$ $>$ S/N$_{\rm [NII]}$. It is also evident that there is an increase of the S/N for the four explored properties when integrating across the entire science bundle. Both results were expected.

More quantitative information is provided by the analysis of the frequency at which a certain feature reaches a S/N that qualifies it as detected (S/N$>$3) or having sufficient S/N to derive its properties (S/N$>$20). Table \ref{tab:real} lists the results of this analysis. The increase in the S/N when coadding entire pointings is reflected in the higher fraction of cases above the defined S/N thresholds for the four properties. It is notable that the increase is not the same for each property. For instance, the cases in which \Ha\ reaches a S/N$>$3 present a very mild increase, while those with S/N$>$20 has almost double its frequency. On the other hand, for the continuum, the number of cases in which S/N$>$3 has increased by a factor six, and for a S/N$>$20 the frequency has multipled by a factor $\sim$35. Similar differences are found for the other two emission lines. This indicates that any spatial binning aimed to increase the S/N should be performed focused on the property to be explored.

Regarding the relative importance of the different analyses included in the DAP, for individual fibers (average pointings) these results indicate that the full stellar synthesis would be performed less than 1\% (15\%) of the time. Despite these relatively low fractions, for the current dataset they corresponds to $\sim$28000 individual fibers ($\sim$600 pointings). At the completion of the survey it is expected that the stellar synthesis would provide reliable physical properties of the stellar population for $\sim$150000 individual fibers ($\sim$3000 pointings). Nevertheless, these results confirm that the primary concern and goal for the LVM DAP should be the analysis of the emission lines, as we indicated in the introduction. In this regards, the parametric analysis, which in the current setup is applied for strong and intermediate emission lines (e.g., \Ha\ and \oiii 5007), would be relevant for $\sim$66\% ($\sim$35\%) of the analyzed fibers (pointings). On the other hand, the use of the less time-consuming non-parametric analysis, that runs for all the fibers (pointings) and the full set of emission lines, is justified as the number of cases in which the weakest emission lines are detected is rather low: $\sim$1\% ($\sim$5\%). 



}

\begin{figure}
 \minipage{0.99\textwidth}
 \includegraphics [width=8.5cm,clip,trim=20 15 25 10]{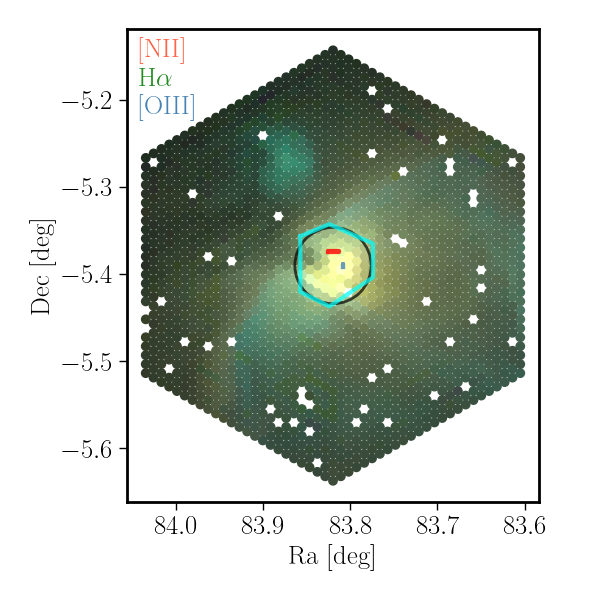}
 \endminipage
 \caption{Spatial distribution of fluxes recovered by the DAP of the \oiii 5007 (blue), H$\alpha$ (green), and \nii 6583 (red) emission lines for the combination of un-saturated pointings covering the central region of the Orion nebulae described in the text, using an arbitrary scale to enhance the contrast. { The approximate location of the slit spectroscopy data presented by O92 and B00, and the IFS  data presented by S07 are indicated with a red and blue line and a cyan polygon, respectively. The black circle corresponds to the location from which we have extracted the integrated spectrum discussed in the text. } }
 \label{fig:O_rgb}
\end{figure}

\subsection{Orion, a showcase of the DAP} 
\label{sec:show}


The Orion Nebula is one of the brightest and closest \hii\ regions in the solar neighborhood, located at a distance of only $\sim$0.41~kpc from the Sun \citep{Binder:18}. For this reason, different studies and observational campaigns have explored it for decades \citep{odell01}. In particular, its central region, M42, ionized by the Trapezium stellar complex \citep{ODell:17b}, has been an archetypal laboratory for the exploration of the physical conditions of the ISM.

Numerous spectroscopic campaigns have aimed to characterize its properties by taking spectra at various ``representative" locations \citep[e.g.,][]{kale76,Peimbert:77, bald91,Mesa-Delgado:2011,Fang2017}. { For instance}, \citet{osterbrock92}  (hereafter O92) compiled high and low dispersion deep spectra of { near the} bright region in the optical-NIR (3000-11000 \AA) wavelength range, identifying 225 emission lines, 88 of which fall within the wavelength range of our IFS data. From the relative intensities of these lines, they derived the relative abundances of several elements, the electron temperature (T $\sim$ 9000 K), electron density ($n_{\rm e} \sim$ 4$\times$10$^3$ cm$^{-3}$), and extinction ($A_V\sim$1.08 mag, derived from the comparison between different \hi~recombination line fluxes from the Balmer and Paschen series, including the classical H$\alpha$/H$\beta$ ratio). { More recently, \citet{bald00}, hereafter B00, obtained high-resolution deep spectra of a nearby location, just $\sim$90$\arcsec$ apart, covering the wavelength range between 3498-7468 range \AA. They recover the flux intensities of 444 emission lines, reporting slightly different line ratios and therefore, different physical conditions}. These properties are often compared with those of distant \hii\ regions within our Galaxy and beyond \citep[e.g.,][]{barnes22,espi22}, though most distant regions are poorly resolved, and their integrated properties are compared to specific areas within the Orion Nebula { (that may present different values, as indicated before.)}.

\begin{figure*}
\minipage{0.99\textwidth}
 \includegraphics[width=18.5cm,clip,trim=180 170 10 80]{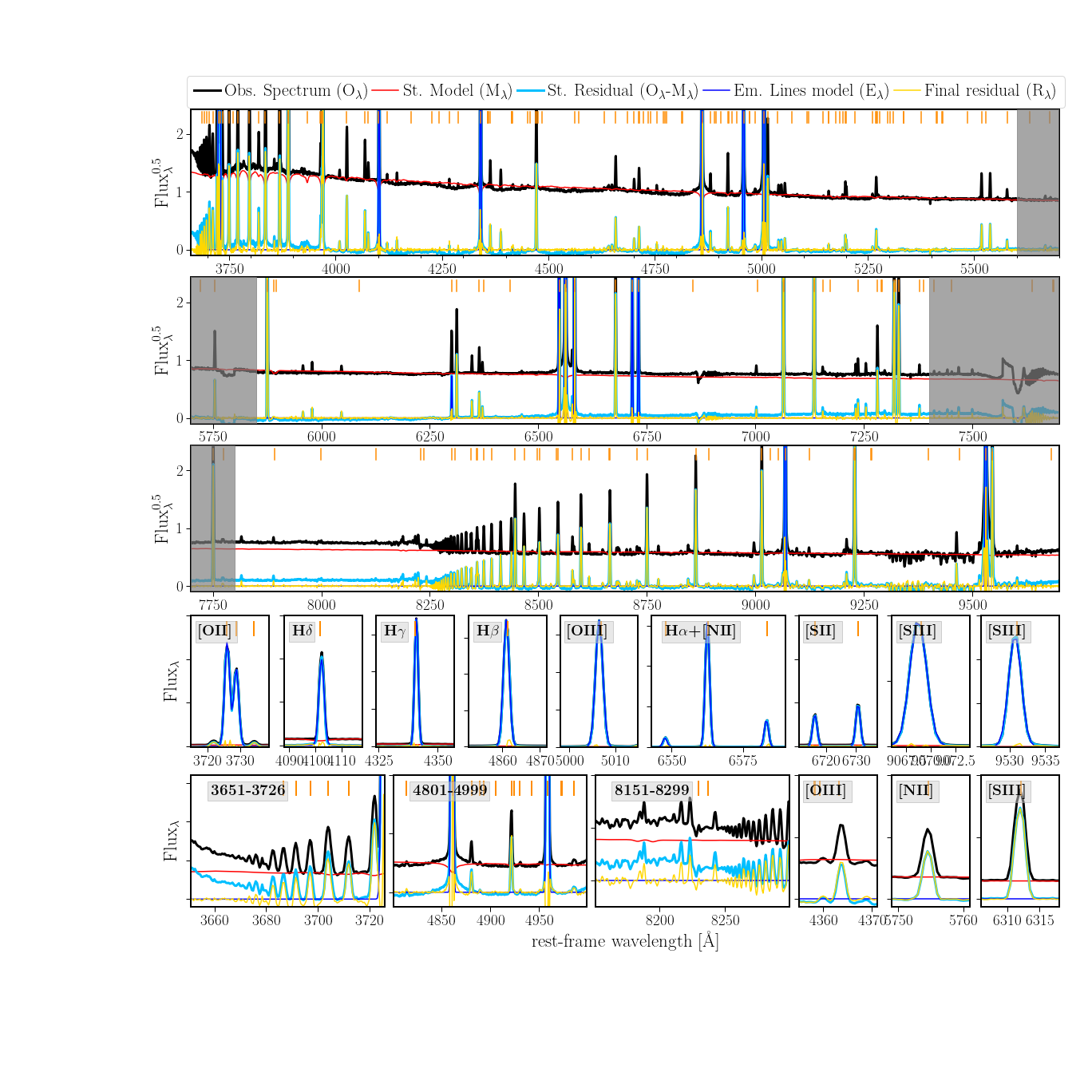}
 \endminipage
 \caption{Integrated spectrum of the central 5'$\times$6' region of the Orion Nebula (black solid line), together with the best-fitted stellar model (red solid line), the model of the strongest emission lines fitted with Gaussian functions (blue solid line), and the residual after subtraction of the stellar model (cyan solid line), and the combination of both models, including a final correction to the shape of the residual (yellow solid line). The first three panels from top to bottom present the wavelength ranges covered by the blue (1st row), green (2nd row), and infrared (3rd row) arms that comprise the LVM spectrograph. Shaded regions correspond to the overlapping regimes between them. Like in the case of figures \ref{fig:lvmsim} and \ref{fig:dapsim}, the insets in the two bottom rows show a set of zoom-ins at certain particular wavelength ranges. The panels in the 4th row illustrate the quality of the modeling of the strongest emission lines, including the same lines already shown in Fig. \ref{fig:lvmsim} and \ref{fig:dapsim}. Finally, the last row of panels includes three wavelength ranges (3651-3726 \AA, 4801-4999 \AA, and 8151-8299 \AA) that illustrate certain particularities of the spectrum discussed in the text. Three additional panels present a zoom around the three auroral lines most frequently used to determine the electron temperature in ionized nebulae: \oiii 4363, \nii 5755, and \siii 6312. The flux ranges in these three last insets are the same to facilitate the comparison between the lines. In the particular case of this very deep spectrum, all three are clearly detected.}
 \label{fig:O_spec}
\end{figure*}

\begin{table*}
\begin{center}
\caption{{ Flux intensities of the EL detected in the core of Orion with a S/N$>$5: non-parametric procedure.}}
\begin{tabular}{lcc|lcc|lcc}\hline\hline
Name & wavelength & flux & name & wavelength & flux & name & wavelengt & flux  \\
     & (\AA) & (10$^{-13}$ erg/s/cm$^2$) &  & (\AA) & (10$^{-13}$ erg/s/cm$^2$) & & (\AA) & (10$^{-13}$ erg/s/cm$^2$)\\
\hline
 \oii & 3726.03   &  1288.47  $\pm$ 13.84  & HeI & 4471.48     & 6.60  $\pm$  3.95 & \oii & 7318.92  & 102.68  $\pm$ 4.60 \\
 \oii & 3728.82   &  985.34  $\pm$ 15.08 & \feii$^1$ & 4474.91  & 62.1   $\pm$   2.52& \caii$^1$ & 7323.88  & 97.23  $\pm$  5.14 \\
 HI   & 3734.37   & 564.75   $\pm$ 18.17 & H$\beta$ & 4861.36  & 1843.04  $\pm$  5.03& \oii & 7329.66  & 78.72  $\pm$  4.76 \\
 \neiii & 3868.75 & 96$\pm$45  $\pm$  6.54 & \oiii & 5006.84 & 3541.54  $\pm$ 196.12& \ariii & 7751.06  & 92.31  $\pm$  1.98 \\
 HeI & 3888.65    & 48.27    $\pm$ 16.65& \nii & 5754.59  & 16.00  $\pm$  2.90& OI & 8446.00  & 30.13  $\pm$  5.06 \\
 HI  & 3889.05  & 247.07 $\pm$ 17.18& HeI & 5876.0  & 216.26  $\pm$ 2.94& HI &8598.39  & 27.30  $\pm$  5.06 \\
 HeI & 3964.73    & 91.16  $\pm$  9.24& \oi & 6300.30  & 19.76  $\pm$  2.67& HI & 8665.02  & 29.71  $\pm$  2.63 \\
 \neiii & 3967.46 & 256.05  $\pm$  9.26& \siii & 6312.06  & 30.97  $\pm$  2.55 & HI & 8750.47  & 43.95  $\pm$  3.75 \\
 CaII & 3968.47   & 256.61  $\pm$  9.14& H$\alpha$ & 6562.85  & 6219.77  $\pm$  1.23& HI & 8862.78  & 57.13  $\pm$  3.42 \\
 H$\epsilon$      & 3970.07  & 254.50  $\pm$  9.30& \nii & 6583.45  & 1281.78  $\pm$ 42.73& \siii & 9069.00  & 1105.70  $\pm$ 13.22 \\
 H$\delta$        & 4101.77  & 355.97  $\pm$  3.54& \sii & 6716.44  & 142.13  $\pm$ 13.83& \feii$^1$ & 9226.60  & 106.43  $\pm$ 2.10 \\
 H$\gamma$        & 4340.49  & 705.01  $\pm$  3.03& \sii & 6730.82  & 187.17  $\pm$ 14.15& HI & 9229.02  & 107.93  $\pm$  3.22 \\
 \oiii & 4363.21  & 11.49  $\pm$  1.45& HeI & 7065.19  & 155.46  $\pm$  2.01& \siii & 9531.10  & 3030.44  $\pm$  4.34 \\
 \feii$^1$ & 4470.29  & 61.60  $\pm$ 2.97& \ariii & 7135.80  & 306.87  $\pm$  2.55& HI & 9545.97  & 102.41  $\pm$  1.61 \\
\hline
\end{tabular}\label{tab:fe_orion}\\
$(1)$ flux intensity potentially polluted with nearby blended lines.\\
\end{center}
\end{table*}

\begin{table}
\begin{center}
\caption{{ Properties of the ELs in the core of Orion: parametric procedure.}}
\begin{tabular}{lcccc}\hline\hline
Name & $\lambda$ & flux & $\sigma$ & velocity  \\
     & (\AA) & (10$^{-13}$ erg/s/cm$^2$) &  (\AA) & (km/s) \\ 
\hline
\oii & 3726.03 & 1269.4 $\pm$ 36.55 & 0.8 $\pm$ 0.06 & 4.46 $\pm$ 5.05 \\
\oii & 3728.82 & 943.12 $\pm$ 36.21 & 0.8 $\pm$ 0.06 & 4.46 $\pm$ 5.05 \\
\hd & 4101.77 & 362.44 $\pm$ 21.93 & 1.14 $\pm$ 0.04 & 3.95 $\pm$ 3.03 \\
\hg & 4340.49 & 722.58 $\pm$ 54.57 & 1.0 $\pm$ 0.01 & 4.35 $\pm$ 2.57 \\
\Hb & 4861.36 & 1842.42 $\pm$ 12.59 & 0.79 $\pm$ 0.01 & -19.25 $\pm$ 1.82 \\
\oiii & 4958.91 & 1183.6 $\pm$ 12.08 & 0.79 $\pm$ 0.01 & -19.25 $\pm$ 1.82 \\
\oiii & 5006.84 & 3554.35 $\pm$ 12.85 & 0.79 $\pm$ 0.01 & -19.25 $\pm$ 1.82 \\
\oi & 6300.30 & 19.56 $\pm$ 4.91 & 0.9 $\pm$ 0.04 & -2.46 $\pm$ 2.35 \\
\nii & 6548.05 & 444.69 $\pm$ 33.52 & 0.76 $\pm$ 0.04 & -10.05 $\pm$ 3.75 \\
\Ha & 6562.85 & 6304.94 $\pm$ 217.98 & 0.76 $\pm$ 0.04 & -10.05 $\pm$ 3.75 \\
\nii & 6583.45 & 1335.41 $\pm$ 64.57 & 0.76 $\pm$ 0.04 & -10.05 $\pm$ 3.75 \\
\sii & 6716.44 & 135.82 $\pm$ 2.72 & 0.7 $\pm$ 0.02 & -4.06 $\pm$ 4.87 \\
\sii & 6730.82 & 180.43 $\pm$ 3.9 & 0.7 $\pm$ 0.02 & -4.06 $\pm$ 4.87 \\
\siii & 9069.00 & 1070.15 $\pm$ 27.75 & 0.8 $\pm$ 0.06 & -10.67 $\pm$ 5.69 \\
\siii & 9531.10 & 3013.59 $\pm$ 22.72 & 0.87 $\pm$ 0.02 & -9.62 $\pm$ 3.54 \\
\hline
\end{tabular}\label{tab:el_orion}\\
\end{center}
\end{table}

The large size of the Orion Nebula projected on the sky ($\sim$5$\times$5 deg$^{2}$) has prevented a detailed exploration of the spatial distribution of the physical properties across the whole nebula or the analysis of its integrated properties over large apertures. Most observations have focused on the Huygens region, located at the core of M42, centered on the Trapezium stellar complex. \citet{1992ApJ...399..147P} (hereafter P92) performed one of the first explorations of the spatially resolved properties of the ionized gas in this region using a Fabry-Perot (FP) interferometer. They sampled the brightest emission lines in the optical range, deriving the line ratios across the sampled region and estimating the dust attenuation, the electron density and temperature, and the helium ionic abundance. \citet{sanchez07c} (hereafter S07) explored a similar region using IFS spectroscopy, sampling a larger number of emission lines and line ratios, re-evaluating the same physical parameters, and including an exploration of the oxygen abundance. More recently, the same area has been covered by MUSE \citep{bacon10}, presenting resolved spectroscopy of unprecedented spatial resolution \citep[][hereafter W15]{Weilbacher2015}. These data have been analyzed in detail by \citet{McLeod2016} and \citet{ODell2017}. In addition, specific areas of this region have been explored by other spectroscopic studies recently \citep[e.g.,][]{Mesa-Delgado:2011,Fang2017}.

\subsection{Dataset}
\label{sec:O_dataset}

The LVM instrumental arrangement was specifically designed to obtain IFS spectroscopy of wide areas in the sky, making it a unique tool to sample large nebulae such as Orion. For this reason, and due to the relevance of this nebula as an archetypical \hii\ region, Orion was selected as one of the first areas to be covered by the LVM survey. During different nights from September to November 2023, several pointings were taken with the goal of covering a continuous area of $\sim$30 square degrees in the sky, centered on the Trapezium complex, and spectroscopically sampling the entire nebula for the first time. An overview of this rich dataset was recently presented by \citet{kreckel24}.

During these observations, it became evident that the standard 900~s exposure time adopted by the LVM in its survey mode causes the strongest emission lines (e.g., H$\alpha$) to saturate. Those observations, although useful for exploring the rather faint emission lines such as the auroral lines (e.g., \oiii 4363, \nii 5755, \siii 6312) and the heavy-element recombination lines, should be complemented with shorter exposure times to capture the properties of the strongest emission lines as well. Furthermore, these shorter observations can be combined to increase the S/N, avoiding saturation while capturing the relevant information of the weak emission lines too. 

The LVM acquired thirteen observations of 5-second exposure time at the same location in the sky centered in the Huygens region, with LVM exposure numbers between 00006555 and 00006567. Each individual exposure was reduced using version 1.0.3 of the LVM DRP, as described in Sec.~\ref{sec:data}. These short exposures are not representative of the depth reached by typical LVM observations. Therefore, we combined them, weighting by the inverse of the variance, to create a single deep but not-saturated exposure that was analyzed using the DAP, following the procedures described in Sec. \ref{sec:method}. We present here an overview of the combination of this set of short exposures, with an emphasis on comparing the integrated properties of the Orion Nebula with results presented in previous IFS studies.

\subsection{Results}
\label{sec:O_results}

Figure \ref{fig:O_rgb} illustrates the result of this analysis, showing the spatial distribution of the flux intensities of the three emission lines adopted by the classical BPT diagnostic diagram \citep{baldwin81} using a color code in which \oiii 5007 is shown in blue, \ha\ in green, and \nii 6583 in red. { We adopted the results from the non-parametric procedure for this figure, but note that the choice of procedure is irrelevant for these emission lines, as the values are  consistent with each other for these emission lines (see Appendix \ref{sec:ec}.} The spatially resolved representation of the ionized gas emission lines was introduced by \cite{ARAA}, being particularly useful for visually exploring the ionization conditions in the ISM at kpc-scales. At those scales most of the variations of the lines ratios for \hii\ regions are deeply connected to the differences in their chemical composition \citep[e.g.][]{sanchez15,mori16,espi22}. However, at the scales sampled by the LVM, these variations are more deeply related to changes in the ionization structure, parameterized by the number of ionizing photons for a certain element, the electron density,  and the distance to the ionizing sources \citep[e.g.][]{mori18}. Despite the somewhat coarse spatial resolution of the data (compared to previous IFS explorations, e.g., P92, S07, and W15), the general structure of the Huygens region is clearly resolved, showing the high-column density region at $\sim$2' northeast from the center affected by heavy dust obscuration (almost black in the figure), and the almost linear/barred structure at $\sim$1.5' southeast from the center showing an increase in the ionization strength (almost white in the figure).

\begin{table*}
\begin{center}
\caption{Balmer and Diagnostic Line Ratios}             
\label{tab:O_phy}      
\begin{tabular}{lllll|llcccc}        
\hline\hline                 
Line ratio & Obs. & S07 & O92 & B00 & Line ratio & Phy. prop. & Obs. & S07 & O92 & B00 \\
\hline   
H$\alpha$/H$\beta$  &  3.37 &3.07 & 2.81 & 3.11 & \sii 6717/6731        & n$_e$ & 0.76  &  0.57 &  0.57 & 0.53 \\
H$\gamma$/H$\beta$  &  0.48 &0.53 & 0.49 & 0.61 &\oii 3729/3727        & n$_e$ & 0.76  &   $^{(3)}$ & 0.52  & 0.47 \\ 
H$\delta$/H$\beta$  &  0.19 &0.29 & 0.26 & 0.34 &\siii (9069+9531)/6312  & $T_e$, n$_e$ & 109.63  &   &  108.0 & \\ 
H$\epsilon$/H$\beta$&  0.14 &0.14$^{(1)}$ & 0.16 & 0.17 & \nii (6548+6583)/5755   & $T_e$, n$_e$ & 100.01  &  111.02 &   91.4   & 86.66\\ 
H8/H$\beta$         &  0.13 &0.10$^{(2)}$& 0.11 & 0.08 & \oiii (4959+5007)/4363  & $T_e$, n$_e$ & 367.74  &  198.18$^{(4)}$ & 290. &  398.89\\ 
H9/H$\beta$         &  0.31 &0.13 &  0.07 & &\oii (3727)/(7319+7330)  & $T_e$, n$_e$ & 10.15  &   & 12.5 & 15.88 \\ 
         &   &  &   & &  HeI5876/\Hb             & He$^+$/H$^+$ & 0.117   & 0.131 & 0.123   & 0.151\\
\hline                                   
\end{tabular}
\\
$(1)$ Blended with  [NeIII];  $(2)$ Blended with  HeI, $(3)$ [OII] lines blended; $(4)$ Blended with FeI\\
\end{center}
\end{table*}

\subsection{Integrated spectrum}
\label{sec:O_int}

{ We compare with the results in the literature, using S07 as a tracer of the integrated properties at the core of the Orion nebula and both O92 and B00 as tracers of two particular locations within this area. For doing so, } we integrated the spectra within an aperture that resembles as closely as possible the area covered by the IFS mosaic presented in S07,  manually selecting the 77 individual fibers that cover the same spatial footprint ($\sim$36 arcmin$^2$ centered in R.A. and DEC.). We should note that an exact match with the previous IFS observations is not feasible as neither IFU observations have full coverage of the FoV (due to the gaps between fibers), the areas covered by each fiber in the sky are considerably different (indeed, one single LVM fiber covers 1/4th of the area covered by the entire IFU adopted in S07), and/or the exact coordinates are not provided in either { S07, O92 or B00}. However, we consider that the matching is good enough for the qualitative comparison foreseen in this paper. Finally, the integrated spectrum within the considered area was analyzed using the DAP.

Figure \ref{fig:O_spec} shows this integrated spectrum, together with the best model for the stellar component and the ionized gas emission lines, following a similar scheme and nomenclature to the one already presented in Figs.~\ref{fig:lvmsim} and \ref{fig:dapsim}. The main difference with respect to those figures is that the full spectral range covered by the LVM observations has been split into three regimes, roughly corresponding to the ranges covered by the three arms of the LVM spectrograph, each one presented in a separate panel. The reason for this was the high quality of this particularly deep combined spectrum, which is rich in features that deserve to be presented in a larger format for discussion. The quality of the integrated spectrum is evident when observing (i) the areas of overlap between the different arms in the spectrograph, which show no deviation from the general trend described by the data { although some data reduction artifacts at the very edge of the spectral channels like the discontinuity near 5800$\AA$ and 7600$\AA$ remain }, and (ii) the lack of the characteristic features produced by the imperfect subtraction of the night-sky lines. This is particularly evident at the location of the strongest night-sky lines in the blue range (e.g., \oi 5577 or NaID 5996) and throughout the infrared range (third row panel).

We should stress that, despite our best efforts, we were not able to fit the continuum component well with only the adopted templates included in the RSP library. The most evident discrepancies are (i) a mismatch in the blue regime, around 3700-4000 \AA, and (ii) the jump observed at $\sim$8200 \AA. The sources of these discrepancies are different. In the first case, the mismatch is due to a combination of an imperfect flux calibration at the blue end of the spectral range, which introduces a change in the shape of the continuum (Fig.~\ref{fig:O_spec}, 1st inset-panel in the bottom row), and the presence of a broad (faint) component in the emission lines (Fig.~\ref{fig:O_spec}, 2nd inset-panel in the bottom row), which produces a pseudo-continuum in this wavelength range with numerous emission lines (both 1st and 2nd inset-panels). { These wing features may correspond to the expansive component of the nebula or to the detection of Raman scattering in this HII region \citep[e.g.][]{dopita16r}. Whatever their origion they have been reported in numerous \hii\ regions before \citep[e.g.,][]{arse86,chu94,rela05}}. An improvement in the data reduction and more complex modeling of the emission lines are required to properly address these issues.

The jump observed in the near-infrared (Fig. \ref{fig:O_spec}, 3rd inset-panel in the bottom row) is the Paschen Break due to the nebular continuum \citep[e.g.,][]{osterbrock89}. Besides the emission lines, the nebular emission produces a continuum due to the photons produced when a free electron of arbitrary kinetic energy is captured into a certain level. For hydrogen, this nebular emission is observed as a truncated smooth component, with jumps at certain wavelengths corresponding to the lower energy level of each transition sequence (e.g., Lyman, Balmer, Paschen). The wavelength range covered by the LVM includes two of these jumps: one corresponding to the Balmer sequence (3646 \AA) and another corresponding to the Paschen sequence (8207 \AA). The first is not evident due to the limitations of the current spectrophotometric calibration in the blue end of the spectrum discussed before, although it may contribute to the observed jump in that regime. The second is clearly the reason for the mismatch in the near-infrared. This nebular continuum is known to affect the stellar component modeling in the presence of ionizing young massive stars \citep[e.g.,][]{byler17}. A method to properly address this feature would be to add a set of nebular continuum models to the fitted templates. However, we should stress that from our visual inspection of the nearly $\sim$4000 LVM pointings observed so far, there is evidence of this nebular continuum in just a very few percent. However, this perception may change when we stack individual observations, as we have done here.

So far, the mismatch between the observed spectra and the model created by the pure combination of RSP templates is well corrected by the final scaling performed by \pyf. This procedure was introduced to correct for the differences in spectrophotometric calibration between the data and the templates \citep{pipe3d,pypipe3d}. This is evident when comparing the residual of the subtraction of the best stellar model from the observed spectra (cyan solid line) and the residual once considering the stellar model, the emission lines, and the correction for the shape of the continuum (yellow solid line), in all panels of Fig. \ref{fig:O_spec}. We must recall that this correction for the shape of the continuum is only considered in the modeling of the emission lines, and therefore does not affect or improve the model of the stellar continuum (besides providing a better residual). { However, it clearly serves its purpose, as appreciated in Fig. \ref{fig:O_spec}, when comparing the best fitted model for the individual emission lines (blue line) and the fitted data (cyan line), that present an almost perfect overlap (for instance, in the insets included in the 3rd row panels).}

\subsection{Detected emission lines}
\label{sec:O_el}

The integrated spectrum has many clearly detected emission lines. This is evident in all panels of Fig. \ref{fig:O_spec}, but particularly in the panel in the third row and the first and third inset panels in the bottom row. There, the high-order emission lines of the Balmer and Paschen series, at $\sim$3700-3730 \AA\ and $\sim$8250-8350 \AA, respectively, are clearly seen. These lines are usually either too faint or are blended and therefore are not frequently considered in any modeling or analysis. Indeed, many of them are not included in the list of emission lines for which we derive the parameters in the current version of the DAP { adopting the non-parametric procedure} (indicated by the vertical orange lines in each panel). Whether it is worth the computational effort to include all these emission lines in the standard analysis performed by the LVM-DAP, or if we could implement a decision tree in the analysis to consider them only in deep observations, is still a decision to be made.

Nevertheless, there is a considerable number of emission lines analyzed both in the integrated spectrum (Fig. \ref{fig:O_spec}) and the combined frame (Fig. \ref{fig:O_rgb}). Of the 192 emission lines analyzed in the integrated spectrum, we detect 42 with a signal-to-noise ratio larger than 5. We list in Table \ref{tab:fe_orion} their integrated fluxes { derived using the non-parametric procedure descrived in Sec. \ref{sec:proc}} . Among them, we see not only the usual strong emission lines frequently explored to determine the ionization conditions in this kind of nebulae, such as \oii 3727,28, H$\beta$, \oiii, H$\alpha$, \nii 6548,83, or \sii 6717,31, but also rather weak emission lines, such as \oiii 4363, \nii 5755, or \siii 6312. This was already noticed when visually exploring the integrated spectrum, as discussed before.
{ For completeness we list the properties of the strongest emission lines derived using parametric procedure (Sec. \ref{sec:proc}) in Table \ref{tab:el_orion}. The flux intensities match those already presented in Table \ref{tab:fe_orion} for most of the emission lines within a few percent, in agreement with the results from the simulations discussed before. We note that this is the case not only for the integrated spectrum but also for the values fiber-to-fiber, as shown in Appendix \ref{sec:ec}. Furthermore, the velocity dispersion derived for the strongest ELs using the parametric fitting  ($\sigma$$\sim$ 0.7$-$1.14 \AA) is on the order of what is expected given the instrumental resolution (e.g., $\sigma_{\rm inst}\sim$0.7\AA\ at the wavelength range of \Ha). Thus, the emission lines are essentially unresolved in the integrated spectrum. In addition, the range of velocities reported for the different emission lines agree with the values previously reported in the literature (e.g. B00). Finally, the errors reported for the velocity ($\sim$2$-$6 \kms) and velocity dispersion ($\sim$1-3 \kms) agree with the expected values based on the simulations  (e.g., Table \ref{tab:el_sim}).}

\begin{figure*}
 \minipage{0.99\textwidth}
 \includegraphics[width=18.25cm,clip,trim=20 140 50 210]{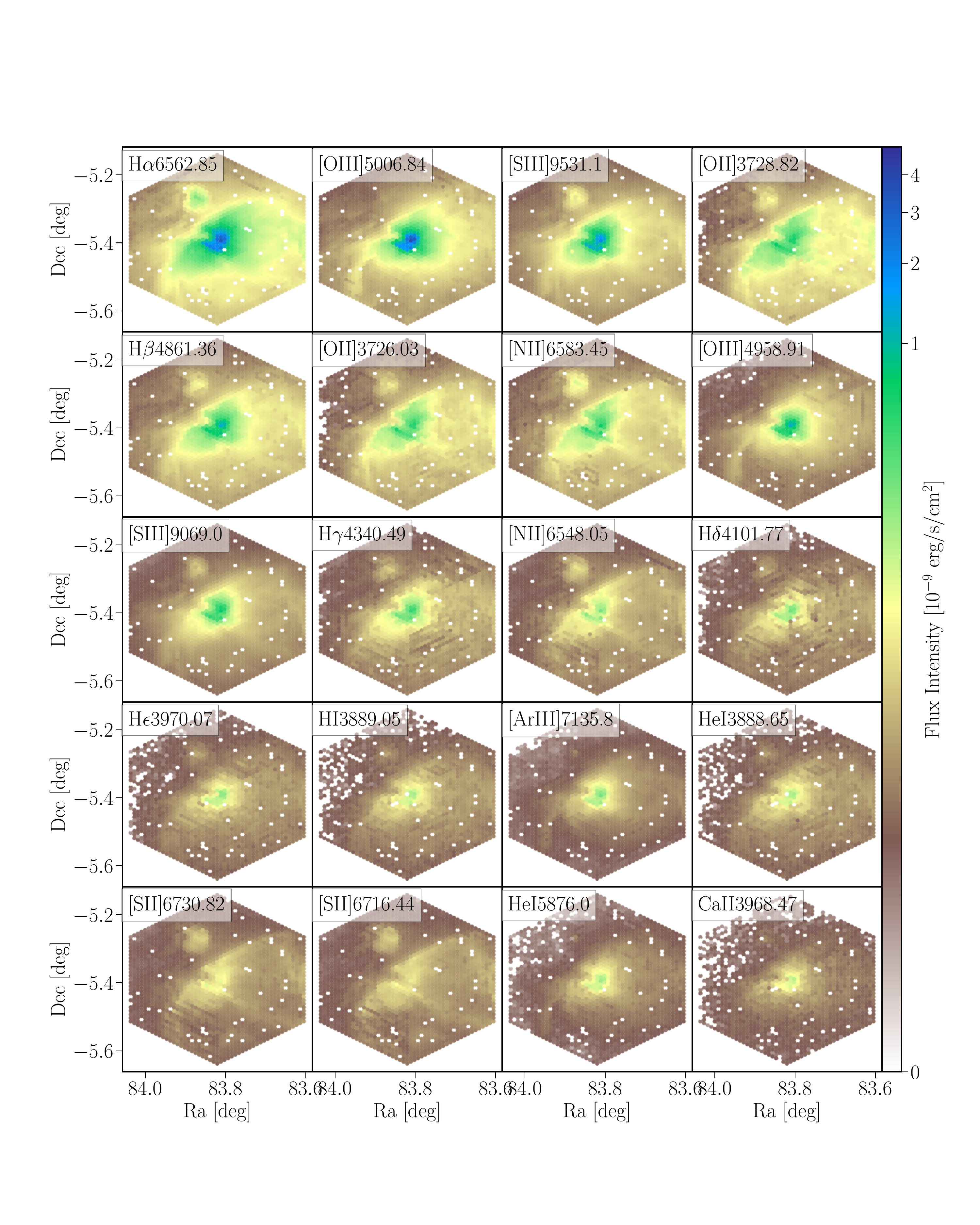}
 \endminipage
 \caption{Example of the analysis performed by LVM-DAP to recover the properties of the ionized gas emission lines. Each panel shows
 the distribution across the FoV of the LVM IFU of the flux intensity estimated by the weighted-moment procedure for the 20 brightest emission lines in the integrated spectrum (Fig. \ref{fig:O_spec}), extracted from the combined frame centred in the core of the Orion nebulae discussed in the text. The emission lines are ordered from the brightest (top-left) to the faintest (bottom-right). The legend in each panel indicates the represented emission line.}
 \label{fig:O_felines}
\end{figure*}

\subsection{Properties derived from the integrated spectrum}
\label{sec:ori_fprop}

It is beyond the scope of this paper to perform a detailed analysis of the physical properties of the ionized gas in the core of the Huygens region. This analysis of the physical and chemical conditions will be addressed in subsequent studies by the LVM collaboration (Méndez-Delgado et al. (in prep)). For now, we refer to previous studies such as \citet{ODell2017} for a more comprehensive view. However, in order to estimate the quality of the current dataset and the analysis performed by the DAP, we have estimated some diagnostic line ratios, comparing them with the results reported by O92, { B00},  S07, and other results from the literature. The data studied by O92 { and B00} were based on classical slit spectroscopy. However, they have better spectral resolution and cover a wider spectral range than the one presented in S07, allowing us to compare emission lines either blended or directly not observed by this latter study.

Table \ref{tab:O_phy}, left columns, shows the line ratios for the Balmer emission lines with respect to \hb\ based on the values listed in Tab. \ref{tab:fe_orion}. The nominal errors are not listed, but they range between 0.04 (for \ha/\hb) and 0.15 (for H9/\hb). All values are very similar to those reported by either S07, { O92 and B00}, excluding H9/\hb. This discrepancy could be due to the fact that H9 is likely to be blended with [NeIII] 3968, at the blue end of the spectral range covered by our data, and that is very strong in M42. In this regime, the spectrophotometric calibration may not be completely accurate, as discussed before. Furthermore, this wavelength range is crowded with other emission lines that could affect the reliability of the measured flux based on the moment analysis. Indeed, this line ratio is the one that presents the largest difference between the values reported in S07 and O92.

Adopting the \citet{cardelli:1989} extinction law with a ratio of total to selective extinction of R$_{V} =$ 5.5 \citep[e.g.][]{1982apj...261..195m,blag07} and assuming an intrinsic value of \ha/\hb~$=$ 2.86 \citep[case B recombination, T$_e\sim$10,000K in][]{1992ApJ...399..147P,osterbrock92}, we derive an extinction of A$_{\rm V}$ $=$ 0.78 mag. This value is considerably larger than the one reported by S07 (0.34 mag) and the one that could be estimated using the values reported by O92 (no extinction) { and B00 ($\sim$0.4 mag)}. However, it is very similar to the value reported by P92, based on FP observations of a similar region in the sky (as indicated before). As already discussed in S07, the comparison of reddening and other properties derived in different locations within M42 may present significant variations { (as illustrated by the differences found between O92 and B00)}, in particular in the vicinity of the Dark Bay at the North-East. Therefore, these differences are expected.

The right columns in Table \ref{tab:O_phy} list a set of line ratios sensitive to different physical properties of the ionized gas. Like in the case of the Balmer line ratios, we include the values derived using the integrated spectrum and those presented by S07, O92 { and B00}. For the line ratios sensitive to the electron density (n$_e$), we find a rough agreement, { with our value being $\sim$30\% larger than the literature ones. However, } P92 also reported a \sii 6717/6731 line ratio higher than S07, O92 { or B00}, more consistent with the value found using our data (0.68). Most probably, this indicates that the LVM dataset is covering a larger area and therefore is integrating a greater amount of diffuse, lower-density gas.


The rest of the line ratios listed in Tab. \ref{tab:O_phy} are { also} very similar to those reported in the literature. The largest discrepancy is found in the \oiii (4959+5007)/4363 line ratio when compared with the S07 value. However, those authors acknowledged that \oiii 4363 may be contaminated by a Hg night-sky pollution line in their spectrum, affecting the derived line ratio. Based on those line ratios, we estimate n$_e$ $\sim$1500 cm$^{-3}$, T$_e$ $\sim$ 7400-9500K, and [He$^+$/H$^+$] $\sim$ 0.08, values similar to those reported in the literature \citep[e.g.,][]{osterbrock92,1992ApJ...399..147P,sanchez07c,ODell2017}.

\subsection{Spatial distribution of the ionized gas emission lines}
\label{sec:ori_spa}

Most of the emission lines listed in Tab. \ref{tab:fe_orion} are also well detected beyond the central region selected to create the integrated spectrum. Figure \ref{fig:O_felines} shows the spatial distribution of the flux intensity of the 20 brightest emission lines among them. All emission lines have been represented using the same color scheme and flux scale for better comparison between them, being ordered from the brightest (top-left) to the faintest (bottom-right). It is apparent that while the faintest ones clearly reach the noise level in the outer regions of the IFU FoV (e.g., for CaII 3968 or HeI 5876), the brightest ones fill the entire area with a considerable S/N. The differences in the relative flux intensities between the different emission lines at different locations reflect the change of ionization conditions and ionized gas properties across the nebula. { These differences are the core of the information provided by the LVM, as it has been recently shown in \citet{kreckel24}.}

%
%

\section{Code Distribution} \label{sec:dist}

The { current version of the } DAP code { (version 1.0.0) } is distributed in the repository \url{https://github.com/sdss/lvmdap}. The distribution includes the main routine used to run the analysis {\sc lvm-dap-conf},  a set of examples of the configuration files required to run it,  basic instructions of how to install and run the code, and several additional Python notebooks to visualize both the LVM data and DAP data products. An example of the simulated dataset discussed in Sec. \ref{sec:sim_real} (Fig. \ref{fig:dapsim}),  the RSP templates described in Sec. \ref{sec:rsp_der}, { and an examples of the DAP dataproducts file described in Sec. \ref{sec:DP}} are available on the following webpage: \url{https://ifs.astroscu.unam.mx/sfsanchez/lvmdap/}.

\section{Summary and conclusions} \label{sec:conc}

In this article, we present the first version of the Local Volume Mapper Data Analysis Pipeline (DAP), a package to extract the stellar and emission line properties of the spectra provided by the LVM (and similar datasets). We describe the analysis sequence, including details on how the algorithm derives (i) the kinematic parameters of the stellar population ($v_\star$ and $\sigma_\star$), (ii) the dust attenuation (A$_{\rm V}$), (iii) the best model for the stellar component, and (iv) the parameters of a pre-defined set of emission lines.

A vast number of the algorithms included in this code have been inherited from the \pyp pipeline and the \pyf package. In particular, we adopted the two different methods to estimate the properties of the emission lines included in these packages: one fitting the emission lines with a set of Gaussian functions (parametric method) and another using a weighted moment analysis (non-parametric method). In parallel, we have introduced a new methodology to explore the stellar component to deal with the wide dynamical range of the number of stars sampled by the variable physical apertures of the LVM spaxels. This methodology performs a decomposition of the stellar spectrum using a template of characteristic stellar spectra well distinguished between them, which have an associated probability distribution function of physical parameters. We call this methodology and the corresponding stellar templates RSP (resolved stellar populations) to distinguish them from the SSP (synthetic stellar populations), usually adopted in the exploration of unresolved stellar spectra. We describe in detail how the RSP templates have been built, distributing the ones currently adopted by the DAP.

The reliability of the introduced procedures has been tested using (i) a set of idealized and realistic simulations, (ii) fitting the spectra of individual stars of known physical parameters, (iii) fitting the spectra resulting from the combination of a controlled number of known stars, and (iv) using a deep exposure in the core of the Orion nebulae to compare with previous results in the literature. As a result of this analysis, we found that, in general, the properties of the emission lines (flux, equivalent width, systemic velocity, and velocity dispersion) are well recovered by both methods ({ see Tables \ref{tab:el_sim} and \ref{tab:fe_orion}}). The main difference is that the currently adopted non-parametric method does not produce reliable results for strongly blended emission lines, i.e., those whose rest-frame central wavelengths are separated by less than the total FWHM of the emission lines (e.g., lines like \oii 3726, 3728). This method will be revised in future distributions of the code. So far, for these lines, we recommend using the results of the parametric analysis if available.

Regarding the average properties of the stellar component, the procedure produces reliable results when the stellar spectrum is dominated by a single star (or a set of stars with similar physical properties). The recovery of the T$_{eff}$ and [$\alpha$/Fe] is accurate and precise (particularly T$_{eff}$, that could be recovered within $\sim$5\%\ of the original value). For log($g$), we find a lack of accuracy in the low-gravity range ($<$3.5 dex). Finally, the recovery of [Fe/H] is accurate in general, but less precise than the rest of the parameters (with a range of errors of $\sim$30\%). 

The results are slightly different when analyzing stellar spectra resulting from the combination of stars of different properties. In this case, we find a constant offset (bias) between the input and recovered values for all the parameters (different for each parameter). The recovery is generally less accurate (and precise) than in the case of a single star. However, as the offsets are constant and follow a well defined linear trend they may be corrected in a statistical sense. Overall, this methodology is less precise in the recovery of the average stellar properties of an individual spectrum when multiple and different stars are included in the same aperture. We illustrate how the methodology can recover the probability distribution functions of the physical parameters of the stellar populations, although the potential of this approach should be more thoroughly investigated in future explorations.

The analysis of a deep spectrum extracted from the core of the bright Orion nebulae, extensively studied in the literature, allowed us to (i) explore the possible systematic artifacts in the LVM spectra themselves, (ii) study the faint spectral features (in general), both emission lines (e.g., auroral lines) and continuum emission (nebular emission), (iii) demonstrate that the current data reduction and analysis allow the production of scientific quality data (and products), and (iv) highlight what remains to be implemented in the current version of the DAP. In summary, we found that when the exposures are deep enough, the spectra lack strong systematic errors, presenting a neat and clean spectrum, without evident instrumental defects (jumps between the overlapping regions covered by each spectrograph) or inaccuracies due to sky subtraction. It is evident that the nebular continuum must be considered somehow in the modeling in future implementations of the DAP. However, even without considering it, the current approach effectively removes the underlying continuum, providing line ratios that generally agree with the results reported in the literature for the showcase examples explored (Orion and Rosetta nebulae).

Finally, we distribute the current version of the code for use by the community either to analyze future distributions of LVM data or to explore its use for similar datasets. We consider that the current version, although it requires clear improvements, is a valid starting point for the exploration of the unique dataset provided by the Local Volume Mapper.


\begin{acknowledgments}

SFS thanks the PAPIIT-DGAPA AG100622 project and CONACYT grant CF19-39578.  This work was supported by UNAM PASPA – DGAPA.

OE, KK, and JEMD gratefully acknowledge funding from the Deutsche Forschungsgemeinschaft (DFG, German Research Foundation) in the form of an Emmy Noether Research Group (grant number KR4598/2-1, PI Kreckel) and the European Research Council’s starting grant ERC StG-101077573 (“ISM-METALS"). OVE acknowledges funding from the Deutsche Forschungsgemeinschaft (DFG, German Research Foundation) -- project-ID 541068876.

G.A.B. acknowledges the support from the ANID Basal project FB210003. 

P. Garc\'ia is sponsored by the Chinese Academy of Sciences (CAS), through a grant to the CAS South America Center for Astronomy (CASSACA). He acknowledges support by the China-Chile Joint Research Fund (CCJRF No. 2312). CCJRF is provided by Chinese Academy of Sciences South America Center for Astronomy (CASSACA) and established by National Astronomical Observatories, Chinese Academy of Sciences (NAOC) and Chilean Astronomy Society (SOCHIAS) to support China-Chile collaborations in astronomy.

Funding for the Sloan Digital Sky Survey V has been provided by the Alfred P. Sloan Foundation, the Heising-Simons Foundation, the National Science Foundation, and the Participating Institutions. SDSS acknowledges support and resources from the Center for High-Performance Computing at the University of Utah. SDSS telescopes are located at Apache Point Observatory, funded by the Astrophysical Research Consortium and operated by New Mexico State University, and at Las Campanas Observatory, operated by the Carnegie Institution for Science. The SDSS web site is \url{www.sdss.org}.

SDSS is managed by the Astrophysical Research Consortium for the Participating Institutions of the SDSS Collaboration, including Caltech, The Carnegie Institution for Science, Chilean National Time Allocation Committee (CNTAC) ratified researchers, The Flatiron Institute, the Gotham Participation Group, Harvard University, Heidelberg University, The Johns Hopkins University, L’Ecole polytechnique f\'{e}d\'{e}rale de Lausanne (EPFL), Leibniz-Institut f\"{u}r Astrophysik Potsdam (AIP), Max-Planck-Institut f\"{u}r Astronomie (MPIA Heidelberg), Max-Planck-Institut f\"{u}r Extraterrestrische Physik (MPE), Nanjing University, National Astronomical Observatories of China (NAOC), New Mexico State University, The Ohio State University, Pennsylvania State University, Smithsonian Astrophysical Observatory, Space Telescope Science Institute (STScI), the Stellar Astrophysics Participation Group, Universidad Nacional Aut\'{o}noma de M\'{e}xico, University of Arizona, University of Colorado Boulder, University of Illinois at Urbana-Champaign, University of Toronto, University of Utah, University of Virginia, Yale University, and Yunnan University.

This work has made use of data from the European Space Agency (ESA) mission
{\it Gaia} (\url{https://www.cosmos.esa.int/gaia}), processed by the {\it Gaia}
Data Processing and Analysis Consortium (DPAC,
\url{https://www.cosmos.esa.int/web/gaia/dpac/consortium}). Funding for the DPAC
has been provided by national institutions, in particular the institutions
participating in the {\it Gaia} Multilateral Agreement.


\end{acknowledgments}

%

\vspace{5mm}






\appendix

\section{List of analyzed emission lines}
\label{app:elines}

Table \ref{tab:fe_long_list} list the emission lines included in the current implementation of the LVM-DAP analysis.
It lists both the emission lines adopted for the non-parametric and parametric explorations, as described in Sec. \ref{sec:proc}.

\begin{table*}
\begin{center}
\caption{Emission lines considered by the LVM-DAP analysis}
\begin{tabular}{lll|lll|lll|lll|lll}\hline\hline
{\tt \#I} &$\lambda$ (\AA) & Id &
{\tt \#I} &$\lambda$ (\AA) & Id &
{\tt \#I} &$\lambda$ (\AA) & Id &
{\tt \#I} &$\lambda$ (\AA) & Id &
{\tt \#I} & $\lambda$ (\AA) & Id \\
\hline
0 &  3686.83  &   HI        & 39 &  4416.27  &   [FeII]  & 78 &  5039.10  &   [FeII]  & 117 &  6087.00  &   [FeVII]  & 156 &  8300.99  &   [NiII]  \\ 
1 &  3691.56  &   HI        & 40 &  4452.11  &   [FeII]  & 79 &  5072.40  &   [FeII]  & 118 &  6300.30  &   {\bf [OI]}  & 157 &  8308.39  &   [CrII]  \\ 
2 &  3697.15  &   HI        & 41 &  4457.95  &   [FeII]  & 80 &  5107.95  &   [FeII]  & 119 &  6312.06  &   [SIII]  & 158 &  8345.55  &   HI  \\ 
3 &  3703.85  &   HI        & 42 &  4470.29  &   [FeII]  & 81 &  5111.63  &   [FeII]  & 120 &  6363.78  &   [OI]  & 159 &  8357.51  &   [CrII]  \\ 
4 &  3711.97  &   HI        & 43 &  4471.48  &   HeI     & 82 &  5145.80  &   [FeVI]  & 121 &  6374.51  &   [FeX]  & 160 &  8359.00  &   HI  \\ 
5 &  3726.03  &   {\bf [OII]} & 44 &  4474.91  &   [FeII]  & 83 &  5158.00  &   [FeII]  & 122 &  6435.10  &   [ArV]  & 161 &  8374.48  &   HI  \\ 
6 &  3728.82  &   {\bf [OII]} & 45 &  4485.21  &   [NiII]  & 84 &  5158.90  &   [FeVII]  & 123 &  6548.05  &   {\bf [NII]}  & 162 &  8392.40  &   HI  \\ 
7 &  3734.37  &   HI        & 46 &  4562.48  &   [MgI]  & 85 &  5176.00  &   [FeVI]  & 124 &  6562.85  &   {\bf H$\alpha$}  & 163 &  8446.00  &   OI  \\ 
8 &  3750.15  &   HI        & 47 &  4571.10  &   MgI]  & 86 &  5184.80  &   [FeII]  & 125 &  6583.45  &   {\bf [NII]}  & 164 &  8467.25  &   HI  \\ 
9 &  3758.90  &   [FeVII]   & 48 &  4632.27  &   [FeII]  & 87 &  5191.82  &   [ArIII]  & 126 &  6678.15  &   HeI  & 165 &  8498.02  &   CaII  \\ 
10 &  3770.63  &   HI       & 49 &  4658.10  &   [FeIII]  & 88 &  5197.90  &   [NI]  & 127 &  6716.44  &   {\bf [SII]}  & 166 &  8502.48  &   HI  \\ 
11 &  3797.90  &   HI       & 50 &  4685.68  &   HeII  & 89 &  5200.26  &   [NI] & 128 &  6730.82  &   {\bf [SII]}  & 167 &  8542.09  &   CaII  \\ 
12 &  3819.61  &   HeI      & 51 &  4701.62  &   [FeIII]  & 90 &  5220.06  &   [FeII]  & 129 &  6855.18  &   FeI  & 168 &  8545.38  &   HI  \\ 
13 &  3835.38  &   HI       & 52 &  4711.33  &   [ArIV]  & 91 &  5261.61  &   [FeII]  & 130 &  7005.67  &   [ArV]  & 169 &  8578.70  &   [ClII]  \\ 
14 &  3868.75  &   [NeIII]  & 53 &  4713.14  &   HeI  & 92 &  5268.88  &   [FeII]  & 131 &  7065.19  &   HeI  & 170 &  8598.39  &   HI  \\ 
15 &  3888.65  &   HeI      & 54 &  4724.17  &   [NeIV]  & 93 &  5270.30  &   [FeIII]  & 132 &  7135.80  &   [ArIII]  & 171 &  8616.96  &   [FeII]  \\ 
16 &  3889.05  &   HI       & 55 &  4733.93  &   [FeIII]  & 94 &  5273.38  &   [FeII]  & 133 &  7155.14  &   [FeII]  & 172 &  8662.14  &   CaII  \\ 
17 &  3933.66  &   CaII     & 56 &  4740.20  &   [ArIV]  & 95 &  5277.80  &   [FeVI]  & 134 &  7171.98  &   [FeII]  & 173 &  8665.02  &   HI  \\ 
18 &  3964.73  &   HeI      & 57 &  4754.83  &   [FeIII]  & 96 &  5296.84  &   [FeII]  & 135 &  7236.00  &   CII  & 174 &  8727.13  &   [CI]  \\ 
19 &  3967.46  &   [NeIII]  & 58 &  4769.60  &   [FeIII]  & 97 &  5302.86  &   [FeXIV]  & 136 &  7281.35  &   HeI  & 175 &  8750.47  &   HI  \\ 
20 &  3968.47  &   CaII     & 59 &  4774.74  &   [FeII]  & 98 &  5309.18  &   [CaV]  & 137 &  7290.42  &   [FeI]  & 176 &  8862.78  &   HI  \\ 
21 &  3970.07  &   H$\epsilon$ & 60 &  4777.88  &   [FeIII]  & 99 &  5333.65  &   [FeII]  & 138 &  7291.46  &   [CaII]  & 177 &  8891.88  &   [FeII]  \\ 
22 &  4026.19  &   HeI         & 61 &  4813.90  &   [FeIII]  & 100 &  5335.20  &   [FeVI]  & 139 &  7318.92  &   [OII]  & 178 &  9014.91  &   HI  \\ 
23 &  4068.60  &   [SII]       & 62 &  4814.55  &   [FeII]  & 101 &  5376.47  &   [FeII]  & 140 &  7323.88  &   [CaII]  & 179 &  9033.45  &   [FeII]  \\ 
24 &  4076.35  &   [SII]       & 63 &  4861.36  &   {\bf H$\beta$}  & 102 &  5411.52  &   HeII  & 141 &  7329.66  &   [OII]  & 180 &  9051.92  &   [FeII]  \\ 
25 &  4101.77  &   {\bf H$\delta$} & 64 &  4881.11  &   [FeIII]  & 103 &  5412.64  &   [FeII]  & 142 &  7377.83  &   [NiII]  & 181 &  9069.00  &   {\bf [SIII]}  \\ 
26 &  4120.81  &   HeI         & 65 &  4889.63  &   [FeII]  & 104 &  5424.20  &   [FeVI]  & 143 &  7388.16  &   [FeII]  & 182 &  9123.60  &   [ClII]  \\ 
27 &  4177.21  &   [FeII]      & 66 &  4893.40  &   [FeVII]  & 105 &  5426.60  &   [FeVI]  & 144 &  7411.61  &   [NiII]  & 183 &  9226.60  &   [FeII]  \\ 
28 &  4227.20  &   [FeV]       & 67 &  4905.35  &   [FeII]  & 106 &  5484.80  &   [FeVI]  & 145 &  7452.50  &   [FeII]  & 184 &  9229.02  &   HI  \\ 
29 &  4243.98  &   [FeII]      & 68 &  4921.93  &   HeI  & 107 &  5517.71  &   [ClIII]  & 146 &  7637.52  &   [FeII]  & 185 &  9266.00  &   OI  \\ 
30 &  4267.00  &   CII         & 69 &  4924.50  &   [FeIII]  & 108 &  5527.33  &   [FeII]  & 147 &  7686.19  &   [FeII]  & 186 &  9267.54  &   [FeII]  \\ 
31 &  4287.40  &   [FeII]      & 70 &  4930.50  &   [FeIII]  & 109 &  5577.34  &   [OI]  & 148 &  7686.90  &   [FeII]  & 187 &  9399.02  &   [FeII]  \\ 
32 &  4340.49  &   {\bf H$\Upsilon$} & 71 &  4942.50  &   [FeVII]  & 110 &  5631.10  &   [FeVI]  & 149 &  7751.06  &   [ArIII]  & 188 &  9470.93  &   [FeII]  \\ 
33 &  4358.10  &   [FeII]        & 72 &  4958.91  &   {\bf [OIII]}  & 111 &  5677.00  &   [FeVI]  & 150 &  7774.00  &   OI  & 189 &  9531.10  &   {\bf [SIII]}  \\ 
34 &  4358.37  &   [FeII]        & 73 &  4972.50  &   [FeVI]  & 112 &  5720.70  &   [FeVII]  & 151 &  7891.80  &   [FeXI]  & 190 &  9545.97  &   HI  \\ 
35 &  4359.34  &   [FeII]        & 74 &  4973.39  &   [FeII]  & 113 &  5754.59  &   [NII]  & 152 &  7999.85  &   [CrII]  & 191 &  9682.13  &   [FeII]  \\ 
36 &  4363.21  &   [OIII]        & 75 &  4985.90  &   [FeIII]  & 114 &  5876.00  &   HeI  & 153 &  8125.22  &   [CrII]  &  &    &  \\ 
37 &  4413.78  &   [FeII]        & 76 &  5006.84  &   {\bf [OIII]}  & 115 &  5889.95  &   NaI  & 154 &  8229.55  &   [CrII]  & &    &  \\ 
38 &  4414.45  &   [FeII]        & 77 &  5015.68  &   HeI  & 116 &  5895.92  &   NaI  & 155 &  8236.77  &   HeII  &   &    &  \\ 
\hline
\end{tabular}\label{tab:fe_long_list}
\end{center}
List of emission lines considered in the non-parametric analysis (i.e., moment analysis), where {\tt \#I} corresponds to the index (1st column) in the tables included in the NP\_EL\_B, NP\_EL\_R and NP\_EL\_I extensions of the DAP file described in Sec. \ref{sec:DP} and Tab. \ref{tab:hdu}. The emission lines considered in the parametric analysis (i.e., those included in the {PM\_ELINES} extension of the DAP file), are indicated with a bold font. {\bf The wavelength of the emission lines have been trimmed to the 2nd decimal to fit into the table. The actual values of the wavelength should be updated electronically in the list distributed within the DAP code.}
\end{table*}

{ 

\section{Comparison between parametric and non-parametric procedures}
\label{sec:ec}

\begin{figure*}
 \minipage{0.99\textwidth}
 \includegraphics[width=8.5cm,clip,trim=0 0 0 0]{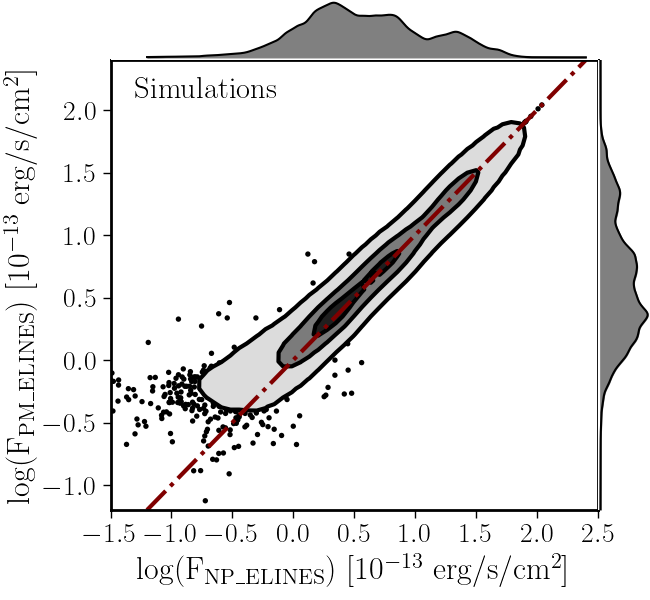}\includegraphics[width=8.5cm,clip,trim=0 0 0 0]{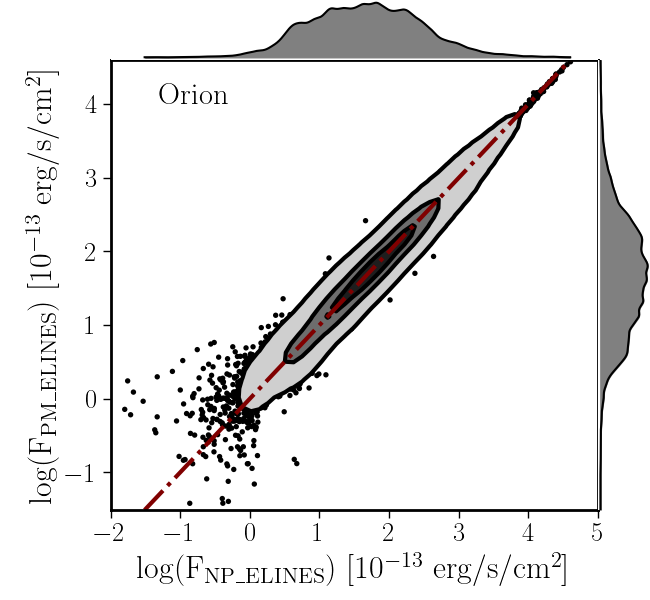}
 \endminipage
 \caption{ Comparison between the flux intensities for the emission lines derived using the non-parametric (NP\_ELINES) and parametric (PM\_ELINES) procedures described in Sec. \ref{sec:proc}, for the ad-hoc realistic simulations described in Sec. \ref{sec:sim_real} (left panel) and the Orion data discussed in Sec. \ref{sec:real}.{ Contours, symbols and lines in each panel have the same meaning as those in Fig. \ref{fig:dap_el}.}}
\label{fig:ec}
\end{figure*}

Throughout this article it has been indicated in different locations that the flux intensities derived using the parametric and non-parametric procedures described in Sec. \ref{sec:proc} are equivalent in most cases. This claim is based on previous experience using both procedures \citep[e.g.]
[]{pypipe3d,sanchez22}, which is confirmed by the different comparisons performed for the set of LVM data analyzed so far. To illustrate this result, we present in Fig. \ref{fig:ec} a comparison between the flux intensities recovered using both methods for the set of emission lines in common (as listed in Tab. \ref{tab:fe_long_list}) for two datasets: (i) the realistic simulation described and discussed in Sec. \ref{sec:sim_real} and (ii) the Orion nebula data described in Sec. \ref{sec:real}. In both cases the two estimations of the emission line fluxes agree within a few percent each other, without any significant systematic offset. 
}


\bibliography{my_bib.bib}{}
\bibliographystyle{aasjournal}



\end{document}